\documentclass[12pt]{article}
\usepackage{amsfonts,amsthm,amsmath,amssymb,mathrsfs,url,pict2e}
\title{The Quantum Formalism and the GRW Formalism}
   
\author{
Sheldon Goldstein\footnote{Departments of Mathematics, Physics and
     Philosophy, Rutgers University, Hill Center,  
     110 Frelinghuysen Road, Piscataway, NJ 08854-8019, USA.
     E-mail: oldstein@math.rutgers.edu},\ 
Roderich Tumulka\footnote{Department of Mathematics,
     Rutgers University, Hill Center, 
     110 Frelinghuysen Road, Piscataway, NJ 08854-8019, USA.
     E-mail: tumulka@math.rutgers.edu},\ 
 and Nino Zangh\`\i\footnote{Dipartimento di Fisica dell'Universit\`a
     di Genova and INFN sezione di Genova, Via Dodecaneso 33, 16146
     Genova, Italy. E-mail: zanghi@ge.infn.it}
}
\date{August 24, 2012}

\newcommand{\be}{\begin{equation}}
\newcommand{\ee}{\end{equation}}
\newcommand{\Hilbert}{\mathscr{H}}

\renewcommand{\Re}{\mathrm{Re}}

\newcommand{\EEE}{\mathbb{E}}

\newcommand{\PPP}{\mathbb{P}}
\newcommand{\SSS}{\mathbb{S}}
\newcommand{\RRR}{\mathbb{R}}
\newcommand{\CCC}{\mathbb{C}}

\newcommand{\scp}[2]{\langle #1|#2 \rangle}
\newcommand{\Bscp}[2]{\Bigl\langle #1\Big|#2 \Bigr\rangle}
\newcommand{\pr}[1]{|#1 \rangle \langle #1|}
\newcommand{\ket}[1]{|#1 \rangle}
\newcommand{\bra}[1]{\langle #1|}
 
\DeclareMathOperator{\tr}{tr}
\newcommand{\sphere}{\mathbb{S}}
\newcommand{\E}{\mathscr{E}} 
\newcommand{\G}{G} 
\newcommand{\salg}{\mathcal{A}} 
\newcommand{\GRW}{\mathrm{GRW}}
\newcommand{\Qu}{\mathrm{Qu}}
\newcommand{\sys}{{\mathrm{sys}}}
\newcommand{\app}{{\mathrm{app}}}
\newcommand{\env}{{\mathrm{env}}}

\newcommand{\Values}{\mathscr{Z}} 
\newcommand{\I}{\mathscr{I}} 
\newcommand{\Times}{\mathscr{T}} 
\newcommand{\cpm}{\mathscr{C}} 
\newcommand{\acpm}{\mathscr{S}} 
\newcommand{\bcpm}{\mathscr{A}} 
\newcommand{\ccpm}{\mathscr{B}} 
\newcommand{\dm}{\rho} 
\newcommand{\trclop}{\rho} 
\newcommand{\Lab}{\mathscr{L}} 
\newcommand{\subLab}{\Lab_\sys} 
\newcommand{\region}{R}
\newcommand{\wf}{\Psi} 
\newcommand{\ti}{s}
\newcommand{\tf}{t}

\newcommand{\Tf}{T}
\newcommand{\tauf}{\tau}

\theoremstyle{plain}\newtheorem{thm}{Theorem}

\newcommand{\y}[1]{{#1}}

\begin{document}
\maketitle
\begin{abstract}
The Ghirardi--Rimini--Weber (GRW) theory of spontaneous wave function collapse is known to provide a quantum theory without observers, in fact two different ones by using either the matter density ontology (GRWm) or the flash ontology (GRWf). Both theories are known to make predictions different from those of quantum mechanics, but the difference is so small that no decisive experiment can as yet be performed. While some testable deviations from quantum mechanics have long been known, we provide here something that has until now been missing: a \emph{formalism} that succinctly summarizes the empirical predictions of GRWm and GRWf. We call it the GRW formalism. Its structure is similar to that of the quantum formalism but involves different operators. In other words, we establish the validity of a general algorithm for directly computing the testable predictions of GRWm and GRWf. We further show that some well-defined quantities \emph{cannot} be measured in a GRWm or GRWf world.

\bigskip

\end{abstract}

\newpage
\tableofcontents

\section{Introduction}

This paper is about the derivation of statistical predictions for macroscopic behavior from a specific microscopic physical model. That is common in statistical physics. A bit unusual, though, is that the microscopic model we study was developed for explaining quantum mechanics. Indeed, in order to obtain a \emph{quantum theory without observers}, and thus to solve the measurement problem and other paradoxes of quantum mechanics, it has been suggested that one should incorporate \emph{spontaneous collapses of the wave function} into the laws of nature by replacing the Schr\"odinger evolution with a stochastic and nonlinear evolution law. The simplest and best known proposal for such a law is due to Ghirardi, Rimini, and Weber (GRW) \cite{GRW86,Bell87} (see \cite{BG03} for a review of collapse theories). This is the framework we are concerned with in this paper. Our goal is to obtain the axioms of quantum mechanics as theorems in the GRW theory.

To complete the GRW theory, one needs to specify a choice of \emph{primitive ontology} (PO) and a law determining how the wave function governs the PO (see \cite{AGTZ06} for a discussion). Two possibilities for the PO and its law have been proposed: the \emph{matter density ontology} and the \emph{flash ontology}, leading to two different theories we shall denote GRWm and GRWf, respectively, in the following. We recall their definitions in Section~\ref{sec:GRWmf}. It is known that GRWm and GRWf are \emph{empirically equivalent}, i.e., that they make exactly and always the same empirical predictions \cite{AGTZ06}; we describe the reasons in Section~\ref{sec:ee}, in fact more carefully than in \cite{AGTZ06}. The first purpose of this paper is to derive what these predictions actually are. By ``empirical predictions'' we mean those predictions that can be tested in experiment; we will see that there are also predictions that cannot be so tested. The totality of all empirical predictions of a theory we also call the \emph{empirical content} of the theory.

While GRWm and GRWf are designed to imitate quantum mechanics, they have been known since their inception to deviate from quantum mechanics, and a number of particular predictions differing from those of quantum mechanics have been identified \cite{GRW86,Rae90,PS94,JPR04,Adl07} (for overviews of proposals to test GRW theories against quantum mechanics, see \cite{BG03,Adl07,FT12}).  Nonetheless, in practice the GRW theories tend to agree extremely well with quantum mechanics: for small systems, collapses are too rare to be noticed, while the breakdown of macroscopic superpositions is hard to test because of decoherence (for explicit figures about how closely GRW theories agree with quantum mechanics, see \cite{BS07}). Thus, the theorems we prove yield not precisely the axioms of quantum mechanics, but something very close.

Is there a general scheme of predictions, or an algorithm for directly calculating the predictions, of the GRW theories, in particular where they differ from quantum mechanics? In this paper, we answer this question in the positive and provide a formalism, which we call the \textit{GRW formalism}, summarizing the empirical predictions of the GRWm and GRWf theories. (Indeed, GRWm and GRWf give rise to the same formalism; they have to, because they are empirically equivalent.) The GRW formalism is analogous to the quantum formalism of orthodox quantum theory that describes the results of quantum experiments in terms of operators as observables, spectral measures, and the like. The main difference between the two formalisms lies in the relevant operators.

We make explicit the \emph{law of operators} for both the quantum and 
the GRW formalism, i.e., the law that determines which operators are 
associated with a given experiment. An analysis of the general 
conditions under which the GRW predictions are close to the quantum 
predictions is provided in Section~\ref{sec:deviations}.

In Section~\ref{sec:Tf} we provide a formulation of both the quantum 
and the GRW formalism that allows for \emph{collapse at random times}, 
i.e., for collapse of the quantum state at the end of an experiment 
whose duration is determined not in advance but by the experiment 
itself. For example, consider a two-stage experiment: in the first stage 
one waits for a detector to click (and measures the time when it 
clicks), in the second stage, right afterwards, one conducts some 
quantum measurement on the particle that triggered the detector; the 
application of the formalism to the second stage requires that the 
quantum state of the particle gets collapsed appropriately in the first 
stage.

Some questions that possess a unique answer in a GRW world  cannot be 
answered by the inhabitants of that world by means of any experiment. 
The following question is presumably of this type: How many collapses 
occurred in a certain system during the time interval $[t_1,t_2]$? We 
discuss this topic in Section~\ref{sec:genuine} and more deeply in a 
future work \cite{grw3C}.

In Appendix~\ref{sec:diagram} we describe a diagram notation well-suited 
for certain types of calculations that arise in this paper, concerning 
the time evolution of the density matrix of composite systems.

An innovation of this paper, besides the formulation of the GRW 
formalism, concerns the nature of the argument used in deriving it: the 
argument is based on the primitive ontology of the theory.

\subsection{A First Look at the GRW Formalism}

The GRW formalism can be formulated in a way similar to the formalism of quantum mechanics using operators in Hilbert space. We will give the complete formulation in Section~\ref{sec:GRWformalism}. Put succinctly, the difference between the quantum and the GRW formalism is
\[
  \text{different evolution, different operators}.
\]
``Different evolution'' means that the unitary Schr\"odinger evolution is replaced by a master equation for the density matrix $\dm_t$ (a Lindblad equation, or quantum dynamical semigroup):
\begin{equation}\label{M}
  \frac{d \dm_t}{d t} = -\tfrac{i}{\hbar} [H,\dm_t] 
  + \lambda \sum_{k=1}^N \int d^3 x \, \Lambda_k (x)^{1/2} \,
  \dm_t \, \Lambda_k(x)^{1/2} - N\lambda \dm_t\,.
\end{equation}
For readers who are not familiar with this type of equation, we note that the term $-\tfrac{i}{\hbar} [H,\dm_t]$ represents the unitary evolution, with $H$ the Hamiltonian, while the further terms, the deviation from the unitary evolution, have the effect that the evolution \eqref{M} transforms ``pure states into mixed states,'' i.e., transform density matrices that are 1-dimensional projections into ones that are not. Equation \eqref{M} holds for the density matrix $\dm_t$ corresponding to the probability distribution of the random GRW wave function $\wf_t$ arising from a fixed initial wave function $\wf_{t_0}$. Concerning the notation, $\lambda>0$ is a constant, and the positive operators $\Lambda_k(x)$ are the collapse rate operators (see Section~\ref{sec:GRWmf} for the definition).

``Different operators'' means that ``observables'' are associated with different operators than in quantum mechanics. This requires some explanation. A precise statement (which forms a crucial part of the GRW formalism) is that \emph{with every experiment $\E$, there is associated a positive-operator-valued measure (POVM) $E(\cdot)$ such that the probability distribution of the random outcome $Z$ of $\E$, when performed on a system with density matrix $\dm$, is given by}
\begin{equation}\label{P}
  \PPP(Z \in B) = \tr \bigl( \dm \, E(B) \bigr)
\end{equation}
for all sets $B$.\footnote{Here $\PPP(Z \in B)$ denotes the probability of the event $Z \in B$; sets are always assumed to be measurable. The notion of ``POVM'' is defined in Section~\ref{sec:POVM}.} This statement, the \emph{main theorem about POVMs}, is valid in quantum mechanics as well as in GRW theories, but the POVM $E^\GRW(\cdot)$ associated with $\E$ in a GRWm or GRWf world is different from the POVM $E^\Qu(\cdot)$ associated with $\E$ in quantum mechanics. We prove this statement in Section~\ref{sec:emergeop}. However, we do not compute any specific operators for specific experiments, but derive only an abstract and general characterization of $E^\GRW(\cdot)$.

When talking about \emph{every} experiment, we mean that any possible future advances of technology are included. The assumptions that define our concept of ``experiment'' are: it involves a \emph{system} (the object on which the experiment is performed) and an \emph{apparatus}; it is possible to consider the same experiment for different states of the system, whereas changing the apparatus counts as considering a different experiment; at the time at which the experiment begins, the system and the apparatus are not entangled.

Some colleagues that we have discussed this topic with have found it difficult to imagine how GRW could lead to different operators. When speaking of different operators, we were asked, does that mean that the momentum operator is no longer $-i\hbar\nabla$? No, it does not mean that. It means that, given any experiment in a quantum world, one can consider the same experiment in a GRWm or GRWf world, and the statistics of the outcome of that experiment are different from those in quantum mechanics---given by a different operator, or different POVM. Which operator should be called the ``momentum operator'' remains a matter of convention, and indeed there are reasons to call $-i\hbar\nabla$ the ``momentum operator'' also in the GRW theories.\footnote{Some ``observables'' of the quantum formalism---the momentum, angular momentum, and energy operators---are the generators of symmetries of the theory, such as translation, rotation, and time translation invariance. By virtue of Noether's theorem, then, they commute with the Hamiltonian. Since GRWm and GRWf, too, are translation, rotation, and time translation invariant (if the interaction potential is), the same self-adjoint operators occur here in the role of generators of symmetries (and commute with the Hamiltonian), even though a particular experiment that ``measures,'' in quantum mechanics, momentum, angular momentum, or energy may, in the GRW formalism, be associated with different operators.} Similarly, \y{it might be convenient to} say that the ``position observable'' is the same in the GRW theories as in quantum mechanics, even though concrete experimental designs for ``measuring position'' may lead to different outcome statistics than in quantum mechanics.

We were also asked, when speaking of different operators, whether we refer to the Heisenberg picture? No, we do not. The question means this: If the time evolution is not unitary then the Heisenberg picture (or whatever replaces it for a master equation such as \eqref{M}) should attribute to all observables different operators than standard quantum mechanics. But the ``different operators'' arise \emph{even} in the Schr\"odinger picture: If the observation of the system (i.e., the period of its interaction with the apparatus) begins at time $\ti$ and ends at $\tf$, then one is supposed, according to the GRW formalism, to evolve the system's density matrix until time $\ti$ using \eqref{M} in the Schr\"odinger picture, and insert into the formula \eqref{P} the resulting $\dm_\ti$, corresponding to what one feeds into the apparatus.\footnote{But some connection with the Heisenberg picture exists indeed: keep in mind that the main theorem about POVMs concerns \emph{any} experiment $\E$; for example, $\E$ could consist of waiting for a while $\Delta t$ and then ``measuring position.'' Then, the quantum operator associated with $\E$ is the Heisenberg-evolved position operator, $\hat{Q}_\E = e^{iH\Delta t}\hat{Q}e^{-iH\Delta t}$, and the reader \y{might well expect} that in GRWm or GRWf there is a different operator (in fact, a POVM) associated with $\E$.}

Maybe the reason why many physicists find it difficult to understand that the GRW formalism involves different operators arises from regarding the operators of quantum mechanics as something that came into the theory by means of a second postulate besides the Schr\"odinger equation, the \emph{measurement postulate}. From such a picture one might expect that the measurement postulate should remain unchanged, and, hence, also the operators, even when the Schr\"odinger equation is modified. The GRW perspective, however, forces us to proceed differently since it contains no measurement postulate, and \y{its} predictions must be derived instead from postulates about the primitive ontology. This makes it evident that the measurement postulate and the Schr\"odinger equation actually never were independent, and that the operators depend on the evolution law, for example because the experiment's outcome depends on the evolution law of the apparatus. The GRW perspective also forces us to make precise what it \emph{means} to say that a certain observable \y{is associated with} operator $A$. We take it to mean that $A$ \emph{encodes the outcome statistics}, in the sense that the relevant experiment has outcome statistics given by \eqref{P} with $E(\cdot)$ the spectral projection-valued measure (PVM) of $A$.

The master equation \eqref{M}, or very similar equations, also arise in the theory of decoherence \cite{Vac07}. As a closely related fact, the GRW formalism would in principle also hold in a hypothetical quantum world in which decoherence is inevitable and affects every system in the same way, corresponding to \eqref{M}. (In practice, of course, decoherence, due to interaction with the environment, cannot correspond to \eqref{M} in exactly the same way for every system because different systems have different environments and interact with their environments in different ways.) Let us underline the difference between deriving the GRW formalism from the quantum formalism together with the right dose of decoherence corresponding to \eqref{M}, and deriving it from GRWm or GRWf: A derivation starting from quantum mechanics would \emph{assume} statements about the outcomes of experiments (the measurement postulate) to deduce other statements about the outcomes of experiments. When starting from GRWm or GRWf, in contrast, we assume statements about the primitive ontology, and derive that, e.g., pointers point in certain directions.

It is an interesting side remark that Bohmian mechanics \cite{Bohm52,Bell87b} can be so modified as to become empirically equivalent to GRWm and GRWf. This modified version is described in \cite{grw3B} under the name ``MBM.'' Its empirical content is also summarized by the GRW formalism. As a consequence, the empirical content of the GRW theories can as well be obtained with a particle ontology, and is not limited to the flash and matter density ontologies.

\subsection{Role of the Primitive Ontology}

What is the connection between empirical predictions and primitive ontology (PO)? \y{The} PO is described by the variables $\xi$ giving the distribution of matter in space and time. \y{Thus,} a statement like ``the experiment $\E$ has the outcome $z$'' should mean that the PO of the apparatus indicates the value $z$. For example, if the apparatus displays the outcome by a pointer pointing to a particular position on a scale, what it means for the outcome to be $z$ is that the matter of the pointer is, according to the PO, in the configuration corresponding to $z$. Thus, the outcome $Z$ is a function of the PO,
\begin{equation}
  Z = \zeta(\xi)\,.
\end{equation}

Precursors of our treatment of the connection between predictions and PO can be found in \cite{Bell87,Gol98,Tum04,Tum05,CDT05,AZ05,AGTZ06,BGS06}, in some of which this connection was implicit, or hinted at, or briefly mentioned. In Bohmian mechanics \cite{Bohm52,Bell87b}, a similar connection between PO and the empirical predictions was explicitly made in \cite{DGZ04}; however, researchers working on Bohmian mechanics have essentially always been aware of this connection---much in contrast to those working on collapse theories, who tended to focus on the wave function and forget about \y{any} PO.

The fact that GRWm and GRWf have the same formalism, despite their difference in PO, may suggest that the PO is not so relevant after all. That is true for practical applications which require working out some predicted values, but not for the theoretical analysis of GRW theories, for their logical structure, or for their definition, as the considerations in this paper exemplify.

\subsection{Status of the Derivation}

It may seem as if the GRW formalism were a rather trivial consequence of the master equation \eqref{M}. So it is perhaps useful to make a list of what is nontrivial about our derivation of the GRW formalism:
\begin{itemize}
\item It is not a priori clear that a GRW formalism should exist.
 \begin{itemize}
 \item The existence of a GRW formalism had not been noticed for 20 years.
 \item Since the predictions of GRWm and GRWf deviate from those of quantum mechanics, it is not obvious that they can be summarized by any small number of simple rules.
 \item The derivation of the GRW formalism has a status similar to that of the quantum formalism from Bohmian mechanics (see, e.g., \cite{DGZ04}), a result implying in particular that there is no possibility of experimentally testing Bohmian mechanics against standard quantum mechanics. If that claim is non-obvious (after all, some authors have claimed the contrary), then so should be the GRW formalism.
 \item The non-linearity of the GRW evolution of the wave function $\wf_t$ might have suggested against the existence of a GRW formalism using linear operators. On the other hand, the master equation \eqref{M} is linear in $\dm_t$, a crucial fact for deriving the GRW formalism. Still, this fact alone does not imply the GRW formalism.\footnote{For example, we do not know of a way of deriving the GRW formalism from GRWm other than exploiting the empirical equivalence to GRWf (or MBM \cite{grw3B}), even though \eqref{M} is valid in GRWm.}
 \end{itemize}
\item Our assertion about the GRW formalism concerns the PO. In detail, it states that the matter density function $m(x,t)$ of GRWm and the set $F$ of flashes in GRWf are such that macroscopic apparatuses display certain results with certain probabilities. 
 \begin{itemize}
 \item Our derivation of the GRW formalism is based on an analysis of the behavior of the PO. Such an analysis was not done in \cite{BGS06,BS07}.
 \item Our derivation applies to the matter density ontology and to the flash ontology. We do not make claims for any other ontology.\footnote{However, there are reasons why every reasonable ontology suitable for the stochastic GRW wave function evolution law should lead to the same empirical predictions. Similarly, the empirical contents of CSLm, the Continuous Spontaneous Localization theory \cite{Pe89,GPR90,BG03} with the matter density ontology, or with any other reasonable ontology, can presumably be summarized by a formalism very similar to the GRW formalism.}
 \item The defining laws of GRWm and GRWf, unlike the ordinary axioms of quantum mechanics, do not refer to observations, but to the wave function and the PO. Thus, the empirical predictions are not immediate from the defining laws of the theory but require a derivation.
 \item To the extent that it is not obvious how the PO variables (\y{such as} $m(x,t)$ and $F$) behave, it is not obvious how macroscopic apparatuses (built out of the elements of the PO) behave.
 \item It has often been noted that there are situations in which \y{the PO variables (such as} $m(x,t)$ and $F$) behave in an unexpected, surprising, or counter-intuitive way. (See, e.g., \cite[p.~347]{BG03}, \cite[footn.~5]{AGTZ06}.)
 \end{itemize}
\item Every physicist knows rules for what can be concluded about measurement results if the wave function is such-and-such. These rules, however, cannot be used in the derivation of the GRW formalism, partly because the GRW theories are not quantum mechanics, and partly because it is the aim of the derivation (and of this paper) to \emph{deduce}, and not to presuppose, rules for the results of experiments. 
 \begin{itemize}
 \item Our derivation makes no use of the rules of standard quantum mechanics for predicting results of experiments given the wave function.
 \item Our derivation makes no use of any customs of standard quantum mechanics for how to interpret or use wave functions.
 \item In particular, operators as observables \emph{emerge} from an analysis of the GRW theories, they are not \emph{postulated}; in fact, they are not even mentioned in the definition of the GRW theories.
 \item Certain wave functions may easily suggest certain macro-states, but this does not mean that the configuration of the PO looks like this macro-state. Our derivation makes no use of such suggestive assumptions. 
 \end{itemize}
\item As a consequence of our analysis, there are severe limitations on the epistemic access to microscopic details of the PO variables $m(x,t)$ or $F$. In other words, there are limitations to the extent to which one can measure $m(x,t)$ or $F$. This fact can be regarded as an instance of surprising behavior of the PO (as mentioned above), and underlines that it is not obvious which functions of the PO are observable.
\end{itemize}

The issue we mentioned in the last item of the list deserves more comment. It turns out to be impossible to measure, with any reasonable microscopic accuracy, the matter density $m(x,t)$ in GRWm (or, presumably, the set $F$ of flashes in GRWf), unless information about the wave function of the system is available. Limitations on the observers' access to $m(x,t)$ were described before in \cite{BGG95}; we describe here several similar limitations. As a particular example, one might wish to measure the number of collapses that occur in a certain system (e.g., a tiny drop of water) during a chosen time interval, in analogy for example to the measurement of the number of radioactive decay events in a sample of radioactive matter. Heuristic considerations suggest, perhaps surprisingly, that it is impossible to measure the number of collapses, with any accuracy and reliability better than what one could estimate without any measurement at all. In other words, the precise number of collapses is \emph{empirically undecidable}, and thus GRWm and GRWf entail sharp \emph{limitations to knowledge}. In a GRWm or GRWf world, certain facts are kept secret from its inhabitants. Note that this situation does not arise from anything like a conspiratorial character of the theory, but simply as a consequence of the defining equations; after all, we do not make \emph{postulates} about what can or cannot be measured but \emph{analyze} the theory. Similar limitations to knowledge are known for Bohmian mechanics, where for example it turns out to be impossible to measure the (instantaneous) velocity of a particle \cite{DGZ04,DGZ08}, unless information about the wave function is available; as another example, it turns out to be impossible to distinguish empirically between certain different versions of Bohmian mechanics (see \cite{GTTZ05b} for a discussion).

A question we do not address here is how to do \emph{scattering theory} for GRW theories. But we briefly state the problem. Normal quantum scattering theory (see, e.g., \cite{scatt}) involves limits $t\to \infty$, which would be inappropriate in GRW theories because one consequence of GRW theories is long-run ``universal warming,'' since every collapse tends to increase energy, as it makes the wave function narrower in the position representation and therefore wider in the momentum representation. In the limit $t\to\infty$, scattered wave packets in a GRW world would therefore always end up with infinite energy, and uniformly distributed over all spatial directions. From a practical point of view, the time scale of free flight in real scattering experiments ($\sim 10^{-2}$ s) is much smaller than the time scale of universal warming ($\sim 10^{15}$ years \cite[p.~481]{GRW86}), usually even much smaller than the time scale of collapse ($\sim 10^{8}$ years), but much larger than the time scale of the interaction process. Thus, a simple and quite appropriate method of predicting the scattering cross section in a GRW world is to take the limit $t\to\infty$ for the \emph{unitary} evolution, which is the dominant part of the evolution of the wave function $\wf_t$ over the relevant time scale. But this is to ignore the difference between the predictions of GRW theories and quantum mechanics for scattering theory, and the question remains how to compute GRW corrections to the quantum formulas for scattering cross sections.

Finally, although the GRW formalism is valid for both GRWm and GRWf, the status of the derivation is very different for the two theories. While we derive the GRW formalism as precise theorems from GRWf, we do not know of a similar derivation from GRWm. In fact, the only way we know of to derive it for GRWm is by exploiting the empirical equivalence with GRWf, and the argument for the empirical equivalence is not as mathematical in character as the derivation of the GRW formalism from GRWf.

\section{The GRWm and GRWf Theories}
\label{sec:GRWmf}

GRWm was essentially proposed by Ghirardi et al.\ \cite{BGG95} and Goldstein \cite{Gol98}, and taken up in \cite{BG03,AZ05,Mau05,CDT05,Tum06c,AGTZ06,BGS06,BS07}. GRWf was proposed by Bell in \cite{Bell87} and taken up in \cite{Bell89,kent,Gol98,Tum04,AZ05,Mau05, CDT05,Tum05, Tum06c,AGTZ06,Tum07}. For a detailed discussion of these two choices of PO see \cite{AGTZ06}. Both GRWm and GRWf are non-relativistic theories. The relativistic GRWf theory proposed in \cite{Tum04} has a more complex mathematical structure than GRWf and is not covered by the considerations in this paper. A discrete version of the flash ontology was proposed for collapse theories on lattices by Dowker et al.\ \cite{Fay02,Fay03,Fay04}.

\subsection{The GRW Jump Process in Hilbert Space}
\label{sec:GRWprocess}

In both GRWm and GRWf the evolution of the wave function follows, instead of the Schr\"odinger equation, a stochastic jump process in Hilbert space, called the GRW process. We shall summarize this process as follows.

Consider a quantum system of (what would normally be called) $N$ ``particles,'' described by a wave function $\wf = \wf(q_1, \ldots,q_N)$, $q_i\in \RRR^3$, $i=1,\dots, N$.  For any point $x$ in $\RRR^3$, define on the Hilbert space of the system  the \emph{collapse rate operator}
\begin{equation}\label{eq:collapseoperator}
  \Lambda_i (x) =\frac{1}{(2\pi \sigma^2)^{3/2}}\, e^{-\frac{( \widehat{Q}_i-x)^2}
  {2\sigma^2}}\,,
\end{equation}
where  $\widehat{Q}_i$ is the position operator of  ``particle'' $i$. Here $\sigma$ is a new constant of nature of order $10^{-7}$m.

Let $\wf_{t_0}$ be the initial wave function, i.e., the normalized wave function at some time $t_0$ arbitrarily chosen as initial time. Then $\wf$ evolves in the following way: 
\begin{enumerate}
\item It evolves unitarily, according to Schr\"odinger's equation, until a random time $T_1= t_0 + \Delta T_1$, so that
\begin{equation}
  \wf_{T_1}= U_{\Delta T_1} \wf_{t_0},
\end{equation}
where $U_t$ is the unitary operator $U_t=e^{-\frac{i}{\hbar}Ht}$ corresponding to the standard Hamiltonian $H$ governing the system, e.g., given, for $N$ spinless particles, by
\begin{equation}\label{eq:H}
  H=-\sum_{k=1}^N\frac{\hbar^2}{2m_k}\nabla^2_k+V,
\end{equation}
where $m_k$, $k=1, \ldots, N$, are the masses of the particles, and $V$ is the potential energy function of the system.  $\Delta T_1$ is a random time distributed according to the exponential distribution with rate $N\lambda$ (where the quantity $\lambda$ is another constant of nature of the theory,\footnote{Pearle and Squires \cite{PS94} have argued that $\lambda$ should be chosen differently for every ``particle,'' with $\lambda_i$ proportional to the mass $m_i$.} of order $10^{-15}$ s$^{-1}$).
\item At time $T_1$ it undergoes an instantaneous collapse with random center 
 $X_1$ and random label $I_1$ according to
\begin{equation}
  \wf_{T_1} \mapsto\wf_{T_1+}= \frac{\Lambda_{I_1} (X_{1})^{1/2}\wf_{T_1}}
  {\| \Lambda_{I_1} (X_{1})^{1/2} \wf_{T_1} \|}.
\end{equation}
$I_1$ is chosen at random in the set $\{1, \ldots, N\}$ with uniform distribution. The center  of the collapse 
$X_1$ is chosen randomly with probability distribution
\begin{equation}\label{p}
  \PPP(X_1\in dx_{1}|  \wf_{T_1}, I_1=i_1) =
  \scp{\wf_{T_1}}{\Lambda_{i_1}(x_1)|\wf_{T_1}} dx_{1}=
  \|\Lambda_{i_1} (x_1)^{1/2} \wf_{T_1}\|^2 dx_{1}.
\end{equation}
\item Then the algorithm is iterated: $\wf_{T_1+}$ evolves unitarily until a random time $T_2 =  T_1 + \Delta T_2$, where  $\Delta T_2$ is a random time (independent of $\Delta T_1$) distributed according to the exponential distribution with rate $N\lambda$, and so on.
\end{enumerate}

Thus, if, between time $t_0$ and any time $t>t_0$, $n$ collapses have occurred at the times  $t_0< T_1 < T_2 < \ldots < T_n < t $, with centers  $X_1, \ldots, X_n$ and labels $I_1, \ldots, I_n$, the wave function at time $t$ will be
\begin{equation}\label{eq:psit}
  \wf_t =  \frac{L_{[t_0,t)}(F_n)\, \wf_{t_0}}{\| L_{[t_0,t)}(F_n) \, \wf_{t_0} \|}\,
\end{equation}
where $F_n = \bigl((X_1,T_1,I_1), \ldots, (X_n,T_n,I_n)\bigr)$, and
\begin{multline}\label{eq:long}
  L_{[t_0,t)}(F_n) =  \lambda^{n/2}  e^{-N\lambda (t-t_0)/2} \:\times\\
  \times\: U_{t-T_n} \Lambda_{I_n}(X_n)^{1/2} 
  \,U_{T_n-T_{n-1}} \Lambda_{I_{n-1}}(X_{n-1})^{1/2} 
  \,U_{T_{n-1}-T_{n-2}}  \cdots \Lambda_{I_1}(X_1)^{1/2} \, U_{T_1-t_0}. 
\end{multline}
(The scalar factor in the first line will be convenient for future use.) Since $T_i$, $X_i$, $I_i$ and $n$ are random, $\wf_t$ is also random. We will also call $\wf_t$ the \emph{collapsed wave function}, particularly when in need to contrast it with the ``uncollapsed'' wave function $U_{t-t_0} \,\wf_{t_0}$.

It should be observed that---unless $t_0$ is the initial time of the universe---also $\wf_{t_0}$ should be regarded as random, being determined by the collapses that occurred at times earlier than $t_0$. However, \emph{given} $\wf_{t_0}$, the statistics of the future evolution of the wave function is completely determined; for example, the joint distribution of the first $n$ collapses after $t_0$, with particle labels $I_1, \ldots, I_n \in \{1,\ldots,N\}$, is
\begin{multline}\label{nflashdist}
  \PPP\bigl( X_1\in d x_1, T_1 \in d t_1, I_1 = i_1, \ldots,
  X_n \in dx_n, T_n \in d t_n, I_n = i_n   |  \wf_{t_0} \bigr) =\\ 
  1_{t_0<t_1<\ldots<t_n} \,\| L(f_n) \, \wf_{t_0}\|^2 \, dx_{1}dt_1 \cdots  dx_{n}dt_n \,,
\end{multline}
where the symbol $1_C$ is 1 if the condition $C$ is satisfied and 0 otherwise, $f_n$ stands for $\bigl((x_1,t_1,i_1), \ldots, (x_n,t_n,i_n)\bigr)$, and 
\begin{multline}\label{Ldef}
  L(f_n) = \lambda^{n/2}  e^{-N\lambda (t_n-t_0)/2} \:\times\\
  \times\: \Lambda_{i_n}(x_n)^{1/2} 
  \,U_{t_n-t_{n-1}} \Lambda_{i_{n-1}}(x_{n-1})^{1/2} 
  \,U_{t_{n-1}-t_{n-2}}  \cdots \Lambda_{i_1}(x_1)^{1/2} \, U_{t_1-t_0}. 
\end{multline}
The expression \eqref{Ldef} equals $\displaystyle \lim_{t\searrow t_n} L_{[t_0,t)}(f_n)$, with $L_{[t_0,t)}(f_n)$ defined in \eqref{eq:long}.

\bigskip

We have described the law for the evolution of the wave function. We now turn to the primitive ontology (PO). In the subsections below we present two versions of the GRW theory, based on two different choices of the PO, namely the \emph{matter density ontology} (in Section~\ref{sec:GRWm}) and the \emph{flash ontology} (in Section~\ref{sec:GRWf}).

\subsection{GRWm}
\label{sec:GRWm}

In  GRWm, the PO is given by a field: We have a variable $m(x,t)$ for every point $x \in \RRR^3$ in space and every time $t\geq t_0$, defined by 
\begin{equation}\label{mdef}
 m(x,t) = \sum_{i=1}^N m_i \int\limits_{\RRR^{3N}}  dq_1 \cdots dq_N \, \delta(q_i-x) \,  \bigl|\wf_t(q_1, \ldots, q_N)\bigr|^2 \,.
\end{equation}
In words, one starts with the $|\wf|^2$--distribution in configuration
space $\RRR^{3N}$, then obtains the marginal distribution of 
the $i$-th degree of freedom $x_i\in \RRR^3$
by integrating out
all other variables $x_j$, $j \neq i$, multiplies by the mass associated with $x_i$, and sums over $i$. Alternatively, \eqref{mdef} can be rewritten as
\begin{equation}
  m(x,t) = \scp{\wf_t}{\tilde\Lambda(x) |\wf_t}
\end{equation}
with $\tilde\Lambda(x) = \sum_{i} m_i \, \delta(\widehat{Q}_i - x)$.

The field $m(\cdot,t)$ is supposed to be understood as the density of matter in space at time $t$. GRWm is a theory about the behavior of matter with density $m(\cdot,t)$ in three-dimensional space.

\subsection{GRWf}
\label{sec:GRWf}

According to  GRWf, the PO is given by ``events'' in space-time called flashes, mathematically described by points in space-time. In GRWf, histories of matter are not made of world lines but of world points. The flashes form the set
\[
  F=\{(X_{1},T_{1}), \ldots, (X_{k},T_{k}), \ldots\}
\]
(with $T_1<T_2<\ldots$), or, when we consider \emph{labeled flashes},
\[
  F=\{(X_{1},T_{1},I_1), \ldots, (X_{k},T_{k},I_k), \ldots\}
\]
with $I_k\in \Lab = \{1,\ldots,N\}$, the set of labels. We often find it convenient to write $F$ as an ordered set,
\[
 F= \bigl( (X_{1},T_{1},I_1), \ldots, (X_{k},T_{k},I_k), \ldots\bigr)\,.
\]
The GRWf law of the flashes asserts that there is a flash at the center $(X,T)$ of every collapse, with the appropriate label. Accordingly, Equation \eqref{nflashdist} gives the joint distribution of the first $n$ flashes, after some initial time $t_0$.

Note that if the number $N$ of the degrees of freedom in the wave function is large, as in the case of a macroscopic object, the number of flashes is also large (if $\lambda=10^{-15}$ s$^{-1}$ and $N=10^{23}$, we obtain a rate of $10^{8}$ flashes per second). Therefore, for a reasonable choice of the parameters of the GRWf theory, a cubic centimeter of
solid matter contains more than $10^8$ flashes per second. Such large collections of flashes can form macroscopic shapes, such as tables and chairs. That is how we find an image of our world in GRWf.

We should add that the mathematical scheme of GRWf that we have introduced here is not the most general one possible. The flash rate operators $\Lambda(x)$ do not have to be of the form \eqref{eq:collapseoperator} but could be other positive operators \cite{Tum05}, they could depend on time, $\Lambda(x) = \Lambda_t(x)$, and they could even be allowed to depend on the previous flashes \cite{Tum07}. (The latter case occurs in the relativistic GRWf theory presented in \cite{Tum04}.) The considerations in this paper are still valid if the $\Lambda(x)$ are other positive operators than in \eqref{eq:collapseoperator} and if they depend on time, but we do not consider the case in which they depend on the previous flashes. For the sake of concreteness readers can simply take $\Lambda(x)$ to be  the multiplication operators \eqref{eq:collapseoperator}.

\subsection{Empirical Equivalence}
\label{sec:ee}

As already remarked, it is known that GRWf and GRWm are empirically equivalent, i.e., they make always and exactly the same predictions \cite{AGTZ06}. In other words, there is no conceivable experiment (\y{even those exploiting} future advances in technology) that could distinguish between GRWf and GRWm. This follows from the following even stronger statement: When applying the flash ontology and the matter density ontology to the same wave function $\Psi$ obtained from the GRW process, the two PO histories are \emph{macro-history equivalent}, i.e., all macroscopic facts come out the same way.

Let us elaborate on this statement. What we mean is to consider a realization of the GRW jump process in Hilbert space as described in Section~\ref{sec:GRWprocess} (that is, $\Psi_t$ for every $t$), and then both the GRWm world and the GRWf world associated with this $\Psi$, defined by $m(x,t)$ as in \eqref{mdef} for every $t$, respectively by putting a flash at the center of every collapse of $\Psi$. What we mean by macro-history equivalence is that the macroscopic world history is the same in both worlds, including, e.g., the weather in a particular place at a particular time, lottery numbers, and more generally the exact sequence of outcomes of any experiment. This is more than empirical equivalence, as the latter requires not that all random events come out the same way in two worlds, but only that the outcome statistics are the same. \y{For example, if the two theories provided different macroscopic histories which, however, are such that one cannot conclude from an analysis of the macroscopic histories alone which one arose from which of the theories, then the two theories would already be empirically equivalent.} Clearly, macro-history equivalence implies empirical equivalence.

For GRWf and GRWm, macro-history equivalence holds \emph{with overwhelming probability}. That is, although there do exist wave functions $\Psi$ for which the macroscopic facts in the GRWf world are different from those in the GRWm world, such wave functions are extremely improbable \y{for} the GRW process.

Here is the argument. It suffices to consider a macroscopic amount of matter, which we call the ``pointer'' (though it could also be, e.g., the shape of ink on paper), that can either be in position 1 or position 2 at time $t$, and a wave function of the form $\Psi_t=c_1\Phi_1 + c_2\Phi_2$, where $\Phi_i$ is concentrated on configurations in which the pointer is in position $i$; we assume $\|\Phi_i\|=1$ and $|c_1|^2+|c_2|^2=1$. If, in GRWm, the matter of the pointer is in position 1, then this means that $m(1,t)\gg m(2,t)$; thus, $|c_1|^2 \gg |c_2|^2$; thus, flashes occur at a much greater rate at position 1 than at position 2; thus, with probability near 1, in GRWf the matter is also in position 1.
To appreciate just how close to 1 this probability is, recall that, as a consequence of the GRW process for $\Psi_t$, it is overwhelmingly likely that either $|c_1|^2$ or $|c_2|^2$ will become exorbitantly small within a fraction of a second (in realistic scenarios, smaller than $10^{-10^{10}}$ in $10^{-9}$ seconds).\footnote{Note also that, in the unlikely event that many flashes occur in position 2 between $t$ and $t+\Delta t$ and thus create a discrepancy between the pointer position in GRWf and that in GRWm, the associated collapses would shrink the size of $c_1$ to a considerable extent; so much indeed, if the number of flashes in position 2 is sufficient, that $|c_1(t+\Delta t)|^2$ is close to zero and $|c_2(t+\Delta t)|^2$ close to 1; as a consequence, $m(1,t+\Delta t) \ll m(2,t+\Delta t)$. That is, even in the unlikely event of a discrepancy, the discrepancy persists only for a limited time---the time it takes the collapses centered at position 2 to make $|c_1(t+\Delta t)|^2$ small.}

\subsection{Systems}
\label{sec:sys}

Since we have not specified, in the definition of the GRW theories, which kinds of systems the defining equations, such as \eqref{eq:psit} through \eqref{Ldef}, apply to, they a priori apply only to the universe as a whole. For any system, being a subsystem of the universe, equations of the same kind may or may not apply, but there is no need, and indeed no room, for postulates about this because the equations for the universe will determine what is true about any subsystem. Hence, the wave function $\wf$ we were talking of is the wave function of the universe. However, in our analysis of the empirical predictions of GRWm and GRWf, we will have to consider \emph{systems}: the system corresponding to those instruments which comprise the apparatus for the experiment and, most importantly, the system upon which the experiment is performed. For this, it will be helpful to formalize the notion of system, as well as that of the \emph{wave function of a system}.

To begin to approach such a notion, note that usually a system corresponds to some of the ``configuration variables'' in the wave function,
\be\label{wfsplit}
\wf=\wf(q) =\wf(q_\sys, q_\env)\,
\ee
where $q=(q_1,\ldots,q_N)$ is the configuration variable of the universe, $q_\sys$ that of the system, and $q_\env$
that of its environment (the rest of the world); defining
a system amounts to splitting the universe into two parts, the system and 
its environment. For example, $q_\sys$ may correspond to a certain collection of ``particle variables'', say
\be\label{sysex}
q_\sys = (q_1,\ldots,q_M) \quad \text{and} \quad q_\env=(q_{M+1},\ldots,q_N)\,.
\ee

Since for the GRW theories, the configuration variables do not play a fundamental role, our mathematical definition of ``system'' is formulated in different terms, namely in terms of the Hilbert space and of the primitive ontology. 

For our purposes, a \emph{system} is defined by two ingredients:
\begin{itemize}
\item A splitting of Hilbert space according to
\be\label{HilbertsysHilbertenv}
\Hilbert = \Hilbert_\sys \otimes \Hilbert_\env\,.
\ee
For example, such a splitting is provided by \eqref{wfsplit} according to $\Hilbert_\sys = L^2(q_\sys)$, $\Hilbert_\env=L^2(q_\env)$, and $\Hilbert=L^2(q)$.
\item A splitting of the PO; this means, in GRWf, a splitting of the flashes according to
\be\label{Fsplit}
F=F_\sys \cup F_\env\,, \quad F_\sys \cap F_\env = \emptyset\,,
\ee
or, in GRWm, a splitting of the matter density according to
\be\label{msplit}
m(x,t) = m_\sys(x,t) + m_\env(x,t)\,.
\ee
\end{itemize}

In both GRWf and GRWm, we assume that the splitting
is defined either through a subset $\subLab \subseteq \Lab$ of the set of labels (corresponding to different types of flashes/collapses), or through a region $\region_\sys\subseteq \RRR^3$ in space, or a combination of both: In GRWf, a flash belongs to $F_\sys$ if and only if it occurs in $\region_\sys$ \emph{and} its label belongs to $\subLab$; $F_\env := F\setminus F_\sys$. In GRWm, $m_\sys$ is the contribution to $m(x,t)$ from labels in $\subLab$ at locations in $\region_\sys$:
\be\label{msysdef}
m_\sys(x,t) = 1_{x\in\region_\sys} \sum_{i\in \subLab}
m_i \int\limits_{\RRR^{3N}}  dq_1 \cdots dq_N \, \delta(q_i-x) \,  \bigl|\wf_t(q_1, \ldots, q_N)\bigr|^2 \,,
\ee
and $m_\env = m-m_\sys$. 
We now define the splitting \eqref{HilbertsysHilbertenv} of Hilbert space in terms of $\subLab$ and $\region_\sys$. For labeled particles, we use that $\Hilbert_{\Lab}=\Hilbert_{\subLab} \otimes \Hilbert_{\Lab\setminus\subLab}$. When using a region $\region_\sys\subset\RRR^3$ of physical space for defining the system, it is best to use Fock spaces (i.e., Hilbert spaces for a variable number of particles) instead of $L^2(\RRR^{3N})$ because, for configurations $(q_1,\ldots,q_n)$, the number of points $q_i$ that lie in $\region_\sys$ varies with the locations of the $q_i$; a natural extension of the GRW theories to Fock spaces was described in \cite{Tum05}. Let $\Hilbert(S)$ be the fermionic or bosonic Fock space over $L^2(S)$. The splitting \eqref{HilbertsysHilbertenv} arises from the fact that if both $\region_\sys$ and $\RRR^3\setminus \region_\sys$ have positive volume then $\Hilbert(\RRR^3)=\Hilbert(\region_\sys) \otimes \Hilbert(\RRR^3\setminus \region_\sys)$.

The set $F_\sys\subseteq F$ of the system's flashes may happen to be empty, but even in that case the definition of the system in terms of $\subLab$ and $\region_\sys$ will be useful. In the example of \eqref{sysex}, $\subLab=\{1,\ldots,M\}$, while $\region_\sys=\RRR^3$ does not play a role. The example provided by \eqref{wfsplit} suggests that everything that could be considered a system in orthodox quantum mechanics also defines a system in the sense of our definition.

We say that the system \emph{has wave function $\psi_\sys$} if the wave function of the universe factorizes according to
\be
\Psi = \psi_\sys \otimes \psi_\env
\ee
with $\psi_\sys \in \Hilbert_\sys$ and $\psi_\env \in \Hilbert_\env$. Since it follows that not every system has a wave function at every time, it will also be useful to say that the system \emph{has reduced density matrix $\dm_\sys$} if
\be
\dm_\sys = \tr_\env \,\pr{\Psi}
\ee
with $\tr_\env$ the partial trace over $\Hilbert_\env$.

We call a system a  {\em GRW system} if it has an autonomous GRW dynamics, i.e., if it behaves as if it were alone in the universe. We postpone the exact definition of what that means to Section~\ref{sec:GRWsystem}; there we will also show that a system is a GRW system if and only if it does not interact with its environment.

\section{Mathematical Tools}

Let $\SSS(\Hilbert)$ denote the unit sphere in Hilbert space,
\be
\SSS(\Hilbert) = \bigl\{ \psi\in\Hilbert: \|\psi\|=1 \bigr\}\,.
\ee

\subsection{POVM}
\label{sec:POVM}

Recall that, while many quantum experiments are associated with self-adjoint operators, this is not the most general case, which corresponds to \emph{positive-operator-valued measures} (POVMs, also known as ``generalized observables''; see \cite{Dav76} and Section~4 of \cite{DGZ04} for an introduction). We recall that a \emph{POVM on the set $\Omega$ acting on $\Hilbert$} is a mapping
\begin{equation}
  E: \mathcal{A} \to \mathcal{L}(\Hilbert)
\end{equation}
from a $\sigma$-algebra $\mathcal{A}$ over $\Omega$ (the family of all subsets of $\Omega$ regarded as ``measurable'') to the space of bounded operators on the Hilbert space $\Hilbert$, with the properties that (i)~$E(B)$ is a positive self-adjoint operator for every $B \in \mathcal{A}$, (ii)~$E(\Omega) = I$, the identity operator, and (iii)~$E(\cdot)$ is $\sigma$-additive, i.e., for pairwise disjoint $B_1,B_2,\ldots \in \mathcal{A}$ 
\begin{equation}
  E\Bigl(\bigcup_{k=1}^\infty B_k\Bigr) = \sum_{k=1}^\infty E(B_k)\,,
\end{equation}
with the infinite sum understood as the weak limit $n\to\infty$ of $\sum_{k=1}^n E(B_k)$. 
(All subsets and functions we consider will be assumed to be measurable with respect to the relevant $\sigma$-algebras. A positive operator $S$ with $S\leq I$ is also called an \emph{effect} in the literature \cite{kraus}, and a POVM also an \emph{effect-valued measure}.) By virtue of the spectral theorem, the self-adjoint operators correspond to special POVMs, the projection-valued measures (PVMs) on the real \y{line}. In many cases relevant to us, $\Omega$ will be a finite or countable set; in that case, the POVM is determined by the operators associated with singleton sets, $E_\omega = E(\{\omega\})$, according to
\be
E(B) = \sum_{\omega\in B} E_\omega\,,
\ee
and any collection of positive operators $(E_\omega)_{\omega \in\Omega}$ such that
\be
\sum_{\omega\in\Omega} E_\omega = I
\ee
defines a POVM. We will thus often identify the POVM with the collection $(E_\omega)_{\omega\in\Omega}$.

The following two very simple observations about POVMs will be used in the course of this paper: 

\textbf{Function Property.} \emph{If the distribution of the random variable $X$ depends on a system's wave function $\psi$ via a POVM $D(\cdot)$, $\PPP(X \in A) = \scp{\psi}{D(A)|\psi}$, and if the random variable $Y$ is a function of $X$, $Y=f(X)$, then the distribution of $Y$ is also given by a POVM:}
\begin{equation}\label{fPOVM}
  \PPP(Y \in B) = \scp{\psi}{E(B)|\psi} \quad \text{with }E(B) = D\bigl(f^{-1}(B)\bigr)\,.
\end{equation}

\textbf{Reduction Property.} \emph{If $D(\cdot)$ is a POVM on $\Omega$ acting on $\Hilbert_1 \otimes \Hilbert_2$, and if $\phi \in \Hilbert_2$ has $\|\phi\|=1$, then}
\be
 \scp{\psi\otimes \phi}{D(B)|\psi\otimes \phi}
 =\scp{\psi}{E(B)|\psi}  \quad \forall \psi\in\Hilbert_1\,,
\ee
\emph{where the partial scalar product}
\begin{equation}\label{partialPOVM}
  E(B) = \scp{\phi}{D(B)|\phi}
\end{equation}
\emph{defines a POVM $E(\cdot)$ on $\Omega$ acting on $\Hilbert_1$. Likewise, if $D(\cdot)$ is as before and $\dm_2$ a density matrix on $\Hilbert_2$ then the partial trace}
\be
  E(B) = \tr_2\bigl([I_1\otimes\dm_2]\, D(B)\bigr)\,,
\ee
\emph{defines a POVM $E(\cdot)$ on $\Omega$ acting on $\Hilbert_1$.}

\subsection{The Distribution of the Flashes}
\label{sec:F}

In GRWf, the joint distribution of all flashes, as a functional of the initial wave function $\wf_{t_0}$, is given by a POVM $\G(\cdot)$, called the \emph{history POVM}. Let us elaborate on this statement.

Reformulating \eqref{nflashdist}, the joint distribution of the first $n$ flashes is given by a POVM $\G_n(\cdot)$ on
\be
\Omega_n = \Bigl\{f_n=((x_1,t_1,i_1),\ldots,(x_n,t_n,i_n))\in(\RRR^3 \times [t_0,\infty) \times \Lab)^n: t_1<\ldots<t_n\Bigr\}
\ee
(where $\RRR^3$ represents space, $[t_0,\infty)$ time, and $\Lab$ is the set of labels),
\begin{equation}
  \PPP(F_n \in df_n) = \scp{\wf_{t_0}}{\G_n(df_n)|\wf_{t_0}}
\end{equation}
with $df_n = dx_1dt_1 \cdots dx_ndt_n$ a ``volume element'' around $f_n\in\Omega_n\subset(\RRR^3 \times \RRR \times \Lab)^n$ and
\be\label{Gndfn}
\G_n(df_n) = L^*(f_n) \, L(f_n) \, df_n\,,
\ee
where $L(f_n)$ was defined in \eqref{Ldef} and $L^*$ denotes the adjoint of $L$. 
To put \eqref{Gndfn} differently, for any measurable set $B\subseteq \Omega_n$,
\begin{equation}\label{Fndef}
  \G_n(B) = \int_B df_n \,  L^*(f_n) \, L(f_n)\,,
\end{equation}
where the measure used is the (Lebesgue) volume measure on each of the $N^n$ $4n$-dimensional sheets of $\Omega_n$ (i.e., integration over $B$ may include summation over labels $i_1,\ldots,i_n$). It is easy to convince oneself that $\G_n(\cdot)$ is a POVM; see \cite{Tum07} for a rigorous proof.

It is no surprise now that also the joint distribution of \emph{all} flashes is given by a POVM $\G(\cdot)$; see \cite{Tum07b} for a rigorous proof. The space on which $\G(\cdot)$ lives is the set $\Omega_{[t_0,\infty)}$ of all countable sequences $(x_n,t_n,i_n)_n$ in $\RRR^3 \times [t_0,\infty) \times \Lab$ with increasing times, $t_n<t_{n+1}$, and $\lim_n t_n=\infty$.

Now consider $F_{[t_0,\tf)}$, the sequence of flashes during the time interval $[t_0,\tf)$ with $t_0<\tf<\infty$. Since $F_{[t_0,\tf)}$ trivially is a function of $F$, the sequence of \emph{all} flashes, by the function property \eqref{fPOVM} its distribution is given by a POVM $\G_{[t_0,\tf)}(\cdot)$ on the space of all histories of flashes in the time interval $[t_0,\tf)$. Since $F_{[t_0,\tf)}$ is almost surely finite, $\G_{[t_0,\tf)}(\cdot)$ is concentrated on the set $\Omega_{[t_0,\tf)}$ of all finite sequences in $\RRR^3 \times [t_0,\tf) \times \Lab$ with increasing times. Put differently,
\be
\Omega_{[t_0,\tf)}=\bigcup_{n=0}^\infty \Omega_{[t_0,\tf)}^n
\ee
with sectors
\begin{equation}
\Omega_{[t_0,\tf)}^n =
 \Bigl\{ ((x_1,t_1,i_1),\ldots,(x_n,t_n,i_n))\in(\RRR^3 \times [t_0,\tf) \times \Lab)^n: 
  t_1<\ldots<t_n\Bigr\}\,.
\end{equation}
We can specify $\G_{[t_0,\tf)}(\cdot)$ explicitly:
\begin{equation}\label{Fttdef}
\G_{[t_0,\tf)}(B) = \int_B df \, L^*_{[t_0,\tf)}(f) \, L_{[t_0,\tf)}(f)
\end{equation}
with $df$ the Lebesgue measure on $\Omega_{[t_0,\tf)}$, defined as being the Lebesgue measure on each sector $\Omega_{[t_0,\tf)}^n\subset (\RRR^3\times\RRR\times\Lab)^n$ as in \eqref{Fndef}.

Finally, we note for later use that there is a natural identification $i:\Omega_{[t_1,t_3)}\to\Omega_{[t_1,t_2)}\times \Omega_{[t_2,t_3)}$ for $t_1<t_2<t_3\leq \infty$: Every pattern $f_{[t_1,t_3)}$ of flashes during $[t_1,t_3)$ defines a pair $(f_{[t_1,t_2)},f_{[t_2,t_3)})$ consisting of a pattern $f_{[t_1,t_2)}$ during $[t_1,t_2)$ and a pattern $f_{[t_2,t_3)}$ during $[t_2,t_3)$. This mapping is bijective and for $t_3<\infty$ measure-preserving, so
\be
df_{[t_1,t_3)}=df_{[t_1,t_2)} df_{[t_2,t_3)}\,.
\ee
As here, we shall often make this identification without explicit use of the symbol $i$.

\subsection{The Conditional Probability Formula}
\label{sec:condprob}

Set, for the ease of notation, $t_0=0$. A simple and important consequence of the distribution law \eqref{nflashdist} of the flashes is the \emph{conditional probability formula}, which asserts that, for $0<\ti<\tf\leq \infty$ and any $B\subseteq \Omega_{[\ti,\tf)}$,
\begin{equation}\label{condprob1}
  \PPP_{\wf_{0}}\Bigl(F_{[\ti,\tf)}\in B\Big|F_{[0,\ti)}\Bigr) = 
  \PPP^{(\ti)}_{\wf_{\ti}}\bigl(F_{[\ti,\tf)}\in B\bigr)\,.
\end{equation}
Here, $\PPP_{\wf_{0}}$ means the distribution obtained starting from the wave function $\wf_{0}$, and $\PPP^{(\ti)}_{\wf_\ti}$ the one obtained starting from $\wf_{\ti}$ at time $\ti$. Note that the dependence on $F_{[0,\ti)}$ of the right hand side is through $\wf_\ti$, which is a function of $F_{[0,\ti)}$. In words, the conditional probability formula asserts that the conditional distribution of the flashes after time $\ti$, given the flashes before $\ti$, coincides with the distribution obtained from starting the universe at time $\ti$ with wave function $\wf_\ti$. 

This formula is the ultimate reason why it is natural in GRWf to regard the collapsed (GRW) wave function $\wf_{\ti}$ as \emph{the} wave function at time $\ti$: because the distribution of the future flashes after $\ti$ (given that the past was what it was) agrees with the distribution arising from $\wf_{\ti}$ as the initial wave function at time $\ti$. 

An algebraic-analytic derivation of the conditional probability formula 
can be found in Appendix~\ref{app:condprob}. Alternatively, the 
conditional probability formula follows from the \emph{Markov property} 
of the stochastic GRW process $\Psi_t$, defined by
\be\label{Markov}
\PPP_{\Psi_0}\Bigl( E \Big| \Psi_{s'}=\psi_{s'} \forall s' \in [0,s] \Bigr)=
\PPP^{(s)}_{\Psi_s}\bigl(E\bigr)
\ee
for every event $E$ concerning only the future of $\Psi_t$ after time 
$s$. For example, $E$ could be the  event 
$(\Psi_{t_1},\ldots,\Psi_{t_k})\in B'$ for $s<t_1<\ldots<t_k$. The 
Markov property means that the process is \emph{memoryless}. That the 
GRW process has the Markov property is more or less clear from its 
definition. To see how the conditional probability formula follows, note 
first that the history of the wave function between 0 and $s$ is 
determined by (and, conversely, determines) the flashes between 0 and 
$s$, so that conditioning on $\Psi_{s'}=\psi_{s'} \forall s' \in [0,s]$ 
amounts to the same thing as conditioning on $F_{[0,s)}=f_{[0,s)}$. 
Similarly, the future history of the wave function is in one-to-one 
correspondence with the future flashes, so that \eqref{condprob1} follows.

\section{How Operators Emerge}
\label{sec:emergeop}

We will formulate and derive the GRW formalism in Section~\ref{sec:GRWformalism}. At this stage, we can already understand, in a particularly easy way, how operators emerge from GRWf, and that is why we present this aspect first.

We give a simple derivation for the \emph{main theorem about POVMs in GRWf}, i.e., for the fact that in GRWf, as in quantum mechanics, there is a POVM $E(\cdot)$ for every experiment, so that the probability distribution of the outcome of the experiment, when performed on a system with wave function $\psi$, is given by $\scp{\psi}{E(\cdot)|\psi}$. To appreciate the substance of this derivation it is relevant to realize that the definition of GRWf did not mention operators as observables. Thus, \emph{operators as observables were not put in, they come out by themselves.}

Many physicists find such a situation hard to imagine, and that is why this point deserves a separate section. Many physicists are used to thinking that the central role of operators in quantum theory, particularly in view of their non-commutativity, constitutes a crucial departure from classical physics, and, even more, from any kind of theory describing an objective reality, or any kind of theory that can be understood as clearly as a classical theory. According to this widespread view, the non-commutativity of operators entails that reality itself is paradoxical and will forever remain incomprehensible to us mortals. This view is often connected to the key word ``complementarity.'' But the same non-commuting operators appear in GRWf, a theory describing an objective reality which indeed allows as clear an understanding as a classical theory!

This is not so surprising since the same can be said of Bohmian mechanics (see, e.g., \cite{Bell87b,DGZ04}), and since it has been clear for 20 years that GRW theories make almost the same predictions as quantum mechanics \cite{GRW86,Bell87}. Nonetheless, it is worthwhile to get a good grasp of how exactly this can be so, how non-commuting operators can emerge from a theory describing non-paradoxical reality.

\bigskip

Here is the derivation. Recall from  
Section~\ref{sec:F} that the joint distribution of all flashes after  
time $t$ is given by a POVM $\G(\cdot)=\G_{[t,\infty)}(\cdot)$ on the appropriate space $\Omega_{[t,\infty)}$ of flash histories and  
the wave function of the universe $\Psi_t$ at time $t$.  Let $t$ be  
the time at which the experiment begins.
Consider splitting the universe into a system (the object of the experiment), the apparatus of the experiment, and the rest of the world. It so happens that for the argument that follows, the division between apparatus and the rest of the world is irrelevant, so we put the two together and call them the environment (of the system). The division between the system and its environment corresponds to a splitting of the Hilbert space into $\Hilbert=\Hilbert_\sys \otimes \Hilbert_\env$; the splitting $F=F_\sys\cup F_\env$ of the flashes is not needed in this section. We assume independence between the  
system and the environment immediately before $t$, so that\footnote{Readers may worry that the factorization condition \eqref{factorize} never holds because of the symmetrization postulate: As soon as both the system and the apparatus contain electrons, the wave function has to be anti-symmetric in the electron variables $q_i$, which conflicts with \eqref{factorize} if the latter is based on a splitting as in \eqref{sysex}, grouping some variables $q_i$ together as ``system variables'' and others as ``environment variables.'' The answer is, \eqref{factorize} can hold nevertheless, as follows: For identical particles, the indices of the variables $q_1,\ldots,q_N$ are mere mathematical labels, and the splitting into system and environment should not be based on these unphysical labels but instead on regions of space. Indeed, \y{as mentioned already}, if $\region_\sys\subseteq \RRR^3$ is a region of space such that both $\region_\sys$ and $\RRR^3\setminus \region_\sys$ have positive volume then $\Hilbert(\RRR^3) = \Hilbert(\region_\sys) \otimes \Hilbert(\RRR^3\setminus \region_\sys)$, where $\Hilbert(S)$ is the fermionic (or bosonic) Fock space over $L^2(S)$. Since a fermionic wave function can be represented by a vector $\Psi\in\Hilbert(\RRR^3)$, it can indeed factorize in the splitting based on $\region_\sys$.}
\begin{equation}\label{factorize}
   \Psi_t = \psi \otimes \phi\,.
\end{equation}
Here $\phi$ is fixed, being part of the characterization of the  
experiment, while $\psi$, the initial wave function of the system upon which  
the experiment is performed, is allowed to vary in the system   
Hilbert space $\Hilbert_\sys$. The outcome $Z$ of the experiment is a function of the pattern $F$ of flashes after time $t$,
\begin{equation}\label{zetaF}
Z=\zeta(F)
\end{equation}
with $\zeta: \Omega_{[t,\infty)} \to \Values$, where $\Values$ is the \emph{value space} of the experiment. That is so because the flashes define where the pointers point, and what the shape of the ink on a sheet of paper is. (It would even be realistic to assume that $Z$ depends only on the flashes of the apparatus, but this restriction is not needed for the further argument.) Therefore, the distribution of the random outcome $Z$ is given by
\begin{equation}\label{PPPZ1}
  \PPP(Z \in B) = \PPP\bigl(F \in \zeta^{-1}(B)\bigr) = 
  \scp{\Psi_t}{\G \circ \zeta^{-1}(B)| \Psi_t}
  = \scp{\psi}{E(B) |\psi} \quad\forall B\subseteq \Values\,,
\end{equation}
where the first scalar product is taken in the Hilbert space of the universe and the second in the Hilbert space of the system (i.e., the object of the experiment), and $E(\cdot)$ is the POVM given by
\begin{equation}\label{Efdef}
  E(B)= \scp{\phi}{\G \circ \zeta^{-1}(B)| \phi}\quad\forall B\subseteq\Values\,,
\end{equation}
where the scalar product is a partial scalar product in the Hilbert space of the environment. Thus, for every experiment in GRWf the distribution of outcomes is given by a POVM $E(\cdot)$ on $\Values$, which is what we wanted to show.

\bigskip

At this point, we would like to go through the derivation again, carefully keeping track of the ingredients in the argument:

\begin{itemize}
\item The distribution of flashes in GRWf is given by a POVM $\G(\cdot)$. In more detail:
  \begin{itemize}
  \item $\G(\cdot)$ is a POVM on the total Hilbert space 
  $\Hilbert = \Hilbert_\sys \otimes \Hilbert_\env$, where $\Hilbert_\sys$ 
  is the Hilbert space of the system and $\Hilbert_\env$ that of its 
  environment, including \y{the} apparatus.
  \item What we really want is, of course, the \emph{conditional} distribution
  of the flashes, \emph{given} what happened up to
  the time $t$ when the experiment begins.
  By the conditional probability formula \eqref{condprob1}, this distribution
  is $\scp{\Psi_t}{\G_{[t,\infty)}(\cdot)|\Psi_t}$ with $\Psi_t$ the (collapsed) 
  wave function at time $t$.
  \end{itemize}
\item The outcome $Z$ of an experiment in a GRWf world must be a function of the flashes (usually, just of the flashes belonging to some apparatus), $Z=\zeta(F)$.
\item By the function property \eqref{fPOVM} of POVMs, the distribution of the outcome is also given by a POVM on $\Hilbert$.
\item Consider a particular setting of the experiment, as encoded in $\phi\in\Hilbert_\env$; ask for the dependence of the distribution of the outcome $Z$ on the wave function $\psi\in\Hilbert_\sys$ of the object. In particular, assume factorization, $\Psi_t = \psi \otimes \phi$.
\item By the reduction property \eqref{partialPOVM} of POVMs, the distribution of $Z$ as a function of $\psi$ is given by a POVM $E(\cdot)$ on $\Hilbert_\sys$.
\end{itemize}

\bigskip

We close this section with a few remarks.

\begin{enumerate}
\item The POVMs corresponding to different experiments may well, and typically will, not commute. Even the single POVM $E(\cdot)$ may be non-commuting, in the sense that $E(B_1)$ does not commute with $E(B_2)$ for suitable sets $B_1,B_2\subseteq\Values$. The simple derivation above, just a few lines long, shows how non-commuting operators can \emph{emerge} from a picture of reality (a random set of flashes) that is completely coherent, clear, easy-to-understand, complementarity-free, and paradox-free. 
Why do different experiments correspond to different POVMs? Because they correspond to different choices of the interaction Hamiltonian between the system and the apparatus, as well as different choices of $\phi$.\footnote{From the point of view of the entire universe, from which the Hamiltonian may be regarded as fixed once and for all, the relevant choice would lie \emph{only} in that of $\phi$.}

\item Since we know that the predictions of GRWf and GRWm are very close to those of quantum mechanics for all presently feasible experiments, for these experiments the POVM $E(\cdot) = E^\GRW(\cdot)$ should be very close to $E^\Qu(\cdot)$, the POVM predicted by quantum mechanics. For a principled consideration see Section~\ref{sec:deviations}.

\item We called the result of our reasoning the ``main theorem about POVMs'' in GRWf. Let us be explicit about the mathematical theorem that is involved here. It was formulated before as Theorem 8 in \cite{Tum07} and asserts the following:

\textit{Let $\Hilbert = \Hilbert_\sys\otimes \Hilbert_\env$ be a separable Hilbert space, $\G(\cdot)$ a POVM on a measurable space $(\Omega,\salg_\Omega)$ acting on $\Hilbert$, $\phi$ a fixed vector in $\Hilbert_\env$ with $\|\phi\|=1$, and $\zeta: (\Omega,\salg_\Omega) \to (\Values,\salg_{\Values})$ a measurable function. For every $\psi\in\Hilbert_\mathrm{sys}$ with $\|\psi\|=1$, let $\Psi_{t} = \psi \otimes \phi$, let $F$ be a random element in $\Omega$ with distribution $\scp{\Psi_{t}}{\G(\cdot) | \Psi_{t}}$, and let $Z=\zeta(F)$. Then there is a POVM $E(\cdot)$ on $(\Values,\salg_\Values)$ acting on $\Hilbert_\sys$ so that the distribution of $Z$ is $\scp{\psi}{E(\cdot) | \psi}$.}

The proof of this theorem is a straightforward application of the function property \eqref{fPOVM} and the reduction property \eqref{partialPOVM} of POVMs. What is important for us here is to appreciate the \emph{relevance} of this theorem as the appropriate mathematical formalization in GRWf of the physical statement that \emph{with every experiment $\E$, there is associated a POVM $E(\cdot)$ such that the probability distribution of the random outcome $Z$ of $\E$, when performed on a system with wave function $\psi$, is given by $\PPP(Z \in B) = \scp{\psi}{E(B)| \psi}$.}

\item If the wave function $\phi$ of the environment were not fixed but random, we would still end up with a POVM, as long as $\phi$ is independent of $\psi$ (at least conditionally on all information available to us about the experimental setup): we would have to replace \eqref{Efdef} by 
\begin{equation}\label{Ephiensemble}
  E(B)= \int \mu(d\phi) \: \scp{\phi}{\G \circ \zeta^{-1}(B)| \phi}\,,
\end{equation}
with $\mu$ the distribution of $\phi$.

\item The reader may find it confusing that part of the characterization of the experiment was the specification of $\phi$, the wave function of the system's environment: After all, it will be practically impossible to repeat the experiment with the same $\phi$, as $\phi$ comprises everything outside the system; for example, when we try to repeat the experiment at a later time, the moons of Jupiter will have moved, and the state of the lab will have changed as it will contain records of the previous experiment. So for practical purposes it is important that $E(\cdot)$ as given by \eqref{Efdef} does not depend on all details of $\phi$, but only on a few features of $\phi$ that we can control---and thus repeat. Mathematically, however, \eqref{Efdef} provides the correct POVM, and \eqref{PPPZ1} the correct distribution, regardless of whether we are able to evaluate or control this expression.\label{rem:robust1}

\item Note that the derivation did not assume any pre-determined time at which the experiment is over. It allows that the time at which the outcome $Z$ can be read off depends on $Z$ itself, a situation that occurs, e.g., in a time-of-arrival measurement, with $Z$ the time when a detector clicks.

\item What if factorization $\Psi_t=\psi\otimes\phi$ is not exactly satisfied, but only approximately? Then the probability distribution of the outcome $Z$ is still approximately given by $\scp{\psi}{E(\cdot)|\psi}$. More precisely, suppose that, instead of \eqref{factorize},
\be\label{almostfactorize}
\Psi_t = c\psi\otimes\phi + \Delta \Psi\,,
\ee
where $\|\Delta \Psi\| \ll 1$, $\|\psi\|=\|\phi\|=1$, and $c=\sqrt{1-\|\Delta \Psi\|^2}$ (which is close to 1). Then for any $B\subseteq \Values$,\footnote{To see this, write $\PPP(Z\in B)$ as $\scp{\Psi_t}{\G \circ \zeta^{-1}(B)| \Psi_t}$; insert \eqref{almostfactorize}; use $0\leq \G \circ \zeta^{-1}(B)\leq I$ to bound the term quadratic in $\Delta \Psi$ by $\|\Delta \Psi\|^2$; use the Cauchy--Schwarz inequality and $|c|<1$ to bound the cross terms by $2\|\Delta \Psi\|$; use that $1-|c|^2=\|\Delta \Psi\|^2$; in total, by the triangle inequality, obtain the bound $2\|\Delta \Psi\| (1+\|\Delta\Psi\|)< 3 \|\Delta\Psi\|$ provided $\|\Delta \Psi\|<1/2$.}
\be
\Bigl| \PPP(Z\in B) - \scp{\psi}{E(B)|\psi} \Bigr| < 3 \|\Delta \Psi\|\,.
\ee
This estimate conveys that the relevant measure for quantifying the size of the deviation from perfect factorization is the \emph{$L^2$ norm} of the deviation $\Delta \Psi$.

\item We do not know of a similar derivation of the main theorem about POVMs from GRWm, mainly because the probability distribution of the random function $m(\cdot,t)$ is not given by a POVM. Nevertheless a derivation from GRWm has been given in \cite{BGS06}, however one that is rather different in character: It requires great effort and yields a limited result, as it assumes a special, idealized type of experiment and, since it allows for small errors in the outcome statistics, does not show that the outcome statistics is \emph{exactly} given by a POVM.
\end{enumerate}

\section{The Quantum Formalism}
\label{sec:Quformalism}

Before we formulate the GRW formalism, we formulate for comparison the standard quantum formalism in the way relevant to us. We begin with the simplified version that one learns in beginner's courses and that suffices for many applications.

\bigskip

\noindent\textbf{The Simplified Quantum Formalism.}
\begin{itemize}
\item A system isolated from its environment has at every time $t$ a density matrix $\dm_t$ which evolves according to the unitary Schr\"odinger evolution,
\begin{equation}\label{SchrDM}
  \frac{d\dm_t}{dt} = -\tfrac{i}{\hbar} [H_\sys, \dm_t]\,.
\end{equation}
\item With suitable experiments $\E$ there is associated a self-adjoint operator $A$ on $\Hilbert_\sys$ (called the ``observable'') with pure point spectrum; let its spectral decomposition be
\begin{equation}\label{Aspectrum}
  A = \sum_{z} z P_z\,,
\end{equation}
with $P_z$ the projection to the eigenspace with eigenvalue $z$.
When the experiment $\E$ is performed on a system with density matrix $\dm$, the outcome $Z$ is random with probability distribution
\begin{equation}\label{RQPPP}
  \PPP(Z=z) = \tr(P_z \, \dm)\,.
\end{equation}
\item In case $Z=z$, the density matrix immediately after the experiment is
\begin{equation}\label{RQC}
  \dm' = \frac{P_z \dm P_z}{\tr(P_z \,\dm)}\,.
\end{equation}
\end{itemize}

\bigskip

The last rule contains the standard kind of collapse of the wave function, induced by ``the observer.''

We will need a more general formulation since the above formalism applies only to a narrow class of experiments, usually called ``ideal measurements.'' And for this we will need some more mathematical notions.

\subsection{Mathematical Tool: Completely Positive Superoperators}

We recall that the \emph{trace class} $TRCL(\Hilbert)$ is (roughly speaking) the space of all operators with finite trace. It contains in particular the density matrices.

By a \emph{superoperator} we mean a 
$\CCC$-linear mapping $\cpm: TRCL(\Hilbert_1) \to TRCL(\Hilbert_2)$.  A superoperator $\cpm$ is called \emph{completely positive} if for every integer $k\geq 1$ and every positive operator $\trclop \in \CCC^{k \times k} \otimes TRCL(\Hilbert_1)$, $(I_k \otimes \cpm) (\trclop)$ is positive, where $I_k$ denotes the identity operator on $\CCC^{k\times k}$ \cite{Cho75,kraus}. (Completely positive superoperators are also often called completely positive maps. If for every density matrix $\dm$, $\tr \cpm(\dm) \leq 1$ (as will be the case for all superoperators that we consider in this paper) then $\cpm$ is also called a \emph{quantum operation} \cite{kraus}.) 

Completely positive superoperators arise as a description of how a density matrix changes under the collapse caused by an experiment: If $\dm$ is the density matrix before the collapse, then $\cpm(\dm)/\tr  \cpm(\dm)$ is the density matrix afterwards. The simplest example of a completely positive superoperator is 
\be
\cpm(\trclop) = P \trclop P\,,
\ee
where $P$ is a projection. Note that for a density matrix $\dm$, $\cpm(\dm)$ is not, in general, a density matrix because completely positive superoperators do not, in general, preserve the trace.

In order to \y{establish} the complete positivity of a given superoperator, the following facts are useful: If $\dm_2$ is a density matrix on $\Hilbert_2$ then the mapping $\cpm: TRCL(\Hilbert_1)\to TRCL(\Hilbert_1\otimes\Hilbert_2)$ given by $\cpm(\trclop) = \trclop\otimes \dm_2$ is completely positive. Conversely, the partial trace $\trclop \mapsto \tr_2 \,\trclop$ is a completely positive superoperator $TRCL(\Hilbert_1\otimes\Hilbert_2) \to TRCL(\Hilbert_1)$. For any bounded operator $R: \Hilbert_1 \to\Hilbert_2$, $\trclop \mapsto R \trclop R^*$ is a completely positive superoperator $TRCL(\Hilbert_1)\to TRCL(\Hilbert_2)$, where $R^*:\Hilbert_2 \to \Hilbert_1$ is the adjoint of $R$. The composition of completely positive superoperators is completely positive. Positive multiples of a completely positive superoperator are completely positive. Finally, when a family of completely positive superoperators is summed or integrated over, the result is completely positive. Indeed, these rules suffice for all cases we will encounter in this paper.

For example, the master equation \eqref{M} of the GRW evolution has the property that the solution $\dm_t$ as a function of the initial datum $\dm_0$ is given by a completely positive superoperator $\acpm_{[0,t)}$, $\dm_t = \acpm_{[0,t)} \dm_0$ (and, in fact, $\acpm_{[0,t)}$ is trace-preserving). 

A canonical form of completely positive superoperators is provided by the theorem of Choi and Kraus \cite{Cho75,kraus} (also sometimes connected with the name of Stinespring), which asserts that \textit{for every bounded completely positive superoperator $\cpm:TRCL(\Hilbert_1) \to TRCL(\Hilbert_2)$ there exist bounded operators $R_{i}:\Hilbert_1 \to \Hilbert_2$ so that}
\begin{equation}\label{choi}
  \cpm (\trclop) = \sum_{i\in \I} 
  R_{i} \, \trclop \, R^*_{i}\,,
\end{equation}
\textit{where $\I$ is a finite or countable index set.}

Another remark concerns notation. Since superoperators are mappings, it is standard to write the composition of superoperators $\bcpm$, $\ccpm$ as $(\bcpm\circ\ccpm)(\dm)=\bcpm(\ccpm(\dm))$. For some calculations involving the composition of many superoperators acting on product spaces $\Hilbert_1 \otimes \cdots \otimes \Hilbert_n$, the standard notation gets cumbersome; for these cases, we propose a more transparent notation using diagrams in Appendix~\ref{sec:diagram}.

\subsection{The Formalism}
\label{sec:fullformalism}

We are now prepared for formulating the quantum formalism in greater generality. Without an essential loss of generality, we only consider experiments with \emph{discrete value space} $\Values$, i.e., experiments for which the set $\Values$ of possible outcomes is finite or countable. The reason why this is essentially no restriction is that every experiment in practice has limited accuracy, and indeed only a finite number of possible outcomes. Nevertheless it is sometimes convenient to consider a continuous variable $z$, and indeed, as far as the main theorem about POVMs, or \eqref{QPPP}, is concerned, we can allow $\Values$ to be any measurable space (i.e., set with a $\sigma$-algebra), including the possibility of a continuous variable $z$. However, when trying to formulate the collapse rule \eqref{QC} for a continuous variable $z$, difficulties arise that lie outside the scope of this paper.

\bigskip

\noindent\textbf{The Quantum Formalism.}
\begin{itemize}
\item A system isolated from its environment has at every time $t$ a density matrix $\dm_t$ which evolves according to the unitary Schr\"odinger evolution \eqref{SchrDM}.
\item With every experiment $\E$ with discrete value space $\Values$, beginning at time $\ti$ and ending at time $\tf$, there is associated a POVM $(E^\Qu_z)_{z\in\Values}$ on $\Values$ acting on $\Hilbert_\sys$. When the experiment $\E$ is performed on a system with density matrix $\dm_\ti$, the outcome $Z$ is random with probability distribution
\begin{equation}\label{QPPP}
  \PPP(Z = z) = \tr \bigl( \dm_\ti \, E^\Qu_z\bigr)\,.
\end{equation}
\item With $\E$ is further associated a family $\bigl(\cpm^\Qu_z\bigr)_{z \in \Values}$ of completely positive superoperators acting on $TRCL(\Hilbert_\sys)$ with the compatibility property that, for all trace class operators $\trclop$,
\begin{equation}\label{ECPM}
  \tr \bigl( \trclop \, E^\Qu_z\bigr) = \tr  \cpm^\Qu_z( \trclop)\,.
\end{equation}
In case $Z=z$, the density matrix of the system at time $\tf$ (immediately after the experiment) is
\begin{equation}\label{QC}
  \dm_\tf = \dm' = \frac{\cpm^\Qu_z(\dm_\ti)}{\tr  \cpm^\Qu_z(\dm_\ti)}\,.
\end{equation}
\end{itemize}

\bigskip

Since readers may not be familiar with this formulation of the quantum formalism, we elucidate it a bit in the following subsections. We begin with a remark.

The assumption that the experiment is over at a fixed time $\tf$ is not in all practical cases satisfied, for example when the experiment measures the time at which a detector clicks. To keep this discussion simple, we postpone the discussion of experiments whose duration is random (i.e., decided upon by the experiment itself) to Section~\ref{sec:Tf}.

\subsection{First Example}

To begin with, the simplified quantum formalism is contained in the full quantum formalism in the following way: Let $\Values$ be the spectrum of the self-adjoint operator $A$ (a finite or countable set since we assume pure point spectrum), $E^\Qu(\cdot)$ the spectral PVM of $A$,
\begin{equation}\label{idealE}
  E^\Qu_z = P_z\,,
\end{equation}
and
\begin{equation}\label{idealCPM}
  \cpm^\Qu_z (\trclop) = P_z \trclop P_z
\end{equation}
for every operator $\trclop$ in the trace class. Then, the compatibility property \eqref{ECPM} is satisfied since
\[
  \tr \bigl( \trclop \, E^\Qu_z\bigr) = \tr \bigl( \trclop \,P_z\bigr) =
  \tr \bigl( P_z \, \trclop \,P_z\bigr) =
  \tr  \cpm^\Qu_z(\trclop)\,.  
\]
Eqs.\ \eqref{QPPP} and \eqref{QC} reduce to \eqref{RQPPP} and \eqref{RQC}.

In general, the set $\Values$ need not be a subset of $\RRR$. For example, an element of $\Values$---an outcome of the experiment---could be a list of numbers ($\Values \subseteq \RRR^n$), or simply a name like ``up'' or ``down''.

\subsection{Compatibility Between Superoperators and POVM}

Using the Choi--Kraus theorem
\begin{equation}\label{choiCoutcome}
  \cpm_z (\trclop) = \sum_{i\in \I_z} 
  R_{z,i} \, \trclop \, R^*_{z,i}
\end{equation}
(where we have dropped the superscript ``$\Qu$'' for ease of notation), we can show that the POVM $E(\cdot)$ associated with $\E$ is completely determined by the $(\cpm_z)_{z\in\Values}$ according to
\begin{equation}\label{POVMCPM}
  E_z = \sum_{i\in \I_z} 
  R^*_{z i} R_{z i}\,.
\end{equation}
To see this, note that the compatibility property \eqref{ECPM} implies, with \eqref{choiCoutcome}, that
\begin{equation}
  \tr \bigl( \trclop \, E_z\bigr) 
  =  \tr  \cpm_z(\trclop)
  =  \tr  \sum_{i\in \I_z} R_{z i} \, \trclop \, R^*_{z i} 
  = \tr  \sum_{i\in \I_z} \trclop\, R^*_{z i} R_{z i} \,. 
\end{equation}
This can hold for all trace class operators $\trclop$ only if \eqref{POVMCPM} holds. Moreover, it follows from \eqref{ECPM} by summing over all $z\in\Values$ that $\sum_{z\in\Values} \cpm_z$ is trace-preserving.

Conversely, suppose the $(\cpm_z)_{z\in\Values}$ are given and that the superoperator $\sum_{z\in\Values} \cpm_z$ is trace-preserving. Then \eqref{POVMCPM} \emph{defines} a POVM $E(\cdot)$ satisfying \eqref{ECPM}: $R^*_{z i} R_{z i}$ is a positive operator, and $E(\Values) =I$ because, for every vector $\psi$ in Hilbert space,
\[
  \langle\psi|E(\Values)|\psi\rangle = 
  \tr \Bigl(\pr{\psi} \sum_{z\in \Values}  
  \sum_{i} R^*_{z i} R_{z i} \Bigr) = 
\]
\[
  = \sum_{z\in \Values} \sum_{i} 
  \tr \Bigl(R_{z i} \pr{\psi} R^*_{z i}\Bigr) = \tr\sum_{z\in \Values}
  \cpm_z \bigl( \pr{\psi} \bigr) = 
  \tr\bigl( \pr{\psi}\bigr) =\|\psi\|^2\,.
\]
To see that \eqref{ECPM} holds, note that
\[
  \tr \bigl(\trclop \, E_z \bigr) = \sum_{i\in \I_z} 
  \tr(\trclop \, R^*_{z i} R_{z i}) = \sum_{i\in \I_z} 
  \tr(R_{z i} \, \trclop \,  R^*_{z i}) 
  = \tr  \cpm_z (\trclop)\,.
\]

\subsection{Another Example: Two Consecutive Experiments}

Here is an example illustrating how the POVM $E(\cdot)$ and the superoperators $\cpm_z$ arise, and how to do calculations with them. Suppose we carry out two experiments $\E_1$ and $\E_2$ in a row on the same system with a lapse of $t$ time units in between, and regard the entire procedure as one experiment $\E$ whose outcome $Z$ is given by the pair $(Z_1,Z_2)$ of outcomes of $\E_1$ and $\E_2$. Suppose we know the POVMs $E_{1,z_1}$ and $E_{2,z_2}$ (for ease of notation, we drop the superscript ``$\Qu$'') as well as the superoperators $\cpm_{1,z_1}$ and $\cpm_{2,z_2}$, and want to determine the POVM $E_z = E_{(z_1,z_2)}$ and the superoperators $\cpm_z = \cpm_{(z_1,z_2)}$ corresponding to $\E$. For example, $\E_1$ and $\E_2$ could be ideal measurements as described in the simplified quantum formalism. We will see that in that case $\E$ is (in general) not itself an ideal measurement, and $E(\cdot)$ is a proper POVM (i.e., not a PVM).

The value space of $\E$ is $\Values = \Values_1 \times \Values_2$. The joint distribution of $Z_1$ and $Z_2$, if the system starts with density matrix $\dm$, is
\[
\PPP(Z_1=z_1,Z_2=z_2) = \PPP(Z_1=z_1) \, \PPP(Z_2=z_2|Z_1=z_1) =
\]
\[
=\tr\bigl(\dm \, E_{1,z_1}\bigr) 
\, \tr\Bigl(e^{-iHt/\hbar}\frac{\cpm_{1,z_1}(\dm)}{\tr  \cpm_{1,z_1}(\dm)} e^{iHt/\hbar} E_{2,z_2} \Bigr) =
\]
[using the compatibility property \eqref{ECPM}]
\[
= \tr\Bigl(e^{-iHt/\hbar}{\cpm_{1,z_1}(\dm)} e^{iHt/\hbar} E_{2,z_2} \Bigr) =
\]
[using the Choi--Kraus theorem for $\cpm_{1,z_1}$]
\[
= \tr \Bigl(e^{-iHt/\hbar}\sum_i R_{1,z_1,i}\, \dm \,R^*_{1,z_1,i}
\, e^{iHt/\hbar} E_{2,z_2}\Bigr) =
\]
\[
=\tr\Bigl(\dm\sum_i R^*_{1,z_1,i}\, e^{iHt/\hbar}\,E_{2,z_2}\,e^{-iHt/\hbar}\, R_{1,z_1,i}\Bigr) =
\]
\begin{equation}
=\tr\bigl(\dm \, E_{(z_1,z_2)} \bigr)
\end{equation}
with
\begin{equation}\label{E1E2E}
E_{(z_1,z_2)} = \sum_i R^*_{1,z_1,i} e^{iHt/\hbar}
E_{2,z_2} e^{-iHt/\hbar}R_{1,z_1,i}\,.
\end{equation}
Note that this expression defines a POVM, since each summand is a positive operator and $E(\Values_1\times\Values_2)=I$:
\[
\sum_{z_1}\sum_{z_2} E_{(z_1,z_2)} =
\sum_{z_1}\sum_i R^*_{1,z_1,i} e^{iHt/\hbar}\underbrace{\sum_{z_2}E_{2,z_2}}_{=I} e^{-iHt/\hbar}R_{1,z_1,i}=
\]
\[
=\sum_{z_1}\sum_i R^*_{1,z_1,i} R_{1,z_1,i}=
\sum_{z_1} E_{1,z_1} =I\,.
\]
In case $\E_1$ and $\E_2$ are ideal measurements, the formula \eqref{E1E2E} reduces to
\begin{equation}
E_{(z_1,z_2)} = P_{1,z_1}\, e^{iHt/\hbar}P_{2,z_2} e^{-iHt/\hbar} P_{1,z_1}\,.
\end{equation}
If $P_{1,z_1}$ commutes with $e^{iHt/\hbar}P_{2,z_2} e^{-iHt/\hbar}$ (equivalently, if the self-adjoint operators $A_1$ and $e^{iHt/\hbar}A_2 e^{-iHt/\hbar}$ commute) then $E(z_1,z_2)$ is itself a projection, and $E(\cdot)$ is a PVM, but in general it is not.

The final density matrix after $\E_2$ is completed, given that the outcomes were $Z_1=z_1$ and $Z_2=z_2$, is
\begin{equation}
\dm'=\dm_2 = \frac{\cpm_{2,z_2}(e^{-iHt/\hbar}\dm_1e^{iHt/\hbar})}
{\tr \cpm_{2,z_2}(e^{-iHt/\hbar}\dm_1e^{iHt/\hbar})} =
\frac{\cpm_{2,z_2}(e^{-iHt/\hbar}\cpm_{1,z_1}(\dm)e^{iHt/\hbar})}
{\tr \cpm_{2,z_2}(e^{-iHt/\hbar}\cpm_{1,z_1}(\dm)e^{iHt/\hbar})}\,.
\end{equation}
That is, the superoperators corresponding to $\E$ are given by the composition law
\begin{equation}
 \cpm_{(z_1,z_2)}(\trclop) = 
 \cpm_{2,z_2}(e^{-iHt/\hbar}\cpm_{1,z_1}(\trclop)e^{iHt/\hbar})\,,
\end{equation}
which is completely positive as a composition of three completely positive superoperators: $\cpm_{1,z_1}$, the unitary evolution, and $\cpm_{2,z_2}$. If $\E_1$ and $\E_2$ are ideal measurements, so that $\cpm_{1,z_1}$ and $\cpm_{2,z_2}$ are of the form \eqref{idealCPM}, then
\begin{equation}
\cpm_{(z_1,z_2)}(\trclop) = P_{2,z_2}e^{-iHt/\hbar}P_{1,z_1}\trclop P_{1,z_1} e^{iHt/\hbar}P_{2,z_2}\,,
\end{equation}
which is not itself of the form \eqref{idealCPM}, unless $t=0$ and $P_{1,z_1}$ commutes with $P_{2,z_2}$. This exemplifies how $\cpm$ can be different from \eqref{idealCPM}.

\subsection{The Law of Operators}
\label{sec:lawop}

How does one know which POVM $(E^\Qu_{z})_{z\in\Values}$ and which family $(\cpm^\Qu_z)_{z\in\Values}$ of superoperators should be associated with $\E$? In practice, this is part of the working knowledge, and it is sometimes obtained by trial and error, or by symmetry arguments, or other methods of guessing. It is also often suggested by ``quantization rules,'' but we prefer here a rule that is generally valid (and does not appeal to classical physics).

\bigskip

\noindent\textbf{The Quantum Law of Operators.}
\begin{itemize}
\item Suppose we are given the density matrix $\dm_\app$ for the ready state of the apparatus, its Hamiltonian $H_\app$, and the interaction Hamiltonian $H_I$. Let
\begin{equation}
  U_t = e^{-\tfrac{i}{\hbar} (H_\sys+H_\app+H_I)t}
\end{equation}
be the unitary Schr\"odinger evolution operator for the composite (system $\cup$ apparatus). Let the experiment $\E$ start at time $\ti$ and be finished at time $\tf$, so that the result can be read off at $\tf$ from the apparatus.\footnote{This assumption is to be understood in an operational sense: It is assumed that we humans can read off the result when looking at the apparatus. This is different from assuming that the result can be read off from the \emph{wave function} of (the system and) the apparatus, which is notoriously not the case, a fact known as the measurement problem of quantum theory.\label{fn:readapp}} Let $P_z^\app$ be the projection to the subspace of apparatus states in which the pointer is pointing to the value $z$. Then 
\be\label{QuPOVM}
E^\Qu_z = \tr_\app 
\Bigl( [I_\sys\otimes \dm_\app] U^*_{\tf-\ti}
[I_\sys\otimes P_z^\app] U_{\tf-\ti} \Bigr)
\ee
and
\begin{equation}\label{QuCPM}
  \cpm^\Qu_z(\trclop) = \tr_\app \Bigl( [I_\sys\otimes P_z^\app] U_{\tf-\ti} 
  [\trclop \otimes \dm_\app] U_{\tf-\ti}^* [I_\sys\otimes P_z^\app] \Bigr) \,,
\end{equation}
where $\tr_\app$ denotes the partial trace over the Hilbert space of the apparatus. We check the compatibility property \eqref{ECPM} in Appendix~\ref{app:check}.
\end{itemize}

\bigskip

In other words, the superoperator $\cpm^\Qu_z$ is obtained by solving the Schr\"odinger equation for the apparatus together with the system, then collapsing the joint density matrix as if applying the collapse rule to a ``quantum measurement'' of the pointer position, and then computing the reduced density matrix of the system.

To obtain that $E^\Qu(\cdot)$ is a POVM, we need that $\sum_{z\in\Values} \cpm_z^\Qu$ is trace-preserving. Indeed,
\[
  \tr  \sum_{z \in \Values} \cpm_z^\Qu(\trclop) = 
  \sum_{z \in \Values}\tr
  \Bigl( U_{\tf-\ti} 
  [\trclop \otimes \dm_\app] U_{\tf-\ti}^* [I_\sys \otimes P_z^\app]^2 \Bigr) =
\]
\[
  = \tr \Bigl(  U_{\tf-\ti} [\trclop \otimes \dm_\app] U_{\tf-\ti}^* 
  [I_\sys\otimes\sum_{z \in \Values}  P_z^\app]\Bigr) = \tr \bigl( U_{\tf-\ti} 
  [\trclop \otimes \dm_\app] U_{\tf-\ti}^* \bigr) = \tr  \trclop\,,
\]
provided
\begin{equation}
\sum_{z \in \Values} P_z^\app = I_\app\,.
\end{equation}
(This equation amounts to the statement that the experiment always has \emph{some} outcome. This is normally not true, as, e.g., the apparatus might get destroyed by some accident with small but nonzero probability. However, we may deal with this trivial problem by assuming that the set $\Values$ of all possible outcomes contains one element representing the possibility that the experiment was not properly carried out.)

\section{The GRW Formalism}
\label{sec:GRWformalism}

\subsection{The Formalism}

The GRW formalism is very similar to the quantum formalism. There are only three differences: (i)~the unitary Schr\"odinger evolution \eqref{SchrDM} between the experiments is replaced with the master equation \eqref{M} \y{with $H=H_\sys$, $N=N_\sys$, and $\Lambda_k = \Lambda^\sys_k$}; (ii)~the POVM $E^\GRW(\cdot)$ associated with an experiment $\E$ as its ``observable'' may be different from $E^\Qu(\cdot)$, and (iii)~the superoperators $\cpm_z^\GRW$ (encoding the ``observer-induced collapse'') may be different from $\cpm^\Qu_z$. Thus, it reads as follows. \y{(Further detail about its precise meaning will be provided in Sections~\ref{sec:isolated}--\ref{sec:applicability} and \ref{sec:dm}--\ref{sec:thm1}.)}

\bigskip

\noindent\textbf{The GRW Formalism.}
\begin{itemize}
\item A system isolated from its environment has at every time $t$ a density matrix $\dm_t$ which evolves according to the master equation \eqref{M}.
\item With every experiment $\E$ with discrete value space $\Values$, beginning at time $\ti$ and ending at time $\tf$, there is associated a POVM $E^\GRW(\cdot)$ on $\Values$ acting on $\Hilbert_\sys$. When the experiment $\E$ is performed on a system with density matrix $\dm_\ti$, the outcome $Z$ is random with probability distribution
\begin{equation}\label{GRWPPP}
  \PPP(Z=z) = \tr \bigl(\dm_\ti\, E^\GRW_{z} \bigr)\,.
\end{equation}
\item With $\E$ is further associated a family $(\cpm_z^\GRW)_{z\in\Values}$ of completely positive superoperators acting on $TRCL(\Hilbert_\sys)$ with the compatibility property that for all trace-class operators $\trclop$, 
\begin{equation}\label{ECPMGRW}
  \tr \bigl( \trclop \, E^\GRW_{z}\bigr) = \tr  
  \cpm_z^\GRW( \trclop)\,.
\end{equation}
In case $Z=z$, the density matrix of the system at time $\tf$ immediately after the experiment $\E$ is
\begin{equation}\label{GRWC}
  \dm_\tf = \dm' = \frac{\cpm_z^\GRW(\dm_\ti)}
  {\tr  \cpm_z^\GRW(\dm_\ti)}\,.
\end{equation}
\end{itemize}

\bigskip

For the same reasons as for the quantum formalism, we assume a discrete value space $\Values$. In theories (such as GRWm and GRWf) with a clear PO, on the other hand, one might consider experiments using an ``analog'' rather than ``digital'' display, for example ones in which the outcome is displayed as the center-of-mass position of a pointer. However, even in this case it is reasonable to regard the outcome as discrete, since it is hard to regard microscopic details of the pointer's PO as a means to display information about the outcome.

Corresponding to the simplified quantum formalism, one can also formulate a simplified GRW formalism: For \emph{suitable} (but not all) experiments $\E$ it so happens that $E^\GRW(\cdot)$ is a PVM (i.e., that $E^\GRW(B)$ is a projection for all subsets $B \subseteq \Values$), that $\Values$ is a subset of $\RRR$, and that $\cpm_z^\GRW(\dm) = P_z \, \dm \, P_z$ for suitable projections $P_z$. In this case, all the data encoding information about $\E$ needed for computing outcomes (i.e., $\Values$, $E^\GRW(\cdot)$, and $(\cpm^\GRW_z)_{z\in\Values}$) can be encoded into a single self-adjoint operator, $A = \sum_{z \in \Values} z P_z$.
\y{The differences between the simplified quantum formalism and the simplified GRW formalism are: the unitary Schr\"odinger evolution is again replaced with the master equation \eqref{M}; the class of experiments $\E$ for which the simplified quantum formalism is appropriate when $\E$ is performed in a quantum world may be different from the class of $\E$s for which the simplified GRW formalism is appropriate when $\E$ is performed in a GRW world; and even if, for an experiment $\E$, both the simplified quantum formalism and the simplified GRW formalism are appropriate then the operator $A^\GRW$ may be different from $A^\Qu$.} 

\bigskip

\noindent\textbf{The GRW Law of Operators.}
\begin{itemize}
\item Suppose we are given the density matrix $\dm_\app$ for the ready state of the apparatus, its Hamiltonian $H_\app$, and the interaction Hamiltonian $H_I$, so that $H = H_\sys + H_\app + H_I$. Let the experiment $\E$ start at time $\ti$ and be finished at time $\tf$, and let $\zeta: \Omega_{[\ti,\tf)} \to \Values$ be the function that reads off the outcome of $\E$ from the flashes between $\ti$ and $\tf$. Then $E^\GRW(\cdot)$ is given by \y{the following generalization of \eqref{Efdef}:} 
\begin{align}\label{GRWE}
E^\GRW_{z}& = \tr_\app\Bigl( [I_\sys\otimes \dm_\app] \,
\G\bigl(\zeta^{-1}(z) \bigr) \Bigr)\\
\label{GRWE2}
&= \tr_\app 
  \int\limits_{\zeta^{-1}(z)} df \: 
  [I_\sys \otimes \dm_\app] \, L^*_{[\ti,\tf)}(f) \, L_{[\ti,\tf)}(f)\,,
\end{align}
where $f=f_{\sys\cup\app}$ and $L_{[\ti,\tf)}=L_{[\ti,\tf)}^{\sys\cup\app}$, and
\begin{equation}\label{GRWCPM}
   \cpm^\GRW_z (\trclop) = \tr_\app  
  \int\limits_{\zeta^{-1}(z)} df \: 
  L_{[\ti,\tf)}(f) \, [\trclop \otimes \dm_\app] \, L^*_{[\ti,\tf)}(f)\,.
\end{equation}
\end{itemize}

\bigskip

We check the compatibility property \eqref{ECPMGRW} in Appendix~\ref{app:check}.

Before we begin the derivation of the GRW formalism, we have to elucidate a bit more what exactly it asserts.

\subsection{Isolated System}
\label{sec:isolated}

The ``system'' is mathematically represented, as described in Section~\ref{sec:sys}, by a splitting $\Hilbert = \Hilbert_\sys \otimes \Hilbert_\env$ of Hilbert space, as well as a splitting $F=F_\sys \cup F_\env$ of the flashes, grounded in either a set $\subLab$ of labels or a region $\region_\sys\subseteq \RRR^3$ (or both) selecting $F_\sys$. When we say that a system is \emph{isolated} or \emph{does not interact with its environment}, we mean two things: First, the Hamiltonian does not contain an interaction term, that is,
\be\label{nointH}
H = H_\sys \otimes I_\env + I_\sys \otimes H_\env\,.
\ee
Second, the collapse operators associated with flashes of the system act only on $\Hilbert_\sys$ but not on $\Hilbert_\env$, and vice versa:
\be\label{Lambdasys}
\Lambda_i(x) =  \begin{cases}
\Lambda_i^\sys(x) \otimes I_\env & \text{if } i\in \subLab \text{ and } x\in \region_\sys\\
I_\sys \otimes \Lambda_i^\env(x) & \text{otherwise.}
\end{cases}
\ee
This second condition, apart from expressing that the splitting $\Hilbert=\Hilbert_\sys \otimes \Hilbert_\env$ is compatible with the splitting $F=F_\sys \cup F_\env$, is necessary because otherwise the system could, despite the absence of an interaction Hamiltonian, interact through collapses with the environment; e.g., an initial product wave function could become entangled.

A basic mathematical fact about isolated systems is the \emph{factorization formula}
\be\label{factformL}
L(f)=L^\sys(f_\sys)\otimes L^\env(f_\env)
\ee
and similarly
\be\label{factformLst}
L_{[s,t)}(f) = L^\sys_{[s,t)}(f_\sys) \otimes L^\env_{[s,t)}(f_\env)\,.
\ee
They are analogs of the formula 
\be\label{factformU}
U_t = e^{-iH_\sys t/\hbar} \otimes e^{-iH_\env t/\hbar} = U_t^\sys \otimes U_t^\env
\ee
for the unitary time evolution, which holds when \eqref{nointH} does. In \eqref{factformL} and \eqref{factformLst}, $f_\sys$ (respectively $f_\env$) is the set of flashes belonging to the system (respectively the environment) and, as the notation suggests,
\begin{multline}
L^\sys\Bigl((x_1,t_1,i_1),\ldots, (x_n,t_n,i_n)   \Bigr) =
 \lambda^{n/2} e^{-N_\sys \lambda(t_n-t_0)/2} \:\times \\
\times \: \Lambda^\sys_{i_n}(x_n)^{1/2} U^\sys_{t_n-t_{n-1}} \cdots
\Lambda^\sys_{i_1}(x_1)^{1/2} U^\sys_{t_1-t_0}\,.
\end{multline}
with $N_\sys = \# \subLab$, and similarly for $L^\env$, $L^\sys_{[s,t)}$, and $L^\env_{[s,t)}$. 

For \eqref{factformLst} it is sufficient that the system ``$\sys$'' be isolated during $[s,t)$. Equations~\eqref{factformL} and \eqref{factformLst} follow from \eqref{factformU}, \eqref{Lambdasys}, the definitions \eqref{eq:long} and \eqref{Ldef} of $L_{[s,t)}$ and $L$, and the fact that $(A \otimes B)(C \otimes D) = (AC)\otimes (BD)$.

\subsection{Density Matrix}
\label{sec:dm1}

Density matrices can arise in two ways: either as representing a statistical mixture (or ensemble) of wave functions, or as the reduced density matrix of a system entangled with another system (which we will call system $b$ in the following, while system $a$ is the system of interest). Both types of density matrices are allowed in the GRW formalism: the system under consideration may be entangled with system $b$ (but not to the apparatus of the experiment), and the wave function (of the two systems together) may be random. It is part of the statement of the GRW formalism that, in this case, (i)~the density matrix $\dm_t$ of the system still evolves according to the master equation \eqref{M} as long as it remains isolated (from system $b$, from the apparatus, and from everything else); (ii)~the statistics of the outcome $Z$ depends only on the density matrix of the system (and not on how it arises); (iii)~in case $Z=z$ the system's reduced density matrix after the experiment is given by \eqref{GRWC}.

We note that the density matrix $\dm_t$ of a system, of which the GRW formalism asserts that it evolves according to the master equation \eqref{M}, does not provide a complete description of the quantum state of the system, even when the initial density matrix $\dm_0$ was pure. After all, the master equation corresponds to averaging over the flashes between the initial time 0 of the system's isolated evolution and the time $\ti$ at which the interaction with an apparatus begins, while the stochastic GRW evolution of the wave function $\wf_t$ corresponds to taking these flashes into account.

\subsection{Conditions of Applicability}
\label{sec:applicability}

Let us make explicit the assumptions we will make in the derivation of the GRW formalism, i.e., the conditions under which the GRW formalism is applicable. The system, called system $a$ in the following, may be entangled with another system called system $b$. We suppose that
\begin{enumerate}
\item the experiment $\E$ involves a splitting of the world into four parts: system $a$ (the ``object'' of $\E$), system $b$, the apparatus of $\E$, and the rest of the world;
\item $\E$ begins at time $\ti$ and ends at time $\tf$;\footnote{This assumption will be relaxed in Section~\ref{sec:Tf}, where we allow that the experiment's run-time is not fixed before the experiment.}
\item system $a$, system $b$, and the apparatus together form a GRW system (i.e., the system is isolated) during the time interval $[\ti,\tf)$, and this system possess a wave function $\Psi_{t'}$, $\ti\leq t'< \tf$;
\item at time $\ti$, the apparatus is not entangled with system $a\cup b$,
\be\label{factorizeab}
\Psi_\ti = \psi_{a\cup b} \otimes \phi \,,
\ee
where $\psi_{a\cup b}$ is the (possibly random) wave function of systems $a$ and $b$ together at time $\ti$, $\phi$ is the (possibly random) wave function of the apparatus at time $\ti$;
\item\label{item:indep} $\psi_{a\cup b}$ and $\phi$ are independent random variables;
\item\label{item:nob} during $[\ti,\tf)$ the apparatus interacts only with system $a$, while system $a$ and the apparatus do not interact with system $b$;
\item\label{item:last} the outcome $Z$ is a function $\zeta$ of the flashes of the apparatus during $[\ti,\tf)$; this assumption can be weakened by allowing that $Z$ is read off from the flashes of both system $a$ and the apparatus, 
\be\label{zetaab}
Z = \zeta(F^{a\cup\app}_{[\ti,\tf)})\,,
\ee
while we need to exclude a direct dependence of $Z$ on the flashes of system $b$.
\end{enumerate}

\subsection{Smallness of Deviations From the Quantum Formalism}
\label{sec:deviations}

In this subsection, we characterize the ``quantum regime'' of the GRW theories, i.e., the regime in which the GRW formalism agrees with the quantum formalism. We do so in a sketchy way by comparing the laws of operators in the quantum and the GRW formalism, \eqref{QuPOVM} and \eqref{GRWE}, which we repeat here for convenience:
\be\label{QuPOVMrepeat}
E^\Qu_z = \tr_\app 
\Bigl( [I_\sys\otimes \dm_\app] U^*_{\tf-\ti}
[I_\sys\otimes P_z^\app] U_{\tf-\ti} \Bigr)\,,
\ee
\begin{equation}\label{GRWErepeat}
  E^\GRW_{z} = \tr_\app  
  \int\limits_{\zeta^{-1}(z)} df \: 
  [I_\sys \otimes \dm_\app] \, L^*_{[\ti,\tf)}(f) \, L_{[\ti,\tf)}(f)\,.
\end{equation}
We take for granted that $\dm_\app$ is the same in both expressions, and that it is sufficient to consider $\dm_\app=\pr{\phi}$. We provide a condition under which 
\be\label{GRWapproxQu}
E^\GRW_z \approx E^\Qu_z\,.
\ee
The condition is the conjunction of the following:

\begin{enumerate}
\item During the experiment $\E$, collapses are likely to occur only in the apparatus, not in the system,
\be\label{Fsysempty}
\PPP(F^\sys_{[\ti,\tf)}=\emptyset) \approx 1\,.
\ee
Equivalently, the average time between collapses in the system is much larger than the duration of $\E$,
\be\label{Nsyslambda}
\frac{1}{N_\sys \lambda} \gg \tf-\ti\,,
\ee
where $\lambda$ is the collapse rate per particle.

\item The pointer states for different $z$, i.e., the vectors in the range of $P_z^\app$, are separated in position space. To be specific, let there be (macroscopic) regions $\region_z\subset \RRR^3$ in position space, mutually disjoint, so that for every $\psi$, the wave function $P_z^\app\psi$ is concentrated on the subset of configuration space with all ``particles'' belonging to the tip of the pointer in $\region_z$. 

\item  The duration $\tf-\ti$ of $\E$ is long enough for macroscopic superpositions of the pointer to decay,
\be
\frac{1}{N_\mathrm{tip}\lambda}\ll \tf-\ti
\ee
with $N_\mathrm{tip}$ the number of ``particles'' at the tip of the pointer.

\item The experiment is such that for every $z\in\Values$ and every $\psi\in\SSS(\Hilbert_\sys)$,
the part $\Phi_z$ of $\Psi_{\ti}=\psi\otimes\phi$ (with $\phi$ the initial wave function of the apparatus) that would lead to outcome $z$ under the unitary evolution,
\be
\Phi_z = U^{-1}_{\tf-\ti}(I_\sys\otimes P_z^\app)U_{\tf-\ti}\Psi_\ti\,,
\ee
evolves under the GRW collapse evolution associated with  the GRW process $\Psi_{t'}$, $t'\geq \ti$, starting from $\Psi_{\ti}$ to a wave function
\be
\Phi_{z,\tf} = L_{[\ti,\tf)}(F_{[\ti,\tf)}) \Phi_z
\ee
that is with probability $\approx 1$ near the range of $P_z^\app$, i.e.,
\be
\Phi_{z,\tf} \approx (I_\sys\otimes P_z^\app) \Phi_{z,\tf}\,.
\ee
(In particular, this holds if $\Phi_{z,\tf}\approx U_{\tf-\ti} \Phi_z$.)

\end{enumerate}

Condition~1 is satisfied by the standard choice $\lambda \approx 10^{-16} \, \mathrm{s}^{-1}$ for microscopic systems (say, $N_\sys\leq 10^5$) if the duration of $\E$ is less than 100 years. Likewise, Conditions 2 and 3 are satisfied if the duration is more than (say) $10^{-5}\,\mathrm{s}$ and the outcome is represented by the position of a pointer that is a macroscopic object.

Condition 4 needs elaboration. Why is any further condition needed besides 1--3? That is because the working of the apparatus might deviate in GRWf from that in quantum mechanics. As an extreme example, the apparatus could contain a device that carries out an empirical test of GRWf versus quantum mechanics; such a device is not feasible with present technology but is in principle; then the apparatus may be so constructed as to do something different with the system ``$\sys$'' depending on whether it finds itself in a GRWf world or in a quantum world. In this case, not excluded by conditions 1--3, $E_z^\GRW$ could be arbitrarily different from $E_z^\Qu$.

Condition 4 holds in particular for an ideal quantum measurement, i.e., if there is an orthonormal basis $\{\psi_n\}$ of $\Hilbert_\sys$ such that for each $\psi_n$ the outcome is deterministic, $Z=f(n)$, and equal in GRW and quantum mechanics.\footnote{In this situation, it can in fact be concluded directly that $E^\GRW_z=E^\Qu_z$. Indeed, if $\scp{\psi_n}{E_z|\psi_n}=\delta_{z,f(n)}$ and $0\leq E_z\leq I$ then $E_z=\sum_{n:f(n)=z} \pr{\psi_n}$ (and thus $E^\GRW_z=E_z=E^\Qu_z$). After all, suppose an off-diagonal entry were nonzero, $c:=\scp{\psi_n}{E_z|\psi_m}\neq 0$ for $n\neq m$, and let $\psi=\alpha\psi_n+\beta\psi_m$ with $|\alpha|^2+|\beta|^2=1$; if $f(n)=z=f(m)$ then $\scp{\psi}{E_z|\psi}= 1+2\,\Re(\alpha^*c\beta)$ can be made $>1$ by suitable choice of $\alpha,\beta$; if $f(n)\neq z \neq f(m)$ then $\scp{\psi}{E_z|\psi}=2\,\Re(\alpha^*c\beta)$ can be made negative; if $f(n)=z\neq f(m)$ then $\scp{\psi}{E_z|\psi} = |\alpha|^2+2\,\Re(\alpha^*c\beta)$ can be made $>1$ and can be made $<0$.}

\bigskip

We now turn to the derivation of $E^\GRW_z \approx E^\Qu_z$ from Conditions~1--4. 

\bigskip

Consider the GRW flash process $F$ for the initial wave function $\Psi_0=\Psi\in\Hilbert$. For any $\Phi\in\Hilbert$, the process
\be
Y^\Phi_t=\frac{\bigl\|L_{[0,t)}(F_{[0,t)})\Phi\bigr\|^2}{\bigl\|L_{[0,t)}(F_{[0,t)})\Psi\bigr\|^2}
\ee
is a martingale, i.e.,
\be
\EEE\bigl( Y^\Phi_t \big| F_{[0,s)} \bigr) = Y^\Phi_s
\quad \forall s<t\,.
\ee
Proof: We emphasize that the distribution of $F$ is governed by $\Psi$, not by $\Phi$. Recall that $L_{[0,t)}(f_{[0,t)}) = L_{[s,t)} (f_{[s,t)}) \, L_{[0,s)}(f_{[0,s)})$. Thus, 
\begin{align}
\EEE\bigl( Y^\Phi_t \big| F_{[0,s)} \bigr) 
&= \EEE \Biggl( \frac{\bigl\|L_{[0,t)}(F_{[0,t)})\Phi\bigr\|^2}{\bigl\|L_{[0,t)}(F_{[0,t)})\Psi\bigr\|^2} \Biggm| F_{[0,s)} \Biggr)\\
&= \int\limits_{\Omega_{[s,t)}} \!\!\! df_{[s,t)} \, \frac{\| L_{[s,t)}(f_{[s,t)}) \,L_{[0,s)} (F_{[0,s)}) \Psi \|^2}{\| L_{[0,s)} (F_{[0,s)})\, \Psi\|^2} \frac{\|L_{[s,t)}(f_{[s,t)}) \, L_{[0,s)}(F_{[0,s)}) \,\Phi\|^2}{ \bigl\|L_{[s,t)}(f_{[s,t)}) \, L_{[0,s)}(F_{[0,s)})\,\Psi\bigr\|^2}\\
&= \frac{1}{\|L_{[0,s)}(F_{[0,s)}) \Psi\|^2}\:\times\nonumber\\
&\quad \times \: \Bscp{L_{[0,s)}(F_{[0,s)}) \Phi}{\underbrace{\int\limits_{\Omega_{[s,t)}} \!\!\! df \, L^*_{[s,t)}(f)\,L_{[s,t)}(f)}_{=I} \Big| L_{[0,s)}(F_{[0,s)}) \Phi}\\
&=\frac{\|L_{[0,s)}(F_{[0,s)}) \Phi\|^2}{\|L_{[0,s)}(F_{[0,s)}) \Psi\|^2}=Y^\Phi_s\,.
\end{align}
This completes the proof of the martingale property.

By the martingale convergence theorem, $Y^\Phi_t$ has a limit as $t\to\infty$, which we call $Y^\Phi_\infty$. Now consider the experiment $\E$, let $\ti$ and $\tf$ denote the times at which $\E$ starts and ends, let $\Hilbert=\Hilbert_\sys\otimes \Hilbert_\app$, regard $\ti$ as the initial time, consider any $\psi\in\SSS(\Hilbert_\sys)$, let $\phi\in\SSS(\Hilbert_\app)$ be the initial wave function of the apparatus, and set $\Psi_\ti=\psi\otimes\phi$. Set 
\be
\Phi_{z} = U^{-1}_{\tf-\ti}(I_\sys\otimes P_z^\app)U_{\tf-\ti} \Psi_\ti
\ee
and let $Y^z_{t'}=Y^{\Phi_z}_{t'}$, $t'\geq \ti$, be the martingale associated with $\Phi_{z}$. By condition 3, the duration of $\E$ is long, so we can approximate the value of $Y^z_{t}$ at the end of $\E$ by $Y^z_{\infty}$. Set 
\be
\Phi_{z,\tf}=L_{[\ti,\tf)}(F_{[\ti,\tf)}) \Phi_z\,.
\ee
By condition 4, $\Phi_{z,\tf}\approx (I_\sys\otimes P_z^\app) \Phi_{z,\tf}$. By condition 2, the pointer states are separated in 3-space, so the $\Phi_{z,\tf}$ are separated in 3-space. Therefore, only for one value $z_0$ of $z$ is $Y^z_{\infty}$ nonzero; otherwise, $\Psi_\tf$ would be a non-trivial superposition of several pointer states (i.e., of contributions from the ranges of $P_z^\app$ for different $z$), and any further flash would change the weights in this superposition; but since the $Y^z_{\infty}$ have already converged they cannot change any more; so $\Phi_{z,\tf}\approx 0$ except for $z=z_0$. Since
\be
\Psi_{\tf}= \frac{\sum_z \Phi_{z,\tf}}{\| L_{[\ti,\tf)}(F_{[\ti,\tf)}) \Psi_\ti\|}\,,
\ee
we have that
\be
\Psi_{\tf}\approx \frac{\Phi_{z_0,\tf}}{\| L_{[\ti,\tf)}(F_{[\ti,\tf)}) \Psi_\ti\|}
\ee
and, as a consequence, $Y^{z_0}_{\tf}\approx 1$. The flashes for the tip-of-the-pointer particles around time $\tf$ will then likely be located in $\region_{z_0}$, so that the outcome is $Z=z_0$. Furthermore, since $Y^z_\tf\approx 1$ for $z=Z$ and $Y^z_\tf\approx 0$ otherwise, the distribution of the outcome is
\be
\PPP(Z=z) \approx \EEE\, Y^z_\tf=Y^z_{\ti} = \| \Phi_{z}\|^2
=\|(I_\sys\otimes P_z^\app)U_{\tf-\ti} \Psi_\ti\|^2\,,
\ee
which is the quantum probability. Since $\psi$ was arbitrary, \eqref{GRWapproxQu} follows.

\section{Derivation of the GRW Formalism}

After some preparatory considerations in Section~\ref{sec:dm}, we derive the GRW formalism from GRWf in Section~\ref{sec:derivation}. 

\subsection{Density Matrix}
\label{sec:dm}

We need to collect some facts about density matrices in GRWf.

\subsubsection{Statistical Density Matrix}

Set, for ease of notation, $t_0=0$. Since the wave function $\wf_t$ is random, with its distribution there is associated the density matrix
\begin{equation}\label{dm}
  \dm_t = \EEE \pr{\wf_t} =
  \int\limits_{\sphere(\Hilbert)} \PPP_{\wf_{0}}(\wf_t \in d \Phi) \, \pr{\Phi}\,,
\end{equation}
where $\sphere(\Hilbert) = \{\wf \in \Hilbert: \|\wf\|=1\}$ is the unit sphere in Hilbert space $\Hilbert$. In other words, \eqref{dm} is the density matrix of a large ensemble of systems, each of which started with the same initial wave function $\wf_{0}$ but experienced collapses independently of the other systems. 

We note \y{without proof} that the density matrix $\dm_t$ obeys the master equation \eqref{M}. But the validity of \eqref{M} is even wider: Suppose that even the initial wave function $\wf_{0}$ is random, with distribution given by any probability measure $\mu_{0}$ on $\sphere(\Hilbert)$. Then, for $t>0$, $\wf_t$ is doubly random, because of the random initial wave function and of the stochastic GRW evolution, with distribution
\begin{equation}
\mu_t(\cdot) = \int \mu_{0}(d\wf_{0}) \: \PPP_{\wf_{0}}(\wf_t \in \cdot)\,.
\end{equation}
Again, the corresponding density matrix
\begin{equation}\label{statdm}
\dm_t = \EEE_{\mu_0} \pr{\Psi_t} =
\int \mu_t(d\wf) \, \pr{\wf}
\end{equation}
obeys \eqref{M}. To see this, note that it satisfies
\begin{equation}
\dm_t = \int \mu_0(d\wf_0) \int \PPP_{\wf_0}(\wf_t \in d\Phi) \: \pr{\Phi}\,,
\end{equation}
where the inner integral obeys \eqref{M}, so that $\dm_t$ is a mixture of solutions of \eqref{M} and therefore is itself a solution of \eqref{M}.

Alternatively, $\dm_t$ can directly be expressed in terms of $\dm_0$ according to
\begin{equation}\label{dm0t}
\dm_t = \acpm_{[0,t)} \dm_0 =
\int\limits_{\Omega_{[0,t)}} df\, L_{[0,t)}(f) \, \dm_{0} \, L^*_{[0,t)}(f)\,.
\end{equation}
From this the master equation \eqref{M} can be obtained by differentiation with respect to $t$. As a by-product, it can be read off from \eqref{dm0t} that the mapping $\acpm_{[0,t)}:\dm_0 \mapsto \dm_t$ obtained by evolving the density matrix $\dm$ according to the master equation \eqref{M} is a completely positive superoperator. It is also clear that $\acpm_{[0,t)}$ is trace-preserving.

The following proposition is a consequence of the fact that the distribution of flashes is given by a POVM: If the initial wave function $\wf_0$ is random with distribution $\mu_0$, then the distribution of the flashes depends only on the density matrix $\dm_0$ associated with $\mu_0$,
\be\label{PPPdm}
\PPP(F\in \cdot) = \int \mu_0(d\wf_0) \PPP_{\wf_0}(F\in \cdot) = \int \mu_0(d\wf_0) \, \scp{\wf_0}{\G(\cdot)|\wf_0} = \tr \bigl( \dm_0 \, \G(\cdot) \bigr)
\ee
with 
\be
\dm_0 = \int_{\sphere(\Hilbert)} \mu_0(d\wf_0) \, \pr{\wf_0}\,.
\ee
In other words, if two probability distributions $\tilde\mu_0$ and $\mu_0$ have the same density matrix, $\tilde\dm_0=\dm_0$, then they lead to the same distribution of the PO. For comparison, this is not true in Bohmian mechanics or GRWm: there, $\tilde\mu_0$ and $\mu_0$ may lead to different trajectories \cite{Bell80,dm} respectively to different probability distributions of the $m$ function \cite{AGTZ06}.

Since $\tilde\mu_0$ and $\mu_0$ lead to the same distribution of flashes, we may write $\PPP_{\dm_0}$ for that distribution. This also means that we can simply talk of the flash process for a given initial density matrix, as opposed to the flash process for a given initial wave function. As time proceeds, the density matrix determining the distribution of the flashes evolves according to the master equation in the sense that
\be\label{PdmM}
\PPP_{\dm_0}(F_{[t,\infty)}\in B) = \PPP^{(t)}_{\dm_t}(F_{[t,\infty)}\in B)\,,
\ee
where the right hand side refers to the distribution of the flashes when starting with $\dm_t$ at time $t$. This fact follows from the conditional probability formula by averaging over $F_{[0,t)}$.

\subsubsection{The Marginal Probability Formula}
\label{sec:margprob}

The \emph{marginal probability formula} expresses that a system which does not interact with its environment is itself governed by GRWf, even if the system is entangled with the environment. (Note that this is not true, e.g., in Bohmian mechanics, where the trajectories of the system's particles depend on the configuration of the environment, even in the absence of interaction. As we will see, it is not true in GRWm either.) 

The marginal probability formula says that for an isolated system,
\be\label{margprob}
\PPP_{\wf_0}(F_\sys \in B) = \PPP_{\dm_\sys}(B) \,.
\ee
Here, $\PPP_{\wf_0}$ is the distribution of the flashes in a universe starting with wave function $\wf_0$ at time $t_0=0$, and $\PPP_{\wf_0}(F_\sys\in \cdot)$ is the marginal distribution of the system's flashes; $\dm_\sys=\tr_\env \, \pr{\wf_0}$ is the reduced density matrix of the system; finally, $\PPP_{\dm_\sys}$ is the distribution of flashes in a universe containing nothing but the system and starting with density matrix $\dm_\sys$ at time $0$ in the sense of equation \eqref{PPPdm}:
\be
\PPP_{\dm_\sys} (\cdot) = \tr\bigl( \dm_\sys \, \G_\sys(\cdot) \bigr)\,.
\ee
We provide a proof of the marginal probability formula in Appendix~\ref{app:margprob}.

The marginal probability formula was first derived by Bell \cite{Bell87} for the purpose of proving a no-signalling theorem for GRWf. To see the connection, suppose the system is Alice's lab, which does not interact with Bob's lab (e.g., because they are, when considering the relevant time intervals, spacelike separated); then the distribution of the flashes in Alice's lab, and thus in particular the distribution of the outcome of any experiment, does not depend on the common wave function $\wf_0$ except through the reduced density matrix $\dm_\sys$, nor on external fields at work in Bob's lab (since $\dm_\sys$ does not).

The marginal probability formula should not be confused with the following simple consequence of the \emph{function property} \eqref{fPOVM}: Since $F_\sys$ is a function of $F$, its distribution is given by a POVM $E(\cdot)$,
\be
\PPP_{\wf_0}(F_\sys \in B) = \scp{\wf_0}{E(B)|\wf_0}\,.
\ee
The marginal probability formula goes further in two respects: First, its right hand side depends only on the reduced density matrix $\dm_\sys$, and not on the entire wave function $\wf_0$; second, the POVM $\G_\sys(\cdot)$ is not just \emph{some} POVM but exactly the one that would govern the flashes if the universe contained nothing but the system.

A related fact is the \emph{independence property}: If a system does not interact with its environment \emph{and} is initially disentangled from its environment, then the flashes of the system and those of the environment are stochastically independent, i.e., their joint distribution is a product:
\be\label{independence}
\PPP_{|\sys\rangle\otimes |\env\rangle}(F_\sys \in B_\sys, F_\env\in B_\env) =
\PPP_{|\sys\rangle}(F_\sys\in B_\sys) \,\PPP_{|\env\rangle}(F_\env \in B_\env)\,.
\ee
Moreover, in that case the wave function $\wf_t$ remains a product at later times.

In GRWm there is a formula that is in a way analogous to the marginal probability formula of GRWf, as it connects $\dm_\sys$ to the PO of the system, namely to $m_\sys$ as introduced in \eqref{msysdef}. However, it is much weaker as it connects $\dm_\sys$ not to the entire future history of the PO, for $t\geq 0$, but just to the PO at $t=0$. This formula reads
\be\label{msysdmsys}
m_\sys(x,t=0) = \sum_{i\in \subLab}
m_i \int  dq_\sys \, \delta(q_{\sys,i}-x) \, \scp{q_\sys}{\dm_\sys|q_\sys}
\ee
assuming, for simplicity, that the system is defined in terms of a label set $\subLab$, not of a region $\region_\sys$. As before, $\dm_\sys=\tr_\env \, \pr{\wf_0}$. The formula implies that a different wave function $\tilde\wf_0\neq \wf_0$ with $\tr_\env \, \pr{\tilde\wf_0} = \tr_\env \, \pr{\wf_0}$ would lead to the same $m_\sys$. An analogous statement holds in Bohmian mechanics: the marginal distribution of $Q_\sys$ at $t=0$ depends only on $\dm_\sys$. Note that in GRWm, $m_\sys$ cannot be obtained from a statistical density matrix.

\label{sec:GRWsystem}

Returning to GRWf, we call a system a \emph{GRW system} if the distribution of the flashes of the system (after time $0$) is given by $\dm_\sys$ (at time $0$), i.e., if \eqref{margprob} holds. The marginal probability formula thus asserts that \emph{every isolated system is a GRW system}---a system whose PO behaves as if the system were alone in the universe. Conversely, if a system is not isolated then it cannot be expected to be a GRW system since the interaction with the environment should affect the pattern of flashes.

Now that we have the concept of GRW system, one conclusion we can draw is that the reasoning of Section~\ref{sec:emergeop} applies not just to the universe as a whole but also when the \y{system (i.e., the object of the experiment)} and the apparatus together form a GRW system: \y{Assuming $\dm_{\sys\cup\app}=\dm_\sys\otimes \dm_\app$ and $Z=\zeta(F_{\sys\cup\app})$, we obtain that 
\be
\PPP(Z\in B) 
= \tr\bigl(\dm_{\sys\cup\app}\, \G_{\sys\cup\app}\circ\zeta^{-1}(B)\bigr)
= \tr(\dm_\sys\, E(B))\,,
\ee
where the POVM $E(\cdot)$ does not depend on anything outside $\sys\cup\app$:}
\be
E(B) = \tr_\app \Bigl( \dm_\app \, \G_{\sys\cup\app}\circ\zeta^{-1}(B)\Bigr)\,.
\ee

A variant of the marginal probability formula asserts the following: If a system is isolated during $[0,t)$ then
\be\label{margprob2}
\PPP_{\wf_0}(F^\sys_{[0,t)} \in B) = \PPP_{\dm_\sys}(F_{[0,t)}\in B) \,.
\ee
Here, a system can stop being isolated because the Hamiltonian or the collapse rate operators are time-dependent, $H=H_t$ and $\Lambda_i(x)=\Lambda_{i,t}(x)$.

The fact \eqref{margprob2} follows from the first version \eqref{margprob} of the marginal probability formula: Consider a hypothetical universe whose time-dependent Hamiltonian $H_t$ and collapse operators $\Lambda_{i,t}(x)$ are whatever we choose. Then, for a fixed initial wave function $\Psi_{0}$, the distribution of flashes during $[0,t)$ will depend on our choices of $H_s$ for all $s\in [0,t)$, but not for $s\geq t$. In particular, if the system is initially isolated, we can turn on the interaction with its environment at time $t$, and the distribution of the flashes up to time $t$ is the same as it would have been if the system were isolated forever, and thus given by \eqref{margprob}.

\subsubsection{The Marginal Master Equation} 
\label{sec:mM}

The \emph{marginal master equation}
\be\label{mM}
(\dm_t)_\sys = (\dm_\sys)_t
\ee
expresses the related fact that also the reduced density matrix of the system, when isolated from but entangled with its environment, evolves according to the master equation \eqref{M}. This is a general fact about the master equation, which can also be expressed by saying that when the system and the environment do not interact, the following diagram commutes:
\be\label{margM}
\begin{array}{rcl}
\dm_0 & \stackrel{\tr_\env}{\longrightarrow} & \dm^\sys_0\\
\eqref{M} \Big\downarrow \:\: && \Big\downarrow \eqref{M}_\sys\\
\dm_t & \stackrel{\tr_\env}{\longrightarrow} & \dm^\sys_t
\end{array}
\ee
Here, \eqref{M}$_\sys$ means the master equation \eqref{M} applied to the system.
In words, the marginal of the master equation is again a version of the master equation: the version that would hold if the universe contained nothing but the system. In another notation, $\acpm^\sys_{[0,t)} \circ \tr_\env = \tr_\env \circ \acpm_{[0,t)}$. The marginal master equation allows us to write $\dm^\sys_t$ instead of either $(\dm_\sys)_t$ or $(\dm_t)_\sys$.

We provide an analytic-algebraic proof of the marginal master equation in Appendix~\ref{app:margM}. Alternatively, here are two derivations from the marginal probability formula: First, since the distribution of the flashes of the system depends only on $(\dm_\sys)_{t=0}$, and for an isolated system the collapses associated with the flashes of the environment (as well as the Hamiltonian evolution) act trivially on $\Hilbert_\sys$, $(\dm_t)_\sys$ depends only on $(\dm_\sys)_{t=0}$, \y{as we see from the evolution law \eqref{M} of $\dm_t$: for example, if $k\notin \subLab$ then $\Lambda_k= I_\sys\otimes\Lambda_k^\env$ by \eqref{Lambdasys}, and thus}
\be
 \tr_\env\Bigl( \lambda\int d^3x \, \Lambda_k^{1/2}(x)\, \dm_t\, \Lambda_k^{1/2}(x) 
 - \lambda \dm_t \Bigr) = 0\,.
\ee
Since for an empty environment $(\dm_t)_\sys$ would trivially equal $(\dm_\sys)_t$, \eqref{mM} must be generally true. Second, the significance of the density matrix associated with the system at time $t$ lies in governing the distribution of the flashes after $t$. Thus if, as the marginal probability formula tells us, the distribution of the system's flashes after $t$ is the same as if the system were alone in the universe and started with $(\dm_\sys)_{0}$, namely $\tr\bigl((\dm_\sys)_t\, \G^\sys_{[t,\infty)}(\cdot)\bigr)$, then the system's density matrix at time $t$ must be $(\dm_\sys)_t$. On the other hand, by the marginal probability formula applied to time $t$, the distribution of the system's flashes after $t$ is $\tr\bigl((\dm_t)_\sys\, \G^\sys_{[t,\infty)}(\cdot) \bigr)$, so the density matrix at time $t$ must be $(\dm_t)_\sys$. (Mathematically, this argument assumes that the family of operators $\{\G^\sys_{[t,\infty)}(B):\text{any }B\}$, is sufficiently rich.)

Here is another derivation of the marginal master equation that readers may find illuminating: If the system ``$\sys$'' is isolated from its environment then
\be\label{acpmfactor}
\acpm^{\sys\cup\env}_{[0,t)} = \acpm^\sys_{[0,t)}\otimes \acpm^\env_{[0,t)}\,.
\ee
Since $\acpm^\env_{[0,t)}$ is trace-preserving, we obtain for $\dm=\dm^{\sys\cup\env}$ that
\be
\tr_\env \acpm^{\sys\cup\env}_{[0,t)}(\dm ) 
= \tr_\env \bigl[  \acpm^\sys_{[0,t)}\otimes \acpm^\env_{[0,t)} (\dm ) \bigr] 
= \acpm^\sys_{[0,t)} \tr_\env \dm\,,
\ee
i.e., $\acpm^\sys_{[0,t)}\circ \tr_\env = \tr_\env\circ \acpm_{[0,t)}$, the marginal master equation.

\subsubsection{Density Matrix and State}

As a conceptual consequence of the marginal probability formula in GRWf, the reduced density matrix $\dm_\sys$ (at time 0) plays the same role for an isolated system as the wave function $\wf_0$ (at time 0) for the universe (or a disentangled isolated system), the role being that of governing the distribution of the flashes.\footnote{If we want to make a similar statement about time $t$, the appropriate density matrix to consider is not the $\dm_t^\sys$ considered in the marginal master equation but rather the random density matrix $\tr_\env \ket{\wf_t}\bra{\wf_t}$, from which $\dm_t^\sys$ is obtained by averaging over the flashes during the time interval $[0,t)$.} This is just a way of re-formulating the marginal probability formula. 

Another way of putting this conceptual consequence is to say that, in GRWf, the reduced density matrix $\dm_\sys$ at time 0 describes the \emph{state} of an isolated system. (Two rather different notions of ``state'' are common in physics, which should not be confused. While in classical mechanics the notion of state at time $t$ used to mean ``phase point,'' i.e., ``a mathematical datum that determines the PO after $t$,'' the meaning has shifted, with the advent of quantum mechanics, to a statistical notion which, in our framework, could be defined as ``a mathematical datum that determines a probability distribution of the PO after $t$.'' For example, in classical mechanics a state in the latter sense would correspond to a probability distribution over states in the former sense.)

It is useful to note that the situation is different in GRWm, where the reduced density matrix is not a ``state'' (though the wave function, if an isolated system possesses one, is): the reduced density matrix (at time 0) of a system that is entangled with its environment is insufficient to determine the probability distribution of $m_\sys(x,t)$ at later times, in spite of \eqref{msysdmsys}.\footnote{To see this, consider for example $\Psi(t=0)=2^{-1/2}(\ket{u}\ket{1} + \ket{d}\ket{2})$, where $\ket{u},\ket{d}$ 
are orthonormal vectors in $\Hilbert_\sys$ and $\ket{1},\ket{2}$ in $\Hilbert_\env$, and suppose that $\Psi$ quickly collapses to either $\ket{u}\ket{1}$ or $\ket{d}\ket{2}$; 
contrast this with $\tilde\Psi(t=0) = 2^{-1/2}(\ket{l}\ket{1} + \ket{r}\ket{2})$, where $\ket{l}=2^{-1/2}(\ket{u} + \ket{d})$ and 
$\ket{r}=2^{-1/2}(\ket{u}-\ket{d})$, and suppose that $\tilde\Psi$ quickly collapses to either $\ket{l}\ket{1}$ or $\ket{r}\ket{2}$. 
Then $\dm_\sys=\tfrac{1}{2} \pr{u}+\tfrac{1}{2} \pr{d} =\tfrac{1}{2}\pr{l}  + \tfrac{1}{2} \pr{r} = \tilde\dm_\sys$, but the $m_\sys$ associated with $\ket{u}$ or $\ket{d}$ may be completely different from that associated with $\ket{l}$ or $\ket{r}$.}

In orthodox quantum mechanics, it is more or less the results of experiments that are regarded as the PO, and a ``state'' is what determines the distribution of the results of experiments. Thus, as in GRWf, the reduced density matrix $\dm$ is a ``state'' for any isolated system, since the distribution of the result of an experiment, acting only on the system and associated with POVM $E^\Qu(\cdot)$ acting on $\Hilbert_\sys$, is given by $\tr(\dm\, E^\Qu(\cdot))$.

\subsection{Theorem~\ref{thm:GRWformalism}}
\label{sec:thm1}

Before we derive the GRW formalism, we summarize what exactly the derivation will show:

\begin{thm}\label{thm:GRWformalism}
Consider a GRWf universe comprising four systems, called $a$, $b$, $\app$, and $\env$. Let the initial wave function of the universe $\Phi_0$ be random with probability distribution $\mu_0$, and let $\Phi_{t}$ for $t\geq 0$ evolve according to the GRW process. Let $t'>0$, let $B_{[0,t')}\subseteq \Omega_{[0,t')}$ be any measurable set of flash histories before time $t'$ with 
\be\label{Bnonzero}
\EEE_{\mu_0}\PPP_{\Phi_0} \bigl( F_{[0,t')}\in B_{[0,t')} \bigr) \neq 0\,, 
\ee
and let us conditionalize on the event $B_{[0,t')}$; explicitly, let $\Psi_{t}$, $t'\leq t$, denote the process with distribution 
\be
\PPP(\Psi\in \,\cdot\,) = \PPP(\Phi \in \,\cdot\,| F_{[0,t')}\in B_{[0,t')} )\,.
\ee
For any $t\geq t'$, define the density matrix of system $a$ (given $B_{[0,t')}$) by
\be\label{dmtdef}
\dm_t = \EEE \tr_{b\cup\app\cup\env} \pr{\Psi_t} 
= \EEE_{\mu_0}\EEE_{\Phi_0} \Bigl( \tr_{b\cup\app\cup\env} \pr{\Phi_t} \Big|  F_{[0,t')}\in B_{[0,t')} \Bigr)\,.
\ee
Then, for as long after $t'$ as system $a$ is isolated (as defined in Section~\ref{sec:isolated}), $\dm_t$ obeys the master equation \eqref{M}.

Furthermore, suppose that during the time interval $[\ti,\tf)$ with $t'\leq\ti<\tf$, the three systems $a\cup \app$, $b$, and $\env$ are mutually isolated, and that $\Psi_{\ti}$ factorizes with probability 1 according to
\be
\Psi_{\ti} = \psi_{\ti}^{a\cup b} \otimes \phi_{\ti}^{\app}\otimes \phi_{\ti}^{\env}\,,
\ee
where $\psi_{\ti}^{a\cup b}$ and $\phi_{\ti}^{\app}$ are independent random variables. Let $\Values$ be a countable set, $\zeta:\Omega^{a\cup\app}_{[\ti,\tf)}\to\Values$ a measurable function, and
\be
Z=\zeta\bigl(F^{a\cup\app}_{[\ti,\tf)}\bigr)\,.
\ee
Finally, define the density matrix of system $a$ at time $t$, conditional on $Z=z$, by
\be\label{thm1:dmtz}
\dm_{\tf|z} = \EEE_{\mu_0} \EEE_{\Phi_0} \Bigl( \tr_{b\cup\app\cup\env} \pr{\Phi_{\tf}} \Big| Z=z, F_{[0,t')}\in B_{[0,t')}\Bigr)\,.
\ee
Then
\be\label{thm1:Z}
\PPP(Z=z)=\tr\bigl(\dm_{\ti} \, E^{\GRW}_z\bigr)
\ee
and
\be\label{thm1:collapse}
\dm_{\tf|z} = \frac{\cpm_z^{\GRW}(\dm_{\ti})}{\tr \cpm_z^{\GRW}(\dm_{\ti})}
\ee
with $E^{\GRW}$ and $\cpm^{\GRW}$ given by the GRW law of operators \eqref{GRWE}, \eqref{GRWCPM} with $\dm_\app= \EEE\pr{\phi_{\ti}^\app}$.
\end{thm}

Remarks:
\begin{enumerate}
\item We can, in fact, allow the Hamiltonian $H$ to be time-dependent and the collapse operators $\Lambda_i(x)$ to be multiplication by a (possibly time-dependent) function $g_i(x-q_i,t)$ other than a Gaussian. We take for granted that the Hamiltonian $H_t$ is self-adjoint and that $g_i(\cdot,t)$ is measurable, bounded, non-negative, and not zero-almost-everywhere. The existence of the GRW process in Hilbert space then follows from standard theorems; see, e.g., \cite{Tum07}.

\item What we regard as ``the'' density matrix $\dm_t$ of the system $a$ at time $t$ depends, according to the definition \eqref{dmtdef}, on the choice of the prior information $B_{[0,t')}$ that we condition on. An extreme possibility is to conditionalize on one pattern of flashes, $B_{[0,t')}=\{f_{[0,t')}\}$ (which violates \eqref{Bnonzero}, but that is not a problem because it is clear enough what is meant by the conditional distribution of $\Phi_{t'}$); if, in addition, $\Phi_0$ is deterministic, i.e., if $\mu_0$ is concentrated on a single point, then $\Psi_{t'}$ is deterministic, and $\dm_{t'}= \tr_{b\cup\app\cup\env} \pr{\Psi_{t'}}$. The other extreme is not to conditionalize at all, i.e., to take $B_{[0,t')}=\Omega_{[0,t')}$. 

A practical possibility in between is to conditionalize on the macroscopic facts about $F_{[0,t')}$ known to the experimenter, including the outcomes of prior experiments. For such choices of $B_{[0,t')}$, we expect that $\PPP(Z=z)$ and $\dm_{t|z}$ are insensitive to variations of $B_{[0,t')}$ such as different choices of what counts as macroscopic, or different choices of which macroscopic facts to include (except about the preparation of the system). In fact, if $\PPP(Z=z)$ and $\dm_{t|z}$ depended sensitively on $B_{[0,t')}$ then the formulas for them would fail to be useful, as one would have to ask of such formulas: Probability given what? Density matrix given what? The robustness of these formulas makes it possible to formulate a GRW formalism that is applicable in practice. (The same is true, e.g., of the quantum formalism in Bohmian mechanics.) See also Remark~\ref{rem:robust1} in Section~\ref{sec:emergeop}. \label{rem:robust2}

\item As an example of how to apply Theorem~\ref{thm:GRWformalism}, consider two consecutive experiments on system $a$, say $\E_2$ after $\E_1$. We describe how Theorem~\ref{thm:GRWformalism} determines the joint probability distribution of the outcomes $Z_1,Z_2$. Suppose $\E_1$ is carried out during the time interval $[\ti_1,\tf_1)$, $\E_2$ during $[\ti_2,\tf_2)$ with $0\leq \ti_1<\tf_1\leq \ti_2<\tf_2<\infty$, and system $a$ is isolated during $[\tf_1,\ti_2)$.\footnote{\label{fn:s2}The assumption that (in particular) $\ti_2$ is fixed in advance is often unrealistic, although such a situation can of course be arranged. Often, the time at which an experimenter begins the second experiment, and even which experiment to perform, will be random; it may depend on the outcome of the first experiment and on other random influences (such as the weather). This scenario is considered in Section~\ref{sec:Ti}.} To compute such probabilities is essentially the purpose of the third rule of the GRW (or quantum) formalism, see \eqref{GRWC}, which leads to the equation
\be\label{PZ2givenZ1}
\PPP(Z_2=z_2|Z_1=z_1) = \tr\Bigl( \acpm^a_{[\tf_1,\ti_2)}\Bigl[\frac{\cpm_{1,z_1}(\dm_{\ti_1})}{\tr  \cpm_{1,z_1}(\dm_{\ti_1})} \Bigr] \, E_2(z_2) \Bigr)
\ee
or
\begin{align}
\PPP(Z_2=z_2,Z_1=z_1) 
&= \tr\Bigl( \acpm^a_{[\tf_1,\ti_2)} \bigl[\cpm_{1,z_1}(\dm_{\ti_1})\bigr] \, E_2(z_2) \Bigr)\label{PZ2Z1}\\
&= \tr\Bigl( 
\cpm_{2,z_2}\bigl(\acpm^a_{[\tf_1,\ti_2)} 
\bigl[\cpm_{1,z_1}(\dm_{\ti_1})\bigr] \bigr)\,  \Bigr)\,,
\end{align}
where $\acpm^a_{[\tf_1,\ti_2)}$ is the superoperator evolving a density matrix according to the master equation \eqref{M} for system $a$ from time $\tf_1$ to time $\ti_2$. Theorem~\ref{thm:GRWformalism} leads to \eqref{PZ2givenZ1} if applied twice, first with $\ti=\ti_1$, $\tf=\tf_1$, $\app=\app_1$ and with $\app_2$ included in $\env$, then with $\ti=\ti_2$, $\tf=\tf_2$, $\app=\app_2$, $t'=\tf_1$, $B_{[0,t')}=B_{1,[0,t_1')} \cap \{Z_1=z_1\}$, and $b=b_1\cup\app_1$ (assuming that $\app_1$ remains isolated after $\tf_1$; otherwise, system $b$ of $\E_2$ would also have to include the part of the environment that $\app_1$ interacts with during $[\tf_1,\tf_2)$). 

A direct calculation of \eqref{PZ2Z1} from the distribution of flashes is carried out in Appendix~\ref{sec:diagram2} using the diagram notation described in Appendix~\ref{sec:diagram}.
\end{enumerate}

\subsection{Proof of Theorem~\ref{thm:GRWformalism}}
\label{sec:derivation}

For the purpose of this proof, we can regard $t'$ as the initial time and the distribution of $\Psi_{t'}$ as the distribution of the initial wave function. That is, it is not relevant for the proof to distinguish between the contributions to this distribution from $\mu_0$, from the GRW process during $[0,t')$, and from conditioning on $B_{[0,t')}$. The reason we distinguished them in the formulation of Theorem~\ref{thm:GRWformalism} is that for the final (``collapsed'') density matrix $\dm_{\tf|z}$, we need to consider conditioning on a different event.

So we now regard $\Psi_{t}$ as the wave function of the universe at time $t$. Then $\dm_t$ as defined in \eqref{dmtdef} is just the partial trace of $\EEE\pr{\Psi_t}$; $\EEE\pr{\Psi_t}$ is a statistical density matrix as in \eqref{statdm} and hence evolves according to the master equation \eqref{M}. 
Now the marginal master equation implies that, as long as system $a$ is isolated, $\dm_t$ evolves according to the appropriate master equation \eqref{M} for system $a$. This yields already the first statement of Theorem~\ref{thm:GRWformalism}, or the first rule of the GRW formalism.

\bigskip

A derivation of the second rule---asserting that the outcome statistics is of the form $\tr(\dm_\ti\,E_z)$---was given in Section~\ref{sec:emergeop}, except for the case of entanglement between system $a$ and system $b$. So let us derive the second rule in this more general situation. 

By the conditional probability formula \eqref{condprob1}, we can regard $\Psi_\ti$ as the (random) initial wave function. By the marginal probability formula (applied to $a\cup \app$ instead of $a$) and the assumption that during $[\ti,\tf)$, $a\cup\app$ does not interact with $b$ or $\env$, the joint distribution of the flashes of $a\cup\app$ (given $\Psi_\ti$) is given by
\be
\PPP\bigl(F^{a\cup\app}_{[\ti,\tf)} \in A\big|\Psi_\ti\bigr) =
\tr\bigl(\dm_{a\cup\app}^{\Psi_\ti} \G^{a\cup\app}_{[\ti,\tf)}(A)\bigr)
\ee
with
\be
\dm_{a\cup\app}^{\Psi_\ti} = \tr_{b\cup\env} \pr{\Psi_\ti} = \bigl(\tr_b \pr{\psi^{a\cup b}_{\ti}}\bigr)\otimes \pr{\phi^\app_\ti}
\ee
and $\G^{a\cup\app}_{[\ti,\tf)}$ the history POVM (as defined in Section~\ref{sec:F}) for $a\cup\app$ during $[\ti,\tf)$. Let $\EEE$ denote averaging over the random wave function $\Psi_\ti$; using that $\phi^\app_\ti$ is stochastically independent of $\psi^{a\cup b}_\ti$, we obtain that
\be
\EEE \dm_{a\cup\app}^{\Psi_\ti} = (\EEE\, \tr_b \pr{\psi^{a\cup b}_\ti})\otimes (\EEE \pr{\phi^\app_\ti}) = \dm_\ti^a \otimes \dm_\ti^\app
\ee
with $\dm_\ti^a$ and $\dm_\ti^\app$ the density matrices---as defined in \eqref{dmtdef}---at time $\ti$ of system $a$ and the apparatus, respectively. As a consequence,
\begin{align}
\PPP(Z =z) &= 
\EEE \PPP\bigl(F^{a\cup\app}_{[\ti,\tf)} \in \zeta^{-1}(z)\big|\Psi_\ti\bigr)\\
&= \EEE \tr\Bigl(\dm_{a\cup\app}^{\Psi_\ti} \G^{a\cup\app}_{[\ti,\tf)}\bigl(\zeta^{-1}(z)\bigr)\Bigr)\\
&= \tr\Bigl([\dm_\ti^{a}\otimes \dm_\ti^{\app}] \G^{a\cup\app}_{[\ti,\tf)}\bigl(\zeta^{-1}(z)\bigr)\Bigr)\\
&= \tr\bigl(\dm_\ti^a \, E^\GRW_z\bigr)
\end{align}
with $E^\GRW_z$ given by \eqref{GRWE}---the GRW law of operators. The operators $E^\GRW_z$ form a POVM because of the function property \eqref{fPOVM} and the reduction property \eqref{partialPOVM}. This completes the derivation of the second rule.

\bigskip

Now we turn to the third rule---the collapse rule, or \eqref{thm1:collapse}. According to the definition \eqref{thm1:dmtz} of $\dm_{\tf|z}$,
\be\label{dmZz}
\dm_{\tf|z} =
\EEE\Bigl(\tr_{b\cup\app\cup\env} \pr{\Psi_\tf}\Big|Z=z \Bigr)
\ee
with $\EEE$ the average over both the random wave function $\Psi_\ti$ before $\E$ and the flashes during $[\ti,\tf)$, conditional on the outcome $z$ of the experiment. Using the $L_{[\ti,\tf)}$ operators defined in \eqref{eq:long}, the expression \eqref{dmZz} can be rewritten as 
\begin{align}
\dm_{\tf|z} 
&=\frac{1}{\mathcal{N}} \,\EEE\Bigl(\tr_{b\cup\app\cup\env} \pr{\Psi_\tf} \, 1_{Z=z} \Bigr)\\
\intertext{[by \eqref{eq:psit} and \eqref{Fttdef}]}
&=\frac{1}{\mathcal{N}} \!\! \int\limits_{\Omega_{[\ti,\tf)}^{a\cup b\cup\app\cup\env}} \!\!\!\!\!\!\! df \, \tr_{b\cup\app\cup\env} \Bigl(L^{a\cup b\cup\app\cup\env}_{[\ti,\tf)}(f)\, \pr{\Psi_\ti} L^{a\cup b\cup\app\cup\env}_{[\ti,\tf)}(f)^*\Bigr) 1_{\zeta(f_{a\cup\app})=z}\\
\intertext{[by \eqref{factformLst} with $\sys\to a\cup\app$ and $\env\to b\cup\env$]}
&=\frac{1}{\mathcal{N}} \, \tr_{\app}\int\limits_{\Omega_{[\ti,\tf)}^{a\cup \app}} \!\! df_{a\cup\app} \int\limits_{\Omega_{[\ti,\tf)}^{b\cup\env}} \!\! df_{b\cup\env} \, \tr_{b\cup\env} \Biggl([L^{a\cup \app}_{[\ti,\tf)}(f_{a\cup\app})\otimes L^{b\cup \env}_{[\ti,\tf)}(f_{b\cup\env})]\:\times\nonumber\\
&\quad \times\: \pr{\Psi_\ti} [L^{a\cup\app}_{[\ti,\tf)}(f_{a\cup\app})^* \otimes L^{b\cup \env}_{[\ti,\tf)}(f_{b\cup\env})^*]\Biggr) 1_{\zeta(f_{a\cup\app})=z}\\
&=\frac{1}{\mathcal{N}} \, \tr_{\app}\int\limits_{\zeta^{-1}(z)} \!\! df_{a\cup\app} \tr_{b\cup\env} \Biggl([L^{a\cup \app}_{[\ti,\tf)}(f_{a\cup\app})\otimes I]\pr{\Psi_\ti} [L^{a\cup\app}_{[\ti,\tf)}(f_{a\cup\app})^* \otimes I] \:\times\nonumber\\
&\quad \times\:  \underbrace{\int\limits_{\Omega_{[\ti,\tf)}^{b\cup\env}} \!\! df_{b\cup\env} \, [I\otimes L^{b\cup \env}_{[\ti,\tf)}(f_{b\cup\env})^*]  [I\otimes L^{b\cup \env}_{[\ti,\tf)}(f_{b\cup\env})]}_{=I}\Biggr) \\
\intertext{[by \eqref{trbrule}, \eqref{trbrule2}]}
&=\frac{1}{\mathcal{N}} \, \tr_{\app}\int\limits_{\zeta^{-1}(z)} \!\!\!\! df_{a\cup\app} \, \Bigl(L^{a\cup\app}_{[\ti,\tf)}(f_{a\cup\app})\, \tr_{b\cup\env}\bigl(\pr{\Psi_\ti}\bigr) L^{a\cup\app}_{[\ti,\tf)}(f_{a\cup\app})^* \Bigr) \\
&=\frac{1}{\mathcal{N}} \, \tr_{\app}
\int\limits_{\zeta^{-1}(z)} \!\!\! df_{a\cup\app} \,L^{a\cup\app}_{[\ti,\tf)}(f_{a\cup\app}) [\dm^a_\ti \otimes \dm^\app_\ti] L^{a\cup\app}_{[\ti,\tf)}(f_{a\cup\app})^*\,,
\end{align}
which agrees with \eqref{GRWCPM}, the GRW law of superoperators.
This completes the proof of Theorem~\ref{thm:GRWformalism}.

\section{Random Run-Time}
\label{sec:Tf}

We now discuss the case in which the duration of the experiment $\E$ is not fixed. Rather, we assume that the time at which $\E$ is finished is itself a random quantity $\Tf$, generated by $\E$ itself. The starting time $\ti$, in contrast, is assumed to be fixed. We assume that $\Tf$ can take values from a finite or countable set $\Times \subset [\ti,\infty)$ (just as we assumed that the value space $\Values$ is discrete). 

As a relevant consequence of the random run-time, we may proceed with the next experiment right after $\Tf$, at a time at which the first experiment could still have been running if it had come out differently. In order to apply the formalism to the second experiment, we need to know the appropriately collapsed density matrix created by the first experiment. It is a crucial part of the formalism for random run-time to tell us what this collapsed density matrix is.

\subsection{Quantum Formalism for Random Run-Time}
\label{sec:QuTf}

Our main concern here is with the GRW formalism for random run-time. 
However, a formulation of the quantum formalism for random run-time has 
rarely, if ever, been explicitly given. We thus begin with that.

\bigskip

\noindent\textbf{The Quantum Formalism for Random Run-Time.}
\begin{itemize}
\item A system isolated from its environment has at every time $t$ a density matrix $\dm_t$ which evolves according to the unitary Schr\"odinger evolution \eqref{SchrDM}.
\item With every experiment $\E$ starting at time $\ti$ with a discrete set $\Values$ of possible outcomes and a discrete set $\Times\subset [\ti,\infty)$ of possible times at which $\E$ is finished, there is associated a POVM $E^\Qu(\cdot)$ on $\Values\times\Times$ acting on $\Hilbert_\sys$. When the experiment $\E$ is performed on a system with density matrix $\dm_\ti$, the outcome $Z$ and the time $\Tf$ at which $\E$ is finished are random with joint probability distribution
\begin{equation}\label{QuPPPT}
  \PPP(Z=z,\Tf=\tf) = \tr \bigl(\dm_\ti\, E^\Qu_{z,\tf} \bigr)\,.
\end{equation}
As a consequence,
\be
  \PPP(Z=z) = \tr \bigl(\dm_\ti \, E^\Qu_z \bigr)
\ee
with
\be\label{EzEztQu}
E_z^\Qu = \sum_{\tf\in\Times} E^\Qu_{z,\tf}\,.
\ee
\item With $\E$ is further associated a family $(\cpm_{z,\tf}^\Qu)_{z \in \Values, \tf\in\Times}$ of completely positive superoperators acting on $TRCL(\Hilbert_\sys)$ with the compatibility property that for all trace-class operators $\trclop$, 
\begin{equation}\label{ECPMQuT}
  \tr \bigl( \trclop \, E^\Qu_{z,\tf}\bigr) = \tr  
  \cpm_{z,\tf}^\Qu( \trclop)\,.
\end{equation}
In case $Z=z$ and $\Tf=\tf$, the density matrix of the system at time $\tf$ immediately after the experiment $\E$ is
\begin{equation}\label{QuCT}
  \dm'=\dm_\tf = \frac{\cpm_{z,\tf}^\Qu(\dm_\ti)}
  {\tr  \cpm_{z,\tf}^\Qu(\dm_\ti)}\,.
\end{equation}
\end{itemize}

\bigskip

\noindent\textbf{The Quantum Law of Operators for Random Run-Time.}
\begin{itemize}
\item Suppose we are given the density matrix $\dm_\app$ for the ready state 
of the apparatus, its Hamiltonian $H_\app$, and the interaction 
Hamiltonian $H_I$, so that $H = H_\sys + H_\app + H_I$; $U_t = 
\exp(-\tfrac{i}{\hbar}Ht)$ is the unitary Schr\"odinger evolution 
operator for system $\cup$ apparatus. We may assume that, like the outcome 
$Z=z$, the event $\Tf=\tf$ corresponds to a pointer on the apparatus 
pointing to $t$.
Let $P_{z,\tf}^\app$ be the projection to the subspace of apparatus 
states in which  the pointer for the outcome is pointing to $z$, and the 
pointer for the time when $\E$ was over is pointing to $\tf$. We may 
also assume that the outcomes $z$ and $\tf$ get recorded permanently, 
i.e., that a vector in the range of $I_\sys\otimes P^\app_{z,\tf}$ stays 
in that space. Then
\begin{equation}\label{QET}
E^\Qu_{z,\tf} =  
\tr_\app \Bigl( [I_\sys\otimes \dm_\app] U^*_{\tf-\ti} \, 
[I_\sys\otimes P^\app_{z,\tf}] U_{\tf-\ti} \Bigr)
\end{equation}
and
\begin{equation}\label{QCPMT}
   \cpm^\Qu_{z,\tf} (\dm) = \tr_\app  
  \Bigl( [I_\sys\otimes P^\app_{z,\tf}] U_{\tf-\ti} [\dm \otimes \dm_\app] U^*_{\tf-\ti} \, 
  [I_\sys\otimes P^\app_{z,\tf}] \Bigr)\,. 
\end{equation}
\end{itemize}

\bigskip

Under the assumptions given just above, the quantum formalism for random 
run-time, together with its law of operators, follows from the standard 
quantum formalism, for fixed run-time. To see that the $E^\Qu_{z,\tf}$ 
form a POVM and $\sum_{z,\tf} \cpm^\Qu_{z,\tf}$ is trace-preserving, use 
that vectors in the range of $I_\sys\otimes P^\app_{z,\tf}$ stay in that 
space, so $U_{\tf-\ti}$ in \eqref{QET} and \eqref{QCPMT} can be replaced 
by $U_{\max \Times-\ti}$, and summing over $z$ and $\tf$ will replace 
$P^\app_{z,\tf}$ by $I_\app$. We check the compatibility condition in 
Appendix~\ref{app:check}. Further  details of the  derivation shall not 
be worked out here.

\subsection{GRW Formalism for Random Run-Time}
\label{sec:GRWTf}

The GRW formalism for random run-time differs from the quantum formalism for random run-time in that (i)~the unitary Schr\"odinger evolution gets replaced by the master equation \eqref{M}, (ii)~the POVM $E^\Qu(\cdot)$ gets replaced by a different one $E^\GRW(\cdot)$, and (iii)~the superoperators $\cpm^\Qu_{z,\tf}$ gets replaced by different ones $\cpm^\GRW_{z,\tf}$. In addition, we need to say a few things about the conditions under which the GRW formalism for random run-time is applicable.

The assumption $Z=\zeta(F)$ now gets complemented by the assumption $\Tf = \tauf(F)$, i.e., that the finishing time $\Tf$ can be read off from the flashes. (In the terminology of Section~\ref{sec:applicability}, we assume that $\tau(F)=\tau(F^{a\cup\app})$ does not depend on $F^{b\cup\env}$.) On top of that, we assume that the random variable $\Tf=\tauf(F)$ is a \emph{stopping time} in the sense of the theory of stochastic processes, i.e., that $\tauf$ is such that
\begin{equation}\label{Tfstoppingtime}
\text{the event }\tauf(F)\leq \tf\text{ depends only on }F_{[\ti,\tf)}\,,
\end{equation}
i.e., on the flashes up to time $\tf$. In other words, we require that it is possible to read off from the flashes up to time $\tf$ whether the experiment is over yet. In the terminology of the theory of stochastic processes, the space $\Omega_{[\ti,\infty)}$ of all possible flash patterns is naturally equipped with a \emph{filtration} $(\salg_{[\ti,\tf)})_{\tf\in (\ti,\infty)}$, where $\salg_{[\ti,\tf)}$ is the $\sigma$-algebra of all events that depend only on the flashes before $\tf$ (i.e., it is the collection of those $B\subseteq\Omega_{[\ti,\infty)}$ such that for any two $f,f'\in \Omega_{[\ti,\infty)}$ with $f_{[\ti,\tf)}=f'_{[\ti,\tf)}$, either both $f,f'\in B$ or both $f,f'\notin B$); our assumption that $\Tf$ is a stopping time means that for each $\tf$, the event $\{\Tf\leq \tf\}$, regarded as the set $\{f:\tauf(f)\leq\tf\}$, belongs to $\salg_{[\ti,\tf)}$.

Moreover, we assume that
\begin{equation}\label{Zadapted}
\text{if }\tauf(F) = \tf\text{ then }\zeta(F)\text{ depends only on }F_{[\ti,\tf)}.
\end{equation}
That is, when the experiment is over it must be possible to read off the result from the flashes so far. In the terminology of the theory of stochastic processes, this assumption is that $Z$ is \emph{adapted} to the $\sigma$-algebra $\salg_{[\ti,\Tf)}$ of all events that depend only on the flashes before $\Tf$ (i.e., it is the collection of those $B\subseteq\Omega_{[\ti,\infty)}$ such that for any two $f,f'\in\Omega_{[\ti,\infty)}$ with $\tauf(f)=\tauf(f')=:\tf$ and $f_{[\ti,\tf)}=f'_{[\ti,\tf)}$, either both $f,f'\in B$ or both $f,f'\notin B$).

Also, when we consider an experiment $\E$ with random run-time that takes place from $\ti$ until $\Tf$, it is understood that the system consisting of system $a$ (the object of $\E$), system $b$ (anything with which system $a$ is entangled) and the apparatus is isolated from its environment \emph{only during the random time interval} $[\ti,\Tf)$.

In this setting, the formalism reads as follows.

\bigskip

\noindent\textbf{The GRW Formalism for Random Run-Time.}
\begin{itemize}
\item A system isolated from its environment has at every time $t$ a density matrix $\dm_t$ which evolves according to the master equation \eqref{M}.
\item With every experiment $\E$ starting at time $\ti$ with a discrete set $\Values$ of possible outcomes and a discrete set $\Times\subset [\ti,\infty)$ of possible times at which $\E$ is finished, there is associated a POVM $E^\GRW(\cdot)$ on $\Values\times\Times$ acting on $\Hilbert_\sys$. When the experiment $\E$ is performed on a system with density matrix $\dm_\ti$, the outcome $Z$ and the time $\Tf$ at which $\E$ is finished are random with joint probability distribution
\begin{equation}\label{GRWPPPT}
  \PPP(Z=z,\Tf=\tf) = \tr \bigl(\dm_\ti\, E^\GRW_{z,\tf} \bigr)\,.
\end{equation}
As a consequence,
\be
  \PPP(Z=z) = \tr \bigl(\dm_\ti \, E^\GRW_z \bigr)
\ee
with
\be\label{EzEzt}
E_z^\GRW = \sum_{\tf\in\Times} E^\GRW_{z,\tf}\,.
\ee
\item With $\E$ is further associated a family $(\cpm_{z,\tf}^\GRW)_{z \in \Values, \tf\in\Times}$ of completely positive superoperators acting on $TRCL(\Hilbert_\sys)$ with the compatibility property that for all trace-class operators $\trclop$, 
\begin{equation}\label{ECPMGRWT}
  \tr \bigl( \trclop \, E^\GRW_{z,\tf}\bigr) = \tr  
  \cpm_{z,\tf}^\GRW( \trclop)\,.
\end{equation}
In case $Z=z$ and $\Tf=\tf$, the density matrix of the system at time $\tf$ immediately after the experiment $\E$ is
\begin{equation}\label{GRWCT}
  \dm'=\dm_\tf = \frac{\cpm_{z,\tf}^\GRW(\dm_\ti)}
  {\tr  \cpm_{z,\tf}^\GRW(\dm_\ti)}\,.
\end{equation}
\end{itemize}

\bigskip

\noindent\textbf{The GRW Law of Operators for Random Run-Time.}
\begin{itemize}
\item Suppose we are given the density matrix $\dm_\app$ for the ready state of the apparatus, its Hamiltonian $H_\app$, and the interaction Hamiltonian $H_I$, so that $H = H_\sys + H_\app + H_I$. Let the experiment $\E$ start at time $\ti$, let $\zeta: \Omega_{[\ti,\infty)} \to \Values$ be the function that reads off the outcome of $\E$ from the flashes, and let $\tauf: \Omega_{[\ti,\infty)} \to \Times$ be the function that reads off the finishing time of $\E$ from the flashes. Then  
\begin{align}\label{GRWET}
E^\GRW_{z,\tf}& = \tr_\app\Bigl( [I_\sys\otimes \dm_\app] \,
\G\bigl(\zeta^{-1}(z)\cap\tauf^{-1}(\tf) \bigr) \Bigr)\\
\label{GRWET2}
&= \tr_\app 
\int\limits_{\zeta^{-1}(z)\cap {\tauf}^{-1}(\tf)} \hspace{-5mm}df\,
[I_\sys\otimes \dm_\app] \, L_{[\ti,\tf)}^*(f) \, L_{[\ti,\tf)}(f)\,,
\end{align}
where $f=f_{\sys\cup\app}$ and $L=L_{\sys\cup\app}$, and
\begin{equation}\label{GRWCPMT}
   \cpm^\GRW_{z,\tf} (\trclop) = \tr_\app  
  \int\limits_{\zeta^{-1}(z)\cap {\tauf}^{-1}(\tf)} \hspace{-5mm}df \: 
  L_{[\ti,\tf)}(f) \, [\trclop \otimes \dm_\app] \, L^*_{[\ti,\tf)}(f)\,.
\end{equation}
As a consequence, $E_z^\GRW$ as defined in \eqref{EzEzt} is given by \eqref{GRWE} (with the difference that $\G$ now must be taken to mean $\G_{[\ti,\infty)}$ and not $\G_{[\ti,\tf)}$).
\end{itemize}

\bigskip

Concerning \eqref{GRWET2}, note that by assumptions \eqref{Tfstoppingtime} and \eqref{Zadapted}, the set $\zeta^{-1}(z) \cap \tauf^{-1}(\tf)\subseteq \Omega_{[\ti,\infty)}$ is of the form $A\times \Omega_{[\tf,\infty)}$ for suitable $A\subseteq \Omega_{[\ti,\tf)}$. We wrote $\zeta^{-1}(z) \cap \tauf^{-1}(\tf)$ for $A$ in the domain of the integral in \eqref{GRWET2} and \eqref{GRWCPMT}; that is, the domain of the integral is to be regarded as a subset of $\Omega_{[\ti,\tf)}$, so that the integration variable $f$ is a history of flashes in the time interval $[\ti,\tf)$ and thus can be inserted into $L_{[\ti,\tf)}$. To see that \eqref{GRWET2} is the same as \eqref{GRWET}, note that 
\[
\G\bigl(\zeta^{-1}(z)\cap\tauf^{-1}(\tf) \bigr) = 
\G(A\times\Omega_{[\tf,\infty)})=
\G_{[\ti,\tf)}(A)=
\int_A df\, L_{[\ti,\tf)}^*(f) \, L_{[\ti,\tf)}(f)
\]
using \eqref{Fttdef}. We check the compatibility condition \eqref{ECPMGRWT} in Appendix~\ref{app:check}.

\subsection{Derivation of the GRW Formalism for Random Run-Time}

The biggest difference from the derivation of the GRW formalism for fixed run-time is that we now have to consider a system that is isolated from its environment \emph{only during the random time interval} $[\ti,\Tf)$. 

In particular, we need a version of the \emph{marginal probability formula for stopping times}: Consider a system ``sys'' (such as $a\cup\app$), let $\Tf$ be a stopping time adapted to sys (i.e., a function $\tauf$ of $F^\sys=F^\sys_{[0,\infty)}$ such that the event $\Tf=\tauf(F^\sys)\leq t$ depends only on $F^\sys_{[0,t)}$), and let $\salg_{[0,\Tf)}^\sys$ be the $\sigma$-algebra of events depending only on the flashes of sys up to time $\Tf$. If the system is isolated during $[0,\Tf)$ then 
\be\label{margprobTf}
\PPP_{\wf_0}(F^\sys \in B) = \PPP_{\dm_\sys}(B) \quad \forall B\in\salg_{[0,\Tf)}^\sys \,.
\ee
Put differently, this means for every $t>0$ and every $B\subseteq \Omega^\sys_{[0,t)}$,
\be\label{margprobTf2}
\PPP_{\wf_0}\Bigl(F^\sys_{[0,t)}\in B, \tauf(F^\sys)=t \Bigr) =
\PPP_{\dm_\sys}\bigl( F_{[0,t)}\in B, \tauf(F)=t \bigr)\,. 
\ee
This fact follows from the marginal probability formula \eqref{margprob} in much the same way as the version \eqref{margprob2} for a system that is isolated during the deterministic interval $[0,t)$.
Consider a hypothetical universe whose time-dependent Hamiltonian $H_t$ and collapse operators $\Lambda_{i,t}(x)$ are whatever we choose, and a fixed initial wave function $\Psi_{0}$, so that the distribution of flashes during $[0,t)$ will not depend on our choices of $H_s$ for $s\geq t$. So we need not specify before $t$ whether we will turn on the interaction at $t$ or not, and we could make this decision depend on the flashes up to time $t$. Since this choice does not affect the distribution of the flashes before $t$, this distribution is the same as it would have been if the system were isolated forever, and thus given by \eqref{margprob}.

With these tools, the derivation of the GRW formalism for random run-time follows the same lines as the derivation of the GRW formalism for fixed run-time.

\section{Random Experiments}
\label{sec:Ti}

So far we have considered a fixed experiment $\E$, carried out at a 
fixed time $\ti$. In practice, as pointed out in \cite[Section~8]{DGZ92} 
and in Footnote~\ref{fn:s2} above, the experiment we carry out---the 
choice of $\E$---is often random, even if we do not have the intention 
to make it random and even though we often do not notice that it is. For 
example, an experiment may depend on the outcomes of previous 
experiments; say, the experimenter may decide to repeat an experiment if 
the previous outcome was unexpected. For another example, $\E$ may 
depend on other random influences such as the weather, traffic 
conditions, or the stock market. (Say, the time at which the 
experimenter arrives in the lab may depend on the traffic; the equipment 
used may depend on the funds available, which in turn depend on the 
economic conditions, represented by the stock market; etc..)

A choice of $\E$ means here a choice of: the time $\ti$ at which it 
starts; the system that will be the object of $\E$; the system that will 
serve as apparatus; its initial state $\dm_\app$; the Hamiltonian 
$H_\sys+H_\app+H_I$; and the calibration function $\zeta$. In this 
section we provide a brief discussion of how the randomness in the 
choice of $\E$ affects the quantum and the GRW formalism.

\subsection{The GRW Case}
\label{sec:TiGRW}

The GRW formalism as we have derived it already includes the possibility of random experiments. As stated in Theorem~\ref{thm:GRWformalism} in Section~\ref{sec:thm1}, we have the possibility to conditionalize on an arbitrary (non-null) event $B_{[0,\ti)}$ prior to the beginning of our experiment, and natural and reasonable practical choices of this event include the information determining $\E$. To ensure that Theorem~\ref{thm:GRWformalism} is applicable, we only need that the probability that $\E$ is a particular experiment $e$ is nonzero; thus, we need to assume that the set of possible experiments to choose between is finite or countable. This assumption is essentially no restriction, as discussed already in the beginning of Section~\ref{sec:fullformalism} for the finite or countable value space $\Values$.

The question remains how to practically obtain the density matrix $\dm_\ti$ of the system at the beginning $\ti$ of $\E$ conditional on $B_{[0,\ti)}$ as in \eqref{dmtdef}. Specifically, we ask under which conditions the following obvious recipe is appropriate: Given the density matrix $\dm_{t_0}$ of the system at a time $t_0\geq 0$ before $\E$, and given that (i.e., conditional on that) $\E$ is a particular experiment $e$ beginning at time $\ti=\ti(e)\geq t_0$, use the master equation \eqref{M} to evolve $\dm$ to time $\ti$,
\be\label{dmsrandomE}
\dm_\ti=\acpm_{[t_0,\ti)} \dm_{t_0}\,.
\ee
(And then apply the GRW formalism for $e$ to $\dm_\ti$.)

To answer this question, partition the world into two systems, ``$\sys$'' and ``$\env$.'' The system that will be the object of the experiment will be system ``$a$,'' a randomly chosen subsystem of $\sys$. Everything entangled with ``$a$'' (such as the apparatus of previous experiments on $a$) also belongs to $\sys$, while the apparatus of $\E$ will be a randomly chosen subsystem of $\env$. Let $b=\sys \setminus a$ be the complement of $a$ in $\sys$, and let $B_{[0,t_0)}$ denote the given information about $\sys \cup \env$ prior to $t_0$. Suppose the following assumptions are satisfied:
\begin{enumerate}
\item Conditionally on $B_{[0,t_0)}$, $\sys$ is not entangled with $\env$ at $t_0$,
\be\label{ranEdisent}
\EEE\Bigl(\pr{\Psi_{t_0}}\Big| F_{[0,t_0)}\in B_{[0,t_0)} \Bigr) = \dm^{\sys}_{t_0} \otimes \dm^\env_{t_0}\,.
\ee
\item\label{ranEass2} The choice of $\E$ depends only on the flashes of the environment,
\be
\E=\varepsilon(F^\env_{[t_0,\infty)})\,.
\ee
\item\label{ranEass3} The decision whether $\E$ is a particular experiment $e$ has been made by the time $\ti(e)$ at which $e$ starts,
\be
\varepsilon^{-1}(e) = A\times \Omega^\env_{[\ti(e),\infty)}
\text{ for some }A\in\Omega^\env_{[t_0,\ti(e))}\,.
\ee
\item System $a=a(e)$ remains isolated during $[t_0,\ti(e))$, and $b=b(e)$ during $[t_0,\tf(e))$, with $\tf(e)$ the time at which $e$ ends.
\end{enumerate}
Then, for $e$ with $\PPP(\E=e|F_{[0,t_0)}\in B_{[0,t_0)})>0$, the density matrix of $a=a(e)$ at $\ti=\ti(e)$ conditional on $B_{[0,t_0)} \cap \{\E=e\}$, defined as
\be
\dm_\ti :=\EEE \Bigl(\tr_{b(e)\cup\env}\pr{\Psi_{\ti}}\Big| F_{[0,t_0)}\in B_{[0,t_0)},\E=e\Bigr)\,,
\ee
is given by the obvious recipe \eqref{dmsrandomE}. See Appendix~\ref{app:recipe} for the proof.

In practice, the information about the history until $\ti$ that one normally and naturally conditions on is not just $B_{[0,t_0)}\cap \{\E=e\}$, but much more. It is the information available to us at $\ti$ and thus includes many facts about the macroscopic world history until $\ti$. However, this further information concerns only $F^\env_{[t_0,\ti)}$, not $F^a_{[t_0,\ti)}$, and thus does not affect the validity of \eqref{dmsrandomE}. This observation is in line with Remark~\ref{rem:robust2} in Section~\ref{sec:thm1}.

\subsection{The Quantum Case}

Like the GRW formalism, the quantum formalism can be taken to provide, with its probability formula \eqref{QPPP}, the \emph{conditional} probabilities of the outcomes, given that the random experiment $\E$ is a particular experiment $e$ beginning at time $\ti=\ti(e)$. Likewise, the collapse rule \eqref{QC} applies in case $\E=e$. 

In particular, for two consecutive experiments $\E_1,\E_2$, the obvious recipe to obtain the conditional distribution of $Z_2$ given $\E_1=e_1,\E_2=e_2$ reads as follows: Use the collapse rule \eqref{QC} to obtain the density matrix $\dm_{t_1}$ of the system at the end $\tf_1$ of $\E_1$, evolve it unitarily to time $\ti_2\geq \tf_1$, and then apply the quantum formalism for $e$ to $\dm_{\ti_2}$. That is,
\begin{multline}\label{Z2givenZ1E}
\PPP(Z_2=z_2|Z_1=z_1,\E_1=e_1,\E_2=e_2) = \\
\tr\Bigl( e^{-iH_\sys(s_2-t_1)}\frac{\cpm^{\Qu,e_1}_{z_1}(\dm_{\ti_1})}{\tr  \cpm^{\Qu,e_1}_{z_1}(\dm_{\ti_1})} e^{iH_\sys(s_2-t_1)}\, E^{\Qu,e_2}_2(z_2) \Bigr)\,.
\end{multline}
We note that in orthodox quantum mechanics these conditional 
probabilities are not like normal  conditional probabilities computed 
from a joint probability distribution. Rather they are the defining 
elements of the quantum formalism, which when combined with other 
ingredients---the probability distributions of $\E_1$ and $\E_2$---yield 
the joint probabilities for the results of quantum experiments.
Those other ingredients are probabilities that themselves do not originate in the quantum formalism but are part of a classical level of description that is required but not derivable from quantum mechanics.

The recipe \eqref{Z2givenZ1E} can only be expected to hold under assumptions analogous to the four assumptions listed in Section~\ref{sec:TiGRW}. Assumption \ref{ranEass2} should mean that in case system $a$ or $b$ is macroscopic, the choice of $\E$ does not depend on the macroscopic state of $a\cup b$. Assumption~\ref{ranEass3} should mean that the choice of whether $\E=e$ is made by the time $\ti(e)$.\footnote{Here is why such assumptions are necessary. If one of the systems ``$a$'' and ``$b$'' can be macroscopic then it makes a difference whether $\E$ can depend only on $\env$ or also on the macro-state of $a$ or $b$. For example, suppose Alice and Bob carry out an EPR experiment (i.e., each performs a Stern--Gerlach experiment in the $z$-direction, with the two particles initially in the singlet state), ``$a$'' consists of Alice's EPR particle, $\env$ consists of Alice's lab, ``$b$'' consists of Bob's EPR particle and his lab, $Z =\zeta$(Alice's outcome), and that the choice of $\zeta$ depends on Bob's outcome---that is where $\E$ depends on $b$. Then it can be arranged that $\PPP(Z=+1)=1$ while the formalism, given $\E$, predicts $\PPP(Z=+1)=\frac12=\PPP(Z=-1)$.

Likewise, without the assumption that the choice $\E=e$ is made by the
time $\ti(e)$ the distribution of $Z$ can change. For example, suppose
Alice carries out 100 Stern--Gerlach experiments in the $z$-direction on
particles in $|x\text{-up}\rangle$ and after the first $+1$ result declares that the last
experiment was $\E$. Then $\PPP(Z=+1)=1-2^{-100}\approx 1$ while the formalism, given $\E$,
predicts $\PPP(Z=+1)=\frac12=\PPP(Z=-1)$.

The distribution of $Z$ can also change if ``$a$'' and the apparatus are initially entangled. For example, consider again the EPR experiment and suppose that ``$a$'' consists of Alice's EPR particle, ``$b$'' is empty, $\env$ consists of Bob's EPR particle, his lab, and Alice's lab, $Z =\zeta$(Alice's outcome), and that the choice of $\zeta$ depends on Bob's outcome. Again, it can be arranged that $\PPP(Z=+1)=1$.

Finally, if ``$a$'' is not isolated until $\ti=\ti(e)$ then there is no reason to believe its density matrix at $\ti$ is $\acpm_{[t_0,\ti)} \dm_{t_0}$, and if ``$b$'' could interact with the apparatus then the same problem would arise as if ``$a$'' was entangled with the apparatus.} 

The quantum formalism, including random experiments, has been derived 
from Bohmian mechanics in sections 8--10 of \cite{DGZ92}. One might be  
tempted to say, in contrast, that formulas such as \eqref{Z2givenZ1E} 
cannot be derived  from the standard quantum formalism for non-random 
experiments. But it is more accurate to say that the very distinction 
between the quantum formalism for non-random experiments and the quantum 
formalism for random ones is not meaningful.

\section{Genuine Measurements}
\label{sec:genuine}

\emph{Genuine measurements} are experiments for determining the values of the variables of the theory, as opposed to \emph{quantum measurements}, which do not actually \emph{measure} anything in the ordinary sense of the word, i.e., do not measure any pre-existing value of a physical quantity. Genuine measurements in GRWm, for example, would be experiments determining $m(x,t)$, or the wave function, or some functional thereof. In this section we discuss the possibilities and limitations of genuine measurements in GRWm and GRWf.
We plan to provide a more thorough discussion in a future work \cite{grw3C}.

\subsection{Limitations to Knowledge}

We show that it is \emph{impossible} to measure, with microscopic accuracy,
\begin{itemize}
\item[(i)] the matter density $m(x,t)$ in GRWm
\item[(ii)] the wave function $\psi_t$ of a system in either GRWm or GRWf.
\end{itemize}
Furthermore, we conjecture that it is also impossible to measure
\begin{itemize}
\item[(iii)] the space-time pattern of flashes $F$ in GRWf or of the collapse centers in GRWm
\item[(iv)] the number $C_{[\ti,\tf)}$ of collapses in a system during $[\ti,\tf)$ in either GRWm or GRWf
\end{itemize}

In other words, the exact values of these variables are \emph{empirically undecidable}. In contrast, it is \emph{possible} to measure the \emph{macroscopic equivalence class} of either $m(\cdot,t)$ in GRWm, or of $F$ in GRWf, or of $\psi_t$ in both GRWm and GRWf. Further genuine measurements are possible, as we will explain in Section~\ref{sec:GMgivenwf}, when information about the wave function is provided or when many systems with the same wave function are provided.

Let us compare this situation to that of Bohmian mechanics. Also Bohmian mechanics entails limitations to knowledge: for example there is no experiment in a Bohmian world that will reveal the velocity of a given particle (unless information about its wave function is given) \cite{DGZ04,DGZ08}. On the other hand, there is no limitation in Bohmian mechanics to measuring the position of a particle, except that doing so will alter the particle's wave function, and thus its future trajectory. Here we encounter a basic difference between Bohmian mechanics and GRWm: the configuration of the PO can be measured in Bohmian mechanics but not in GRWm. (In Bohmian mechanics, the configuration of the PO corresponds to the positions of all particles, while in GRWm it corresponds to the $m(x,t)$ function for all $x\in\RRR^3$.) In GRWf, for comparison, there is nothing like a configuration of the PO \emph{at time $t$}, of which we could ask whether it can be measured. There is only a space-time \emph{history} of the PO, which we may wish to measure. Bohmian mechanics provides an example of a world in which the history of a system cannot be measured without disturbing its course, and indeed disturbing it all the more drastically the more accurately we try to measure it. This suggests that also in GRWf, measuring the pattern of flashes may entail disturbing it---and thus finding a pattern of flashes that is different from what would have occurred naturally (i.e., without intervention). This kind of measurement is not what was intended when wishing to measure the history of flashes. 

The conjectures to the effect that the pattern of flashes can only be measured on the macroscopic level, but not with microscopic accuracy, imply that the calibration function $\zeta$, which provided the outcome $Z$ of an experiment as a function of the set $F$ of flashes, cannot be an arbitrary function but must be suitably coarse. While our derivation of the GRW formalism did not require any assumptions on $\zeta$, its coarseness now becomes relevant.

About (iv), we conjecture further that, unless information about the system's wave function is given, no experiment can reveal any information at all about the random number $C_{[\ti,\tf)}$. This means the following. Without any experiment we can say that $C_{[\ti,\tf)}$ has a Poisson distribution with expectation $N\lambda (\tf-\ti)$, i.e.,
\be\label{Poisson}
\PPP(C_{[\ti,\tf)} = n) = \frac{1}{n!} (N\lambda (\tf-\ti))^n \, e^{-N\lambda (\tf-\ti)}\,,
\ee
where $N\lambda$ is the collapse rate of the system (see after \eqref{eq:H}). The conjecture is that no experiment on the system can produce an outcome $Z$ such that the conditional distribution $\PPP(C_{[\ti,\tf)}=n|Z)$ would be narrower than \eqref{Poisson}, or indeed in any way different from \eqref{Poisson}.

\subsection{The Quadratic Function Argument}
\label{sec:QuadFuncArg}

There is a simple argument, the \emph{quadratic function argument}, that will prove the impossibility claims (i) and (ii). This argument was first used, to our knowledge, in \cite{DGZ04} in the context of Bohmian mechanics, and goes as follows. Measuring a quantity $Z$ pertaining to a system without knowing the system's initial wave function $\psi$ requires an experiment that measures $Z$ for every system regardless of its initial wave function. As a consequence, the probability distribution of $Z$ is a quadratic function of $\psi$, i.e., $\PPP(Z=z) = \scp{\psi}{E(z)|\psi}$ for some POVM $E(\cdot)$, since, by the GRW formalism, for every experiment the distribution of its results is a quadratic function of $\psi$. This allows us to conclude that a quantity whose distribution is not a quadratic function of $\psi$ cannot be measured.

We list some such quantities: 
\begin{itemize}
\item The wave function $\psi$ itself, since its distribution depends on $\psi$ like a Dirac $\delta$ function. (More explicitly, $Z=\psi$ would have the distribution $\PPP(Z\in B) = 1_B(\psi)$, which is $1$ if $\psi\in B$ and $0$ otherwise, for any subset $B\subset \Values = \sphere(\Hilbert)$ of the unit sphere in Hilbert space. Unlike $\psi\mapsto \scp{\psi}{E(B)|\psi}$, the step function $1_B$ is not a quadratic function.) \y{More generally, any quantity that is deterministic in $\psi$, i.e., given by a non-constant function of $\psi$, has a distribution that depends on $\psi$ like a $\delta$ function, not quadratically.}
\item Also the distribution of the wave function $\psi_\tf$ at a later time $\tf$, arising from the initial $\psi = \psi_{\ti}$ through the GRW evolution, is not a quadratic function of $\psi$. \y{(For the unitary evolution, the corresponding statement follows from the previous remark, as $\psi_\tf$ then is a function of $\psi_\ti$. The statement for the GRW evolution thus seems more or less clear when we regard it as a kind of perturbation of the unitary evolution. Also, since for $\tf=\ti$ the distribution of $\psi_\tf$ is not a quadratic function, it seems more or less clear that it will not suddenly be one for $\tf>\ti$. To appreciate better that there is no reason why the dependence on $\psi_\ti$ should be quadratic, it may be helpful to note that the fact that the distribution of $F_{[\ti,\tf)}$ is a quadratic function of $\psi_\ti$ does \emph{not} imply the same for $\psi_\tf$: To be sure, by the function property \eqref{fPOVM} of POVMs, any function of $F_{[\ti,\tf)}$ has distribution quadratic in $\psi_\ti$; however, $\psi_\tf$ is not a function of $F_{[\ti,\tf)}$ alone but rather one of $F_{[\ti,\tf)}$ and $\psi_\ti$ together, namely $L(F_{[\ti,\tf)})\psi_\ti/\| L(F_{[\ti,\tf)})\psi_\ti\|$.)}
\item The distribution of $m(x,t)$ is not quadratic in $\psi_\ti$; in fact, for $t=\ti$ it is a $\delta$ distribution. 
\end{itemize}

We have thus proved statements (i) and (ii). Since the distribution of $F$ \emph{is} in fact a quadratic functional of $\psi$, the quadratic functional argument does not yield statement (iii). If we could measure wave functions, we would be able to detect collapses by measuring the wave function before and after; but we cannot.

\bigskip

\label{sec:measureF}

Let us now turn to the heuristic behind the conjecture that flashes cannot be measured.
Here is a very simple, non-rigorous argument suggesting this. Suppose we had an apparatus capable of detecting flashes in a system. Think of the wave function of system and apparatus together as a function on configuration space $\RRR^{3N}$. There is a region in configuration space containing the configurations in which the apparatus display reads ``no flash detected so far,'' and another region, disjoint from the first, containing the configurations in which the display reads ``one flash detected so far.'' Recall that a flash in the system leads to a change in the wave function of the form $\psi\to\psi'=\Lambda_i(x)\psi/\|\Lambda_i(x)\psi\|$, where $\Lambda_i(x)$ is a multiplication operator. But with such a change it is impossible to push the wave function from the first region to the second.

Here is a somewhat similar argument concerning the wish to measure the location $X$ of a flash. For simplicity, let us assume the system consists of a single ``particle,'' and let us further assume we are given the following information about a system's wave function: It is a superposition of finitely many disjoint packets $\psi_\ell$, each so narrow that its width is much smaller than $\sigma$, and any two so very disjoint that their distance is much greater than $\sigma$. Then a collapse will essentially remove all but one of these packets. Now a collapse acting on the system can indeed force the apparatus into a particular state, for example if the wave function of system and apparatus together before the collapse was
\be\label{psiellphiell}
\sum_\ell \psi_\ell \otimes \phi_\ell\,,
\ee
where $\phi_\ell$ may be a state in which the apparatus displays the location of $\psi_\ell$ as the location of the flash. The state \eqref{psiellphiell} may arise from the initial state $(\sum\psi_\ell)\otimes \phi_0$ by means of the interaction between the system and the apparatus. However, in case no flash occurs, the reduced density matrix of the system arising from the state \eqref{psiellphiell} would be $\sum_\ell \pr{\psi_\ell}$, which leads to a different distribution of flashes than the pure state $\pr{\sum \psi_\ell}$. This means that the presence of the apparatus has altered the distribution of the future flashes. Moreover, the state \eqref{psiellphiell} represents essentially the wave function resulting from a quantum position measurement, and will collapse most probably because of flashes associated with the apparatus, thus forcing the first system flash to occur at the location that was the outcome of the position measurement.

\subsection{If Information About the Initial Wave Function Is Given}
\label{sec:GMgivenwf}

Further genuine measurements are possible when information about the wave function is provided. What does that mean? For example, while there \y{exists no experiment, according to the above conjecture,} that could measure the number $C_{[\ti,\tf)}$ of collapses between $\ti$ and $\tf$ \emph{on any given system with any (unknown) wave function}, there do exist experiments that work \emph{for one particular wave function} $\psi$ and can, for a system with initial wave function $\psi_{\ti} = \psi$, disclose at least partial information about this number. We describe a concrete example of an experiment with which, though it does not reliably determine $C_{[\ti,\tf)}$, one can estimate $C_{[\ti,\tf)}$ better than one could without performing any experiment. 

Suppose $\psi$ is a wave function of a single electron which is a superposition of two wave packets,
\be\label{superposition}
\psi = \tfrac{1}{\sqrt{2}} |\text{here}\rangle + \tfrac{1}{\sqrt{2}} |\text{there}\rangle\,,
\ee
as may result from a double-slit setup. Suppose, for simplicity, that the Hamiltonian of the system vanishes, so that the quantum time evolution is trivial, and that the time span $\tf-\ti$ is of the order $1/N\lambda$, so that $\PPP(C_{[\ti,\tf)}=0)$ is neither close to 1 nor close to 0. We ask whether $C_{[\ti,\tf)}$ is zero or nonzero, i.e., whether a collapse has occurred. Without any experiment, one can only say that the probability that a collapse has occurred is
\be
p= 1-e^{-N\lambda (\tf-\ti)}\,.
\ee
The following experiment provides further information. The task is equivalent to determining whether $\psi_{\tf} =  \tfrac{1}{\sqrt{2}} |\text{here}\rangle + \tfrac{1}{\sqrt{2}} |\text{there}\rangle$ or $\psi_{\tf} = |\text{here}\rangle$ or $\psi_{\tf} = |\text{there}\rangle$, since any collapse would effectively reduce \eqref{superposition} to either $|\text{here}\rangle$ or $|\text{there}\rangle$. To this end, carry out a ``quantum measurement of the observable'' $O$ given by the projection to the 1-dimensional subspace spanned by \eqref{superposition}.\footnote{Actually, it is not obvious that such an experiment is possible in GRWf. Strictly speaking, that is a gap in our argument for the possibility of obtaining probabilistic information about $C_{[\ti,\tf)}$. However, it seems plausible that such an experiment is possible in GRWf if it is in ordinary quantum mechanics, and there it is commonly taken for granted that every self-adjoint operator on a 2-dimensional Hilbert space (such as span$\{|\text{here}\rangle,|\text{there}\rangle\}$) is the observable for some experiment.} If the result $Z$ is zero, then it can be concluded that a collapse has occurred, or $C_{[\ti,\tf)} > 0$. If the result $Z$ is $1$, nothing can be concluded with certainty (since also $|\text{here}\rangle$ and $|\text{there}\rangle$ lead to a probability of $1/2$ for the outcome to be 1). However, in this case the (Bayesian) conditional probability that a collapse has occurred is less than $p$ (and thus $Z$ is informative about $C_{[\ti,\tf)}$):
\begin{equation}
  \PPP(C_{[\ti,\tf)}>0|Z=1)=\frac{\PPP(C_{[\ti,\tf)}>0,Z=1)}{\PPP(C_{[\ti,\tf)}>0,Z=1)
  +\PPP(C_{[\ti,\tf)}=0,Z=1)}=
\end{equation}
\begin{equation}
  =\frac{\PPP(Z=1|C_{[\ti,\tf)}>0)\,\PPP(C_{[\ti,\tf)}>0)}{\PPP(Z=1|C_{[\ti,\tf)}>0)\,
  \PPP(C_{[\ti,\tf)}>0)+
  \PPP(Z=1|C_{[\ti,\tf)}=0)\,\PPP(C_{[\ti,\tf)}=0)}= 
\end{equation}
\begin{equation}
  = \frac{\tfrac{1}{2}p}{\tfrac{1}{2}p + 1\cdot(1-p)} = \frac{p}{2-p}<p\,.
\end{equation}
Thus, in every case with the experiment we can retrodict $C_{[\ti,\tf)}$ with greater reliability than could have been achieved a priori. 

This leads us to the question whether it is possible, for a known initial wave function, to determine \emph{reliably} whether a collapse has occurred or not. We conjecture that the answer is no, and that indeed with no other experiment can we retrodict whether $C_{[\ti,\tf)}>0$ for the initial wave function \eqref{superposition} with greater reliability than with the quantum measurement of $O$.

\section{Conclusions}

We have formulated a GRW formalism that is analogous to, but not the same as, the quantum formalism and summarizes the empirical content of both GRWm and GRWf. We have given a derivation of the GRW formalism based on the primitive ontologies (POs) of GRWm and GRWf. We have further shown that several quantities that are real in the GRWm or GRWf worlds cannot be measured by the inhabitants of these worlds. These were the main contributions of this paper. Derivations of empirical predictions of GRW theories have been given before in \cite{GRW86,Bell87, Rae90,PS94, BG03,JPR04, BGS06,Adl07,BS07}, but with two gaps: First, these derivations did not pay attention to the role of the PO; see \cite{grw3B} for discussion. Second, these derivations either focused on how to obtain the quantum probabilities from GRW theories, ignoring the (usually tiny) differences between the empirical predictions of GRW theories and those of quantum mechanics, or focused on particular experiments but did not provide a general formalism \cite{GRW86,Rae90,PS94,BG03,JPR04,Adl07}.

It has played an important role for our analysis that the GRW theories are given by explicit equations. Other collapse theories, for example that of Penrose \cite{Pen00,Pen04}, are formulated in a more vague way that still permits one to arrive at concrete testable predictions deviating from quantum mechanics but does not yield any general theorems about arbitrary experiments. The concreteness of the GRW theories also has (what may seem like) disadvantages, as it gives the theory a flavor of arbitrariness, and that of being ``merely a toy model,'' as opposed to a serious theory. For example, arbitrariness may be seen in the existence of the two parameters $\lambda$ and $\sigma$ (whose values must remain unknown until experiments confirm deviations from quantum mechanics), or in the choice of the Gaussian \eqref{eq:collapseoperator} (could it not be another function instead of a Gaussian?), or in the assumption that collapses are instantaneous, or in other aspects. But at the end of the day it is the concreteness of the GRW theories, or their explicit character, that paves the way for their successful analysis. In this paper in particular, theorems are established about the GRW theories, and this would not have been possible if the GRW theories had not been defined by unambiguous mathematics. Since we are dealing with concrete equations, we can derive precisely what predictions these equations entail---with rather unexpected results, such as the emergence of a simple operator formalism.

It has also played an important role to be explicit about the PO, i.e., to say clearly what the PO is and to specify an equation governing the PO, namely \eqref{nflashdist} respectively \eqref{mdef}. To provide such an equation is somewhat unusual; instead it is often silently assumed that when $\psi_t$ is the wave function of a live cat then there is a live cat. Our derivation of the GRW formalism relied on this equation \eqref{nflashdist}, which makes the structure of the argument explicit and simple.

Another question arises once the GRW formalism is formulated: Should we not, given that the GRW formalism summarizes the empirical content of GRWm/GRWf, keep only the GRW formalism as a physical theory and abandon GRWm and GRWf? From a positivistic point of view it would seem that we should because in this view only empirical predictions are regarded as scientific, meaningful statements. But our answer is no. In our view, the positivistic position is too extreme. The goal of science is not only to summarize empirical observations but also to explore explanations of the observations. It is entirely reasonable to ask for a theory that speaks about reality and not only about observations, i.e., for a quantum theory without observers. But even if the positivistic position were 
not extreme, one could hardly help noticing that the GRW formalism, given by \eqref{M}, \eqref{GRWPPP}, \eqref{GRWC}, \eqref{GRWE}, and \eqref{GRWCPM}, is considerably 
more complicated that the theory GRWf itself.

\bigskip

\noindent\textit{Acknowledgments.} 
We thank David Albert (Columbia University), Detlef D\"urr (LMU M\"unchen), Tim Maudlin (New York University), Daniel Victor Tausk (S\~ao Paulo), and Bassano Vacchini (Milano) for helpful discussions. 
S.~Goldstein is supported in part by NSF Grant DMS-0504504.
R.~Tumulka is supported in part by NSF Grant SES-0957568 and by the Trustees Research Fellowship Program at Rutgers, the State University of New Jersey. 
N.~Zangh\`\i\ is supported in part by INFN. 

\appendix

\section{Proof of the Conditional Probability Formula \eqref{condprob1}}
\label{app:condprob}

Note first that, for $\tf<\infty$,
\begin{equation}\label{Lttt}
L_{[0,\tf)}(f) = L_{[\ti,\tf)}(f_{[\ti,\tf)}) \, L_{[0,\ti)}(f_{[0,\ti)})
\end{equation}
and
\begin{equation}
\int_{\Omega_{[\ti,\tf)}} df \, L^*_{[\ti,\tf)}(f) \, L_{[\ti,\tf)}(f) = \G_{[\ti,\tf)}(\Omega_{[\ti,\tf)}) = I\,,
\end{equation}
the identity operator. As a consequence,
\begin{align}
  &\PPP_{\wf_{0}}\Bigl(F_{[\ti,\tf)}\in B\Big|F_{[0,\ti)}=f_{[0,\ti)}\Bigr)\nonumber\\[4mm]
  &=\frac{\Bscp{\wf_{0}}{\int_B df_{[\ti,\tf)} \, L^*_{[0,\tf)}(f_{[0,\ti)}\cup f_{[\ti,\tf)}) 
  \, L_{[0,\tf)}(f_{[0,\ti)}\cup f_{[\ti,\tf)}) \Big|\wf_{0}}}
  {\Bscp{\wf_{0}}{\int_{\Omega_{[\ti,\tf)}} df_{[\ti,\tf)} \, 
  L^*_{[0,\tf)}(f_{[0,\ti)}\cup f_{[\ti,\tf)}) 
  \, L_{[0,\tf)}(f_{[0,\ti)}\cup f_{[\ti,\tf)}) \Big|\wf_{0}}} \\[4mm]
 & =  \frac{\Bscp{\wf_{0}}{L^*_{[0,\ti)}(f_{[0,\ti)}) 
  \Bigl(\int_B df_{[\ti,\tf)} \, L^*_{[\ti,\tf)}(f_{[\ti,\tf)}) 
  \, L_{[\ti,\tf)}(f_{[\ti,\tf)})\Bigr) L_{[0,\ti)}(f_{[0,\ti)}) \Big|\wf_{0}}}
  {\Bscp{\wf_{0}}{L^*_{[0,\ti)}(f_{[0,\ti)}) 
  \, L_{[0,\ti)}( f_{[\ti,\tf)}) \Big|\wf_{0}}} \\[4mm]
  &=\Bscp{\wf_{\ti}}{\int_B df_{[\ti,\tf)} \, L^*_{[\ti,\tf)}(f_{[\ti,\tf)}) 
  \, L_{[\ti,\tf)}(f_{[\ti,\tf)}) \Big|\wf_{\ti}} \\[3mm]
  &= \PPP^{(\ti)}_{\wf_{\ti}}\bigl(F_{[\ti,\tf)}\in B\bigr)
\end{align}
with $\wf_{\ti}=L_{[0,\ti)}(f_{[0,\ti)}) \wf_0/\|L_{[0,\ti)}(f_{[0,\ti)}) \wf_0\|$.

This proves the conditional probability formula for $\tf<\infty$. The one for $\tf=\infty$ follows from the one for finite $\tf$ because in the $\sigma$-algebra of $\Omega_{[\ti,\infty)}$, the family $\mathcal{A}_\mathrm{finite}$ of events depending only on a finite amount of time form a $\cap$-stable generator, and thus the two measures $\PPP_{\wf_{0}}\Bigl(F_{[\ti,\infty)}\in \cdot\big|F_{[0,\ti)}=f_{[0,\ti)}\bigr)$ and $\PPP_{\wf_{\ti}}\bigl(F_{[\ti,\infty)}\in \cdot \bigr)$ coincide since they coincide on $\mathcal{A}_\mathrm{finite}$.

\section{Check of Compatibility Conditions \eqref{ECPM}, \eqref{ECPMGRW}, \eqref{QET}, and \eqref{ECPMGRWT}}
\label{app:check}

We provide the proofs of the equations expressing the compatibility property between the POVM $E(\cdot)$ and the superoperator $\cpm_z$ as defined in the various versions of the law of operators. We often use the following mathematical fact: If $\Hilbert_{a\cup b}=\Hilbert_a\otimes \Hilbert_b$, $S_a$ is an operator on $\Hilbert_a$, and $T_{a\cup b}$ is an operator on $\Hilbert_{a\cup b}$ then
\be\label{trbrule}
S_a \, \tr_b\, T_{a\cup b} = \tr_b \bigl( [S_a\otimes I_b] T_{a\cup b} \bigr)\,,
\ee
where $\tr_b$ denotes the partial trace; the mirror image of \eqref{trbrule} holds as well,
\be\label{trbrule2}
\Bigl(\tr_b\, T_{a\cup b}\Bigr) S_a = \tr_b \bigl(T_{a\cup b}  [S_a\otimes I_b] \bigr)\,.
\ee

\bigskip

To check the compatibility property \eqref{ECPM} between \eqref{QuPOVM} and \eqref{QuCPM}, note that
\begin{align}
\tr(\trclop \, E^\Qu_z) 
&= \tr \Bigl( [\trclop \otimes I_\env] [I_\sys\otimes \dm_\app] U^*_{\tf-\ti}
[I_\sys \otimes P_z^\app] U_{\tf-\ti} \Bigr)\\
&= \tr \Bigl( [\trclop \otimes \dm_\app] U^*_{\tf-\ti}
[I_\sys \otimes P_z^\app] U_{\tf-\ti} \Bigr)\\
&=\tr \Bigl(  U_{\tf-\ti} [\trclop \otimes \dm_\app] U^*_{\tf-\ti}
[I_\sys \otimes P_z^\app] \Bigr)\\
&=\tr \Bigl( [I_\sys \otimes P_z^\app] U_{\tf-\ti} [\trclop \otimes \dm_\app] U^*_{\tf-\ti}
[I_\sys \otimes P_z^\app] \Bigr)\\
&= \tr  \cpm^\Qu_z (\trclop)\,,
\end{align}
where $\tr$ always means the trace, sometimes on $\Hilbert_\sys$ and sometimes on $\Hilbert_\sys \otimes \Hilbert_\env$.

\bigskip

To check the compatibility property \eqref{ECPMGRW} between \eqref{GRWE} and \eqref{GRWCPM}, note that
\begin{align}
  \tr \bigl( \trclop \, E^\GRW_z\bigr) 
  &= \tr \Bigl( [\trclop \otimes I_\app] \int\limits_{\zeta^{-1}(z)} df \: 
  [I_\sys \otimes \dm_\app] \, L^*_{[\ti,\tf)}(f) \, L_{[\ti,\tf)}(f) \Bigr)\\
  &= \tr \int\limits_{\zeta^{-1}(z)} df \: 
  L_{[\ti,\tf)}(f) \, [\trclop \otimes \dm_\app] \, L^*_{[\ti,\tf)}(f) \\
  &= \tr \cpm_z^\GRW( \trclop)\,.
\end{align}

\bigskip

To check the compatibility condition between \eqref{QET} and \eqref{QCPMT}, note that
\begin{align}
\tr \bigl( \trclop \, E^\Qu_{z,\tf}\bigr) 
&= \tr \Bigl( [\trclop \otimes I_\app]  [I_\sys\otimes \dm_\app] U^*_{\tf-\ti} \, [I_\sys\otimes P^\app_{z,\tf}] U_{\tf-\ti}\Bigr)\\
&=\tr \Bigl( [\trclop \otimes \dm_\app] U^*_{\tf-\ti} \, [I_\sys\otimes P^\app_{z,\tf}] U_{\tf-\ti}\Bigr)\\
&=\tr \Bigl( [I_\sys\otimes P^\app_{z,\tf}] U_{\tf-\ti} [\trclop \otimes \dm_\app] U^*_{\tf-\ti} \, 
[I_\sys\otimes P^\app_{z,\tf}] \Bigr)\\
&=\tr  \cpm^\Qu_{z,\tf}(\trclop)\,.
\end{align}

\bigskip

To check the compatibility condition \eqref{ECPMGRWT} between \eqref{GRWET2} and \eqref{GRWCPMT}, note that
\begin{align}
\tr\bigl(\trclop \, E^\GRW_{z,\tf}\bigr)
&=\tr\Bigl( [\trclop \otimes I_\app] \int\limits_{\zeta^{-1}(z)\cap {\tauf}^{-1}(\tf)}\hspace{-5mm}df\,
[I_\sys\otimes \dm_\app] L_{[\ti,\tf)}^*(f) \, L_{[\ti,\tf)}(f)\Bigr)\\
&=\tr \int\limits_{\zeta^{-1}(z)\cap {\tauf}^{-1}(\tf)}\hspace{-5mm}df\,
[\trclop\otimes \dm_\app] L_{[\ti,\tf)}^*(f) \, L_{[\ti,\tf)}(f)\\
&=\tr   \cpm^\GRW_{z,\tf} (\trclop)\,.
\end{align}

\section{Proof of the Marginal Probability Formula \eqref{margprob}}
\label{app:margprob}

As a consequence of the factorization formula \eqref{factformL},
\be\label{FFsysFenv}
\G(B_\sys \times B_\env) = \G_\sys(B_\sys) \otimes \G_\env(B_\env)\,,
\ee
where $\G_\sys$ (respectively $\G_\env$) is the POVM that would govern the system (respectively the environment) if it were alone in the universe. (In particular, the marginal of $\G_\sys$ for the first $n$ flashes is given by
\be
\G_{\sys,n}(B) = \int_B df \, L_\sys(f)^* \, L_\sys(f)\quad \forall B \subseteq \Omega_n\,,
\ee
in parallel to \eqref{Fndef}.) From \eqref{FFsysFenv} we obtain the marginal probability formula:
\begin{align}
\PPP_{\wf_0} \bigl( F_\sys \in B_\sys \bigr)  
&=\scp{\wf_0}{\G(B_\sys \times \Omega_\env) |\wf_0}\\
&= \scp{\wf_0}{\G_\sys(B_\sys) \otimes I_\env| \wf_0} \\
&=\tr \bigl( \dm_\sys \, \G_\sys(B_\sys) \bigr)
\end{align}
with $\dm_\sys = \tr_\env \pr{\wf_0}$.

\section{Proof of the Marginal Master Equation \eqref{mM}}
\label{app:margM}

We now provide a proof of the fact, described around \eqref{margM}, that for two non-interacting but entangled systems $a$ and $b$, also the reduced density matrix of system $a$ evolves according to the appropriate version of the master equation \eqref{M}.
This follows from two ingredients: the factorization formula \eqref{factformLst} and the fact that the solution to the master equation \eqref{M} can be expressed in terms of the $L$ operators as
\be
\dm_t^{a\cup b} = \int\limits_{\Omega_{[0,t)}} df\, L_{[0,t)}(f) \, \dm_{0}^{a\cup b} \, L^*_{[0,t)}(f)\,,
\ee
see \eqref{dm0t}. Now it follows, using $f=f_a \cup f_b$, that
\begin{align}
\dm_t^a 
&= \tr_b\, \dm_t^{a\cup b} \\
&= \tr_b\int df_a \int df_b \, [L_{[0,t)}^a(f_a) \otimes L_{[0,t)}^b(f_b)] \, \dm_{0}^{a\cup b} \, [L^a_{[0,t)}(f_a)^* \otimes L^b_{[0,t)}(f_b)^*]\\
&= \int df_a L_{[0,t)}^a(f_a) \, \tr_b\Bigl[ \dm_{0}^{a\cup b} \, I_a \otimes 
\underbrace{\int df_b \, L_{[0,t)}^b(f_b) L^b_{[0,t)}(f_b)^*}_{=I_b} 
\Bigr] L^a_{[0,t)}(f_a)^* \\
&=\int df_a L_{[0,t)}^a(f_a) \, (\tr_b\, \dm_{0}^{a\cup b} )\, L^a_{[0,t)}(f_a)^* \\
&=\int df_a L_{[0,t)}^a(f_a) \, \dm_0^a \, L^a_{[0,t)}(f_a)^*\,,
\end{align}
which means that the reduced density matrix $\dm_t^a$ satisfies the appropriate version of the master equation \eqref{M}.

\section{Proof of \eqref{dmsrandomE}}
\label{app:recipe}

Indeed,
\begin{align}
\dm_\ti 
&=\frac{1}{\mathcal{N}}\tr_{b(e)\cup\env}
\int\limits_{\Omega_{[0,\ti)}} df_{[0,\ti)}
L_{[0,\ti)}(f_{[0,\ti)})\:\times\nonumber\\
&\quad\times\:\pr{\Psi_0} 
L^*_{[0,\ti)}(f_{[0,\ti)}) 
1_{f_{[0,t_0)}\in B_{[0,t_0)}}
1_{\E=e}\\[4mm]
&=\frac{1}{\mathcal{N}}\tr_{b(e)\cup\env}
\int\limits_{\Omega_{[t_0,\ti)}} df_{[t_0,\ti)}
\int\limits_{\Omega_{[0,t_0)}} df_{[0,t_0)}
L_{[t_0,\ti)}(f_{[t_0,\ti)})
L_{[0,t_0)}(f_{[0,t_0)})\:\times\nonumber\\
&\quad\times\:\pr{\Psi_0} 
L^*_{[0,t_0)}(f_{[0,t_0)}) 
L^*_{[t_0,\ti)}(f_{[t_0,\ti)}) 
1_{f_{[0,t_0)}\in B_{[0,t_0)}}
1_{\E=e}\\[4mm]
&=\frac{1}{\mathcal{N'}}\tr_{b(e)\cup\env}
\int\limits_{\Omega_{[t_0,\ti)}} df_{[t_0,\ti)}
L_{[t_0,\ti)}(f_{[t_0,\ti)})
\:\times\nonumber\\&\quad\times\:
\EEE\Bigl(\pr{\Psi_{t_0}}\Big| F_{[0,t_0)}\in B_{[0,t_0)} \Bigr)
L^*_{[t_0,\ti)}(f_{[t_0,\ti)}) 
1_{\E=e}\\
\intertext{[by \eqref{ranEdisent}]}
&=\frac{1}{\mathcal{N'}}\tr_{b(e)\cup\env}
\int\limits_{\Omega_{[t_0,\ti)}} df_{[t_0,\ti)}
L_{[t_0,\ti)}(f_{[t_0,\ti)})
\:\times\nonumber\\&\quad\times\:
\bigl[\dm^{\sys}_{t_0} \otimes \dm^\env_{t_0}\bigr]
L^*_{[t_0,\ti)}(f_{[t_0,\ti)}) 
1_{\E=e}\\
\intertext{[by \eqref{factformLst}]}
&=\frac{1}{\mathcal{N'}}\tr_{b(e)}\tr_\env
\int\limits_{\Omega^{\sys}_{[t_0,\ti)}} \!\!\! df^{\sys}_{[t_0,\ti)}
\int\limits_{\Omega^\env_{[t_0,\ti)}} \!\!\! df^\env_{[t_0,\ti)} \,
\bigl[L^{\sys}_{[t_0,\ti)}(f^{\sys}_{[t_0,\ti)})\otimes
L^\env_{[t_0,\ti)}(f^\env_{[t_0,\ti)})\bigr]
\:\times\nonumber\\&\quad\times\:
\bigl[\dm^{\sys}_{t_0} \otimes \dm^\env_{t_0}\bigr]
\bigl[L^{\sys}_{[t_0,\ti)}(f^{\sys}_{[t_0,\ti)})^*\otimes
L^\env_{[t_0,\ti)}(f^\env_{[t_0,\ti)})^*\bigr]
1_{\varepsilon(f^\env_{[t_0,\ti)})=e}\\
\intertext{[carry out $\tr_\env$]}
&=\frac{1}{\mathcal{N''}}\tr_{b(e)}
\int\limits_{\Omega^{\sys}_{[t_0,\ti)}} \!\!\! df^{\sys}_{[t_0,\ti)} \:
L^{\sys}_{[t_0,\ti)}(f^{\sys}_{[t_0,\ti)})
\dm^{\sys}_{t_0} L^{\sys}_{[t_0,\ti)}(f^{\sys}_{[t_0,\ti)})^*\\
\intertext{[by \eqref{dm0t}]}
&=\frac{1}{\mathcal{N''}}\tr_{b(e)} \acpm^{\sys}_{[t_0,\ti)} 
\dm^{\sys}_{t_0}\\
\intertext{[by \eqref{acpmfactor}]}
&=\frac{1}{\mathcal{N''}}\tr_{b(e)} \Bigl( \bigl[\acpm^{a(e)}_{[t_0,\ti)}\otimes \acpm^{b(e)}_{[t_0,\ti)}\bigr] 
\dm^{a(e)\cup b(e)}_{t_0}\Bigr)\\
&=\frac{1}{\mathcal{N''}}\acpm^{a(e)}_{[t_0,\ti)}
(\tr_{b(e)} \,\dm^{a(e)\cup b(e)}_{t_0})\\
&=\frac{1}{\mathcal{N''}}\acpm^{a(e)}_{[t_0,\ti)}\dm_{t_0}\,,
\end{align}
and since $\acpm$ is trace-preserving, the normalization factor $\mathcal{N}''$ must be 1. This completes the proof of \eqref{dmsrandomE}.

\section{Diagram Notation}
\label{sec:diagram}
\setlength{\unitlength}{0.13\unitlength}
\newcommand{\un}{\unitlength}

\newcommand{\sdma}{\begin{minipage}[b]{4\un}
\begin{picture}(4,6)
\put(0,6){\line(0,-1){6}}
\put(0,6){\line(2,-1){4}}
\put(4,4){\line(-2,-1){4}}
\end{picture}
\end{minipage}}

\newcommand{\sdmc}{\begin{minipage}[b]{4\un}
\begin{picture}(4,6)
\put(0,6){\line(0,-1){6}}
\put(0,6){\line(1,0){4}}
\put(4,6){\line(0,-1){3}}
\put(4,3){\line(-1,0){4}}
\end{picture}
\end{minipage}}

\newcommand{\sdmd}{\begin{minipage}[b]{4\un}
\begin{picture}(4,6)
\put(0,6){\line(0,-1){6}}
\put(0,6){\line(1,0){4}}
\put(4,6){\line(-4,-3){2}}
\put(4,3){\line(-4,3){2}}
\put(4,3){\line(-1,0){4}}
\end{picture}
\end{minipage}}

\newcommand{\sdme}{\begin{minipage}[b]{6\un}
\begin{picture}(6,6)
\put(0,6){\line(0,-1){6}}
\put(0,6){\line(1,0){4}}
\put(4,6){\line(4,-3){2}}
\put(4,3){\line(4,3){2}}
\put(4,3){\line(-1,0){4}}
\end{picture}
\end{minipage}}

\newcommand{\sdmab}{\begin{minipage}[b]{20\un}
\begin{picture}(20,6)
\put(4,2){\line(0,-1){2}}
\put(16,2){\line(0,-1){2}}
\put(4,6){\line(1,0){12}}
\put(0,2){\line(1,0){20}}
\put(4,6){\line(-1,-1){4}}
\put(16,6){\line(1,-1){4}}
\end{picture}
\end{minipage}}

\newcommand{\sSone}{\begin{minipage}[b]{8\un}
\begin{picture}(8,8)
\put(4,8){\line(0,-1){8}}
\put(1,4){\line(1,-1){1}}
\put(2,3){\line(1,1){2}}
\put(4,5){\line(1,-1){2}}
\put(6,3){\line(1,1){1}}
\end{picture}
\end{minipage}}

\newcommand{\sStwo}{\begin{minipage}[b]{20\un}
\begin{picture}(20,8)
\put(4,8){\line(0,-1){8}}
\put(16,8){\line(0,-1){8}}
\put(1,4){\line(1,-1){1}}
\put(2,3){\line(1,1){2}}
\put(4,5){\line(1,-1){2}}
\put(6,3){\line(1,1){2}}
\put(8,5){\line(1,-1){2}}
\put(10,3){\line(1,1){2}}
\put(12,5){\line(1,-1){2}}
\put(14,3){\line(1,1){2}}
\put(16,5){\line(1,-1){2}}
\put(18,3){\line(1,1){1}}
\end{picture}
\end{minipage}}

\newcommand{\sSthree}{\begin{minipage}[b]{32\un}
\begin{picture}(32,8)
\put(4,8){\line(0,-1){8}}
\put(16,8){\line(0,-1){8}}
\put(28,8){\line(0,-1){8}}
\put(1,4){\line(1,-1){1}}
\put(2,3){\line(1,1){2}}
\put(4,5){\line(1,-1){2}}
\put(6,3){\line(1,1){2}}
\put(8,5){\line(1,-1){2}}
\put(10,3){\line(1,1){2}}
\put(12,5){\line(1,-1){2}}
\put(14,3){\line(1,1){2}}
\put(16,5){\line(1,-1){2}}
\put(18,3){\line(1,1){2}}
\put(20,5){\line(1,-1){2}}
\put(22,3){\line(1,1){2}}
\put(24,5){\line(1,-1){2}}
\put(26,3){\line(1,1){2}}
\put(28,5){\line(1,-1){2}}
\put(30,3){\line(1,1){1}}
\end{picture}
\end{minipage}}

\newcommand{\sI}{\begin{minipage}[b]{4\un}
\begin{picture}(4,4)
\put(2,4){\line(0,-1){4}}
\end{picture}
\end{minipage}}

\newcommand{\str}{\begin{minipage}[b]{4\un}
\begin{picture}(4,4)
\put(2,4){\line(0,-1){2}}
\put(2,2){\circle*{2}}
\end{picture}
\end{minipage}}

\newcommand{\sTone}[1]{\begin{minipage}[b]{16\un}
\begin{picture}(16,16)
\put(4,14){\line(1,0){8}}
\put(4,2){\line(1,0){8}}
\put(8,14){\line(0,1){2}}
\put(8,2){\line(0,-1){2}}
\put(4,14){\line(-2,-3){4}}
\put(12,14){\line(2,-3){4}}
\put(4,2){\line(-2,3){4}}
\put(12,2){\line(2,3){4}}
\put(4,4){\makebox(8,8){{\scriptsize $#1$}}}
\end{picture}
\end{minipage}}

\newcommand{\sTtwo}[1]{\begin{minipage}[b]{28\un}
\begin{picture}(28,16)
\put(4,14){\line(1,0){20}}
\put(4,2){\line(1,0){20}}
\put(8,14){\line(0,1){2}}
\put(20,14){\line(0,1){2}}
\put(8,2){\line(0,-1){2}}
\put(20,2){\line(0,-1){2}}
\put(4,14){\line(-2,-3){4}}
\put(24,14){\line(2,-3){4}}
\put(4,2){\line(-2,3){4}}
\put(24,2){\line(2,3){4}}
\put(4,4){\makebox(20,8){{\scriptsize $#1$}}}
\end{picture}
\end{minipage}}

\newcommand{\sTthree}[1]{\begin{minipage}[b]{40\un}
\begin{picture}(40,16)
\put(4,14){\line(1,0){32}}
\put(4,2){\line(1,0){32}}
\put(8,14){\line(0,1){2}}
\put(20,14){\line(0,1){2}}
\put(32,14){\line(0,1){2}}
\put(8,2){\line(0,-1){2}}
\put(20,2){\line(0,-1){2}}
\put(32,2){\line(0,-1){2}}
\put(4,14){\line(-2,-3){4}}
\put(36,14){\line(2,-3){4}}
\put(4,2){\line(-2,3){4}}
\put(36,2){\line(2,3){4}}
\put(4,4){\makebox(32,8){{\scriptsize $#1$}}}
\end{picture}
\end{minipage}}

\newcommand{\sTfour}[1]{\begin{minipage}[b]{52\un}
\begin{picture}(52,16)
\put(4,14){\line(1,0){44}}
\put(4,2){\line(1,0){44}}
\put(8,14){\line(0,1){2}}
\put(20,14){\line(0,1){2}}
\put(32,14){\line(0,1){2}}
\put(44,14){\line(0,1){2}}
\put(8,2){\line(0,-1){2}}
\put(20,2){\line(0,-1){2}}
\put(32,2){\line(0,-1){2}}
\put(44,2){\line(0,-1){2}}
\put(4,14){\line(-2,-3){4}}
\put(48,14){\line(2,-3){4}}
\put(4,2){\line(-2,3){4}}
\put(48,2){\line(2,3){4}}
\put(4,4){\makebox(44,8){{\scriptsize $#1$}}}
\end{picture}
\end{minipage}}

\newcommand{\sCa}[1]{\begin{minipage}[b]{16\un}
\begin{picture}(16,16)
\put(4,14){\line(1,0){8}}
\put(4,2){\line(1,0){8}}
\put(8,14){\line(0,1){2}}
\put(8,2){\line(0,-1){2}}
\put(4,14){\line(2,-3){4}}
\put(12,14){\line(2,-3){4}}
\put(4,2){\line(2,3){4}}
\put(12,2){\line(2,3){4}}
\put(7,4){\makebox(8,8){{\scriptsize $#1$}}}
\end{picture}
\end{minipage}}

\newcommand{\sCb}[1]{\begin{minipage}[b]{16\un}
\begin{picture}(16,16)
\put(4,14){\line(1,0){8}}
\put(4,2){\line(1,0){8}}
\put(8,14){\line(0,1){2}}
\put(8,2){\line(0,-1){2}}
\put(4,14){\line(-2,-3){4}}
\put(12,14){\line(-2,-3){4}}
\put(4,2){\line(-2,3){4}}
\put(12,2){\line(-2,3){4}}
\put(1,4){\makebox(8,8){{\scriptsize $#1$}}}
\end{picture}
\end{minipage}}

The kind of calculations relevant to the derivation of the GRW formalism involve combinations of superoperators, some of which act on several systems, as well as the operations of tensor product and partial trace. When such calculations become more complicated, the standard notation often becomes hard to follow, as exemplified by the direct calculation in Section~\ref{sec:diagram2} below of the joint distribution of the outcomes of two consecutive experiments. Here we introduce a diagram notation that is better suited than standard notation for this type of calculation because the terms involved can be arranged more clearly in two dimensions (as in a diagram) than in one (as in standard notation). Of the two dimensions, one represents time and the other is used for listing several systems (such as $a$, $b$, $\app_1$, $\app_2$, etc.). This notation is based on a similar diagram notation developed by Penrose and Rindler \cite{PR84} for the tensors of general relativity.

\subsection{Diagrams for Superoperators}

Each diagram represents either a completely positive superoperator or a non-normalized density matrix (i.e., a positive trace-class operator). The composition $\ccpm\circ\bcpm$ of two superoperators is represented by drawing the diagram of $\ccpm$ below that of $\bcpm$ and drawing a line connecting the two. To this end, the diagrams have outward lines (``legs'') on top and at the bottom. For example, the following symbols can represent superoperators $\bcpm,\ccpm$ on $\Hilbert$:
\be
\bcpm=\begin{minipage}{16\un}\sCa{}\end{minipage}\,,\qquad
\ccpm=\begin{minipage}{16\un}\sCb{}\end{minipage}\,,
\ee
and their composition is
\be\label{compdiag}
\ccpm\circ\bcpm =
\begin{minipage}{16\un}
\begin{picture}(16,32)
\put(0,16){\sCa{}}
\put(0,0){\sCb{}}
\end{picture}
\end{minipage}\:.
\ee
The symbol of a (possibly non-normalized) density matrix on $\Hilbert$ has only a leg at the bottom, e.g.,
\be
\dm = \sdma\,.
\ee
To apply the superoperator $\bcpm$ to the density matrix $\dm$, we write the symbol of $\bcpm$ below that of $\dm$ and connect the outward lines:
\be
\bcpm(\dm) = 
\begin{minipage}{16\un}
\begin{picture}(16,22)
\put(8,16){\sdma}
\put(0,0){\sCa{}}
\end{picture}
\end{minipage}\:.
\ee
To take the trace of an operator, we add a bullet $\str$ to its bottom leg, e.g.,
\be\label{trdmdiag}
\tr\dm = 
\begin{minipage}{6\un}
\begin{picture}(6,10)
\put(2,4){\sdma}
\put(0,0){\str}
\end{picture}
\end{minipage}
\:.
\ee
A diagram without legs, such as the right hand side of \eqref{trdmdiag}, represents a (non-negative) number. (It could be regarded as a completely positive superoperator on $\CCC$, just as a density matrix could be regarded as a completely positive superoperator from $\CCC$ to $\Hilbert$, i.e., $TRCL(\CCC)\to TRCL(\Hilbert)$.)

A superoperator on $\Hilbert_1\otimes \Hilbert_2$ has two upper and two lower legs, one for $\Hilbert_1$ and one for $\Hilbert_2$, e.g., 
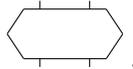
\[
\sTtwo{}\:.
\]
The symbol of a density matrix $\dm_{12}$ on $\Hilbert_1 \otimes \Hilbert_2$ has only two lower legs, as in $\sdmab$. The partial trace is represented by
\be
\tr_2 \dm_{12} = 
\begin{minipage}{20\un}
\begin{picture}(20,10)
\put(0,4){\sdmab}
\put(14,0){\str}
\put(4,0){\line(0,1){4}}
\end{picture}
\end{minipage}
\:.
\ee
The tensor product of two superoperators, $\bcpm\otimes\ccpm$, is denoted by drawing the symbol of $\ccpm$ next to that of $\bcpm$:
\be
\bcpm\otimes\ccpm = 
\begin{minipage}{33\un}
\sCa{}\:\sCb{}
\end{minipage}
\:.
\ee
Since $[\ccpm_1\otimes \ccpm_2] \circ [\bcpm_1\otimes \bcpm_2] = (\ccpm_1\circ\bcpm_1)\otimes (\ccpm_2\circ \bcpm_2)$, it is unambiguous which superoperator on $\Hilbert_1\otimes \Hilbert_2$ the diagram
\be
\begin{minipage}{33\un}
\begin{minipage}{16\un}
\begin{picture}(16,32)
\put(0,16){\sCa{1}}
\put(0,0){\sCb{1}}
\end{picture}
\end{minipage}\:\begin{minipage}{16\un}
\begin{picture}(16,32)
\put(0,16){\sCa{2}}
\put(0,0){\sCb{2}}
\end{picture}
\end{minipage}
\end{minipage}
\ee
represents. The identity superoperator $I$, $I(\dm)=\dm$, is represented by just a straight vertical line $|$ so that legs can be extended arbitrarily.

The legs of a diagram can be thought of as representing indices of a matrix representation of a superoperator relative to a basis of the trace class (of each relevant Hilbert space). For example, let $\{B^{(1)}_{\alpha_1}:\alpha_1=1,2,\ldots\}$ be a basis of $TRCL(\Hilbert_1)$ and $\{B^{(2)}_{\alpha_2}:\alpha_2=1,2,\ldots\}$ a basis of $TRCL(\Hilbert_2)$. Then a density operator $\dm\in TRCL(\Hilbert_1)$ can be expanded in the appropriate basis,
\be
\dm = \sum_{\alpha_1} \dm^{\alpha_1} B^{(1)}_{\alpha_1}\,,
\ee
and thus expressed by the coefficients $\dm^{\alpha_1}$. The index $\alpha_1$ corresponds to the leg of the diagram $\sdma$ for $\dm$. (Note though, that an \emph{upper} index corresponds to a \emph{lower} leg. This is because it is common, particularly in relativity theory, to write the index of an expansion coefficient as an upper index, while our convention about lower legs makes sure that, in a chain of superoperators such as in \eqref{compdiag}, the superoperators get executed from top to bottom.) 

A superoperator $\bcpm:TRCL(\Hilbert_1)\to TRCL(\Hilbert_1)$ can be represented by a matrix $\bcpm_{\alpha_1}^{\beta_1}$ according to
\be
\bcpm(B^{(1)}_{\alpha_1}) = \sum_{\beta_1} \bcpm_{\alpha_1}^{\beta_1} B^{(1)}_{\beta_1}\,.
\ee
The upper leg of the symbol for $\bcpm$ corresponds to the index $\alpha_1$, the lower leg to $\beta_1$, and connecting two legs as in \eqref{compdiag} to summing over the corresponding index as in 
\be
\sum_{\beta_1} \ccpm_{\beta_1}^{\gamma_1}\bcpm_{\alpha_1}^{\beta_1} \,.
\ee
The coefficients of a superoperator on $\Hilbert_1\otimes \Hilbert_2$ are of the form $\bcpm_{\alpha_1\alpha_2}^{\beta_1\beta_2}$ corresponding to four legs, and the coefficients of a density operator on $\Hilbert_1\otimes\Hilbert_2$ are of the form $\dm^{\alpha_1\alpha_2}$, corresponding to two legs. The trace (or partial trace) symbol $\str$ corresponds to the sequence of coefficients $\tr B^{(1)}_{\alpha_1}$ or $\tr B^{(2)}_{\alpha_2}$, whichever is appropriate.

\subsection{Diagram Notation Applied to GRW Theories}

In GRW theories, the time evolution of the density matrix from $t_1$ to $t_2$ is given by a completely positive superoperator $\acpm_{[t_1,t_2)}$, for which we introduce the symbol
\be
\acpm_{[t_1,t_2)}=\begin{minipage}{8\un}\sSone\end{minipage}\:.
\ee
Correspondingly, for the time evolution of two or three systems together we write $\sStwo$ or $\sSthree$. The fact that $\acpm_{[t_1,t_2)}$ is trace-preserving can be expressed as follows:
\be\label{sStr}
\begin{minipage}{8\un}
\begin{picture}(8,12)
\put(0,4){\sSone} 
\put(2,0){\str}
\end{picture}
\end{minipage}
=
\begin{minipage}{8\un}
\begin{picture}(8,12)
\put(4,4){\line(0,1){8}} 
\put(2,0){\str}
\end{picture}
\end{minipage}
\:.
\ee
For two mutually isolated systems,
\be\label{sSisolated}
\begin{minipage}{20\un}\sStwo\end{minipage}
=
\begin{minipage}{8\un}\sSone\end{minipage}
\begin{minipage}{8\un}\sSone\end{minipage}
\:.
\ee
That is, $\acpm_{[t_1,t_2)}$ for both systems is the tensor product of one such superoperator for each system. Note that in \eqref{sSisolated}, the two symbols $\sSone$ may actually represent two \emph{different} superoperators; we take the symbol $\sSone$ always to mean the ``appropriate'' time evolution superoperator.  From \eqref{sStr} and \eqref{sSisolated}, we immediately obtain the marginal master equation: for two mutually isolated systems,
\be
\begin{minipage}{20\un}
\begin{picture}(20,12)
\put(0,4){\sStwo}
\put(2,0){\sI}
\put(14,0){\str}
\end{picture}
\end{minipage}
=
\begin{minipage}[b]{16\un}
\begin{picture}(16,8)
\put(0,0){\sSone} 
\put(10,0){\str}
\put(10,4){\sI}
\end{picture}
\end{minipage}
\:.
\ee

Another superoperator that comes up frequently is
\be
\begin{minipage}{16\un}
\sTone{A}
\end{minipage}
:= \Bigl[\rho \mapsto \int_A df \, L_{[t_1,t_2)}^*(f) \, \rho\, L_{[t_1,t_2)}(f) \Bigr] \:,
\ee
where $A\subseteq \Omega=\Omega_{[t_1,t_2)}$ is a set of flash histories.
We observe the general fact that
\be
\begin{minipage}{16\un}
\sTone{\Omega}
\end{minipage}
= 
\begin{minipage}{8\un}
\begin{picture}(8,16)
\put(2,12){\sI}
\put(0,4){\sSone}
\put(2,0){\sI}
\end{picture}
\end{minipage}
\:.
\ee
The distribution of flashes can be expressed as follows:
\be
\text{If }\rho=\sdma \text{ then }
\mathbb{P}_{\rho}(F\in A) = 
\begin{minipage}{16\un}
\begin{picture}(16,26)
\put(8,20){\sdma}
\put(0,4){\sTone{A}}
\put(6,0){\str}
\end{picture}
\end{minipage}
\:.
\ee
Moreover, for two mutually isolated systems,
\be
\begin{minipage}{28\un}
\sTtwo{B\times B'}
\end{minipage} 
= 
\begin{minipage}{16\un}
\sTone{B}
\end{minipage} 
\, 
\begin{minipage}{16\un}
\sTone{B'}
\end{minipage}
\:.
\ee

With the notation
\be
\mathscr{C}_{z} = \begin{minipage}{16\un}\sCa{z}\end{minipage}\:,
\ee
the GRW formalism implies that
\be
\mathbb{P}(Z=z) = 
\begin{minipage}{16\un}
\begin{picture}(16,26)
\put(8,20){\sdma}
\put(0,4){\sCa{z}}
\put(6,0){\str}
\end{picture}
\end{minipage}
\ee
and the GRW law of operators says that
\be
\begin{minipage}{16\un}
\begin{picture}(16,28)
\put(8,22){\line(0,1){6}}
\put(0,6){\sCa{z}}
\put(8,0){\line(0,1){6}}
\end{picture}
\end{minipage}
=
\begin{minipage}{28\un}
\begin{picture}(28,28)
\put(8,22){\line(0,1){6}}
\put(20,22){\sdmc}
\put(0,6){\sTtwo{\zeta^{-1}(z)}}
\put(8,0){\line(0,1){6}}
\put(18,2){\str}
\end{picture}
\end{minipage}
\text{ with }\sdmc=\rho_{\app}\,.
\ee

\subsection{Example: Two Consecutive Experiments}
\label{sec:diagram2}

As an example for the use of the diagram notation, we carry out the calculation that yields the formula \eqref{PZ2Z1} for the joint distribution of the outcomes of two consecutive experiments $\E_1,\E_2$ on the same system $a$. This calculation amounts more or less to another derivation of the third rule of the GRW formalism.

In the diagrams that follow, the columns correspond to different systems (such as system $a$, system $b$, the apparatus), and different rows correspond to different times (with the time axis pointing downward).
\be
\mathbb{P}\Bigl( \zeta_2\bigl(F_{[\ti_2,\tf_2)}\bigr)=z_2, \zeta_1\bigl(F_{[\ti_1,\tf_1)} \bigr)=z_1\Bigr) 
\:=\:
\begin{minipage}{70\un}
\begin{picture}(70,62)
\put(12,54){\makebox(10,8){\scriptsize$b_1$}}
\put(24,54){\makebox(10,8){\scriptsize$a$}}
\put(36,54){\makebox(10,8){\scriptsize$\app_1$}}
\put(48,54){\makebox(10,8){\scriptsize$\app_2$}}
\put(60,54){\makebox(10,8){\scriptsize$\env$}}
\put(12,44){\sdmab}
\put(40,44){\sdmc}
\put(64,44){\sdme}
\put(0,40){\makebox(8,8){\scriptsize$\ti_1$}}
\put(8,28){\sTthree{\zeta_1^{-1}(z_1)}}
\put(64,40){\line(0,1){4}}
\put(60,32){\sSone}
\put(64,28){\line(0,1){4}}
\put(0,24){\makebox(8,8){\scriptsize$\tf_1$}}
\put(12,20){\sSthree}
\put(52,20){\sdmd}
\put(60,20){\sSone}
\put(0,16){\makebox(8,8){\scriptsize$\ti_2$}}
\put(8,4){\sTfour{\zeta_2^{-1}(z_2)}}
\put(64,14){\line(0,1){6}}
\put(60,6){\sSone}
\put(64,4){\line(0,1){2}}
\put(0,2){\makebox(8,8){\scriptsize$\tf_2$}}
\put(14,0){\str}
\put(26,0){\str}
\put(38,0){\str}
\put(50,0){\str}
\put(62,0){\str}
\end{picture}
\end{minipage}
\:=\:
\ee
[the ``env'' column is equal to 1, and using that $b_1$ is isolated from $a\cup\app_1\cup\app_2$]
\be
\:=\:
\begin{minipage}{60\un}
\begin{picture}(60,62)
\put(12,54){\makebox(10,8){\scriptsize$b_1$}}
\put(24,54){\makebox(10,8){\scriptsize$a$}}
\put(36,54){\makebox(10,8){\scriptsize$\app_1$}}
\put(48,54){\makebox(10,8){\scriptsize$\app_2$}}
\put(12,44){\sdmab}
\put(40,44){\sdmc}
\put(0,40){\makebox(8,8){\scriptsize$\ti_1$}}
\put(16,40){\line(0,1){4}}
\put(12,32){\sSone}
\put(20,28){\sTtwo{\zeta_1^{-1}(z_1)}}
\put(0,24){\makebox(8,8){\scriptsize$\tf_1$}}
\put(16,28){\line(0,1){4}}
\put(12,20){\sSone}
\put(24,20){\sStwo}
\put(52,20){\sdmd}
\put(0,16){\makebox(8,8){\scriptsize$\ti_2$}}
\put(16,16){\line(0,1){4}}
\put(12,8){\sSone}
\put(16,4){\line(0,1){4}}
\put(20,4){\sTthree{\zeta_2^{-1}(z_2)}}
\put(0,2){\makebox(8,8){\scriptsize$\tf_2$}}
\put(14,0){\str}
\put(26,0){\str}
\put(38,0){\str}
\put(50,0){\str}
\end{picture}
\end{minipage}
\:=\:
\begin{minipage}{60\un}
\begin{picture}(60,62)
\put(12,54){\makebox(10,8){\scriptsize$b_1$}}
\put(24,54){\makebox(10,8){\scriptsize$a$}}
\put(36,54){\makebox(10,8){\scriptsize$\app_1$}}
\put(48,54){\makebox(10,8){\scriptsize$\app_2$}}
\put(12,44){\sdmab}
\put(40,44){\sdmc}
\put(0,40){\makebox(8,8){\scriptsize$\ti_1$}}
\put(14,40){\str}
\put(20,28){\sTtwo{\zeta_1^{-1}(z_1)}}
\put(0,24){\makebox(8,8){\scriptsize$\tf_1$}}
\put(24,20){\sStwo}
\put(52,20){\sdmd}
\put(0,16){\makebox(8,8){\scriptsize$\ti_2$}}
\put(20,4){\sTthree{\zeta_2^{-1}(z_2)}}
\put(0,2){\makebox(8,8){\scriptsize$\tf_2$}}
\put(26,0){\str}
\put(38,0){\str}
\put(50,0){\str}
\end{picture}
\end{minipage}
\:=\:
\ee
[introducing the abbreviation $\sdma=\begin{minipage}{20\un}\begin{picture}(20,10)\put(0,4){\sdmab}\put(2,0){\str}\put(16,0){\line(0,1){4}}\end{picture}\end{minipage}$]
\be
\:=\:
\begin{minipage}{48\un}
\begin{picture}(48,62)
\put(12,54){\makebox(10,8){\scriptsize$a$}}
\put(24,54){\makebox(10,8){\scriptsize$\app_1$}}
\put(36,54){\makebox(10,8){\scriptsize$\app_2$}}
\put(16,44){\sdma}
\put(28,44){\sdmc}
\put(0,40){\makebox(8,8){\scriptsize$\ti_1$}}
\put(8,28){\sTtwo{\zeta_1^{-1}(z_1)}}
\put(0,24){\makebox(8,8){\scriptsize$\tf_1$}}
\put(12,20){\sStwo}
\put(40,20){\sdmd}
\put(0,16){\makebox(8,8){\scriptsize$\ti_2$}}
\put(8,4){\sTthree{\zeta_2^{-1}(z_2)}}
\put(0,2){\makebox(8,8){\scriptsize$\tf_2$}}
\put(14,0){\str}
\put(26,0){\str}
\put(38,0){\str}
\end{picture}
\end{minipage}
\:=\:
\ee
[changing the order of the columns]
\be
\:=\:
\begin{minipage}{48\un}
\begin{picture}(48,62)
\put(12,54){\makebox(10,8){\scriptsize$\app_1$}}
\put(24,54){\makebox(10,8){\scriptsize$a$}}
\put(36,54){\makebox(10,8){\scriptsize$\app_2$}}
\put(16,44){\sdmc}
\put(28,44){\sdma}
\put(0,40){\makebox(8,8){\scriptsize$\ti_1$}}
\put(8,28){\sTtwo{\zeta_1^{-1}(z_1)}}
\put(0,24){\makebox(8,8){\scriptsize$\tf_1$}}
\put(12,20){\sStwo}
\put(40,20){\sdmd}
\put(0,16){\makebox(8,8){\scriptsize$\ti_2$}}
\put(8,4){\sTthree{\zeta_2^{-1}(z_2)}}
\put(0,2){\makebox(8,8){\scriptsize$\tf_2$}}
\put(14,0){\str}
\put(26,0){\str}
\put(38,0){\str}
\end{picture}
\end{minipage}
\:=\:
\ee
[using that $\app_1$ is isolated from $a\cup\app_2$ after $\tf_1$]
\be
\:=\:
\begin{minipage}{48\un}
\begin{picture}(48,62)
\put(12,54){\makebox(10,8){\scriptsize$\app_1$}}
\put(24,54){\makebox(10,8){\scriptsize$a$}}
\put(36,54){\makebox(10,8){\scriptsize$\app_2$}}
\put(16,44){\sdmc}
\put(28,44){\sdma}
\put(0,40){\makebox(8,8){\scriptsize$\ti_1$}}
\put(8,28){\sTtwo{\zeta_1^{-1}(z_1)}}
\put(0,24){\makebox(8,8){\scriptsize$\tf_1$}}
\put(12,20){\sSone}
\put(24,20){\sSone}
\put(40,20){\sdmd}
\put(0,16){\makebox(8,8){\scriptsize$\ti_2$}}
\put(16,16){\line(0,1){4}}
\put(12,8){\sSone}
\put(16,4){\line(0,1){4}}
\put(20,4){\sTtwo{\zeta_2^{-1}(z_2)}}
\put(0,2){\makebox(8,8){\scriptsize$\tf_2$}}
\put(14,0){\str}
\put(26,0){\str}
\put(38,0){\str}
\end{picture}
\end{minipage}
\:=\:
\begin{minipage}{48\un}
\begin{picture}(48,62)
\put(12,54){\makebox(10,8){\scriptsize$\app_1$}}
\put(24,54){\makebox(10,8){\scriptsize$a$}}
\put(36,54){\makebox(10,8){\scriptsize$\app_2$}}
\put(16,44){\sdmc}
\put(28,44){\sdma}
\put(0,40){\makebox(8,8){\scriptsize$\ti_1$}}
\put(8,28){\sTtwo{\zeta_1^{-1}(z_1)}}
\put(0,24){\makebox(8,8){\scriptsize$\tf_1$}}
\put(14,24){\str}
\put(24,20){\sSone}
\put(40,20){\sdmd}
\put(0,16){\makebox(8,8){\scriptsize$\ti_2$}}
\put(20,4){\sTtwo{\zeta_2^{-1}(z_2)}}
\put(0,2){\makebox(8,8){\scriptsize$\tf_2$}}
\put(26,0){\str}
\put(38,0){\str}
\end{picture}
\end{minipage}
\:=\:
\begin{minipage}{24\un}
\begin{picture}(24,62)
\put(12,54){\makebox(10,8){\scriptsize$a$}}
\put(16,44){\sdma}
\put(0,40){\makebox(8,8){\scriptsize$\ti_1$}}
\put(8,28){\sCa{z_1}}
\put(0,24){\makebox(8,8){\scriptsize$\tf_1$}}
\put(12,20){\sSone}
\put(0,16){\makebox(8,8){\scriptsize$\ti_2$}}
\put(8,4){\sCb{z_2}}
\put(0,2){\makebox(8,8){\scriptsize$\tf_2$}}
\put(14,0){\str}
\end{picture}
\end{minipage}
\:=\:
\tr \cpm_{2,z_2}\circ\acpm^a_{[\tf_1,\ti_2)}\circ \cpm_{1,z_1}(\dm )\,,
\ee
which is what we wanted to show, as it agrees with \eqref{PZ2Z1}.


\begin{thebibliography}{29}

\bibitem{Adl07} Adler, S. L.:
  Lower and Upper Bounds on CSL Parameters from Latent Image 
  Formation and IGM Heating.
  \textit{Journal of Physics A: Mathematical and Theoretical} 
  \textbf{40}: 2935--2957 (2007).
  arXiv:quant-ph/0605072.

\bibitem{AAV} Aharonov, Y., Anandan, J., Vaidman, L.:
	Meaning of the wave function.
	\textit{Physical Review A} \textbf{47}: 4616--4626 (1993).

\bibitem{AZ05} Allori, V., Dorato, M., Laudisa, F., Zangh\`\i, N.: 
	\textit{La natura delle cose, introduzione ai fondamenti e alla 
	filosofia della fisica.} Rome: Carocci (2005).

\bibitem{AGTZ06} Allori, V., Goldstein, S., Tumulka R., Zangh\`\i, N.:
  On the Common Structure of Bohmian Mechanics and the
  Ghirardi--Rimini--Weber Theory. 
  \textit{British Journal for the Philosophy of Science} \textbf{59}: 353--389 (2008).
  arXiv:quant-ph/0603027.

\bibitem{grw3B}
  Allori, V., Goldstein, S., Tumulka, R., Zangh\`\i, N.:
  Predictions and Primitive Ontology in Quantum Foundations: A Study of Examples.
  arXiv:1206.0019.

\bibitem{BG03} Bassi, A., Ghirardi, G.C.: Dynamical Reduction Models.
  \textit{Physics Reports} \textbf{379}: 257--426 (2003). arXiv:quant-ph/0302164.
  
\bibitem{BGS06} Bassi, A., Ghirardi, G.C., Salvetti, D. G. M.:
  The Hilbert-Space Operator Formalism within Dynamical Reduction Models.
  \textit{Journal of Physics A: Mathematical and Theoretical} 
  \textbf{40}: 13755--13772 (2007).
  arXiv:0707.2940.

\bibitem{BS07}
  Bassi, A., Salvetti, D. G. M.:
  The Quantum Theory of Measurement Within Dynamical Reduction Models.
  \textit{Journal of Physics A: Mathematical and Theoretical} 
  \textbf{40}: 9859--9876 (2007).
  arXiv:quant-ph/0702011.

\bibitem{Bell87} Bell, J. S.: Are There Quantum Jumps? In C. W. Kilmister (ed.)
  \textit{Schr\"odinger. Centenary Celebration of a Polymath.} Cambridge:
  Cambridge University Press (1987), pp.~41--52. Reprinted as chapter 22 of
  \cite{Bell87b}.
  
\bibitem{Bell80} Bell, J. S.: De Broglie--Bohm, Delayed-Choice
   Double-Slit Experiment, and Density Matrix.
   \textit{International Journal of Quantum Chemistry} \textbf{14}:
   155--159 (1980). Reprinted as
   chapter 14 of \cite{Bell87b}.

\bibitem{Bell87b} Bell, J. S.: \textit{Speakable and Unspeakable in
    Quantum Mechanics}. Cambridge: Cambridge University Press (1987).

\bibitem{Bell89} Bell, J. S.: Toward An Exact Quantum
  Mechanics. In \textit{Themes in Contemporary Physics, II},
  S.~Deser and R.~J.~Finkelstein (eds.), p.~1--26.
  Teaneck, NJ: World Scientific (1989).
  
\bibitem{BGG95} Benatti, F., Ghirardi, G.C., Grassi, R.: 
	Describing the macroscopic world: closing the circle within
	the dynamical reduction program.
	\textit{Foundations of Physics} {\bf 25}: 5--38 (1995).

\bibitem{Bohm52} Bohm, D.: A Suggested Interpretation of the Quantum
  Theory in Terms of ``Hidden'' Variables, I and II. \textit{Physical
  Review} \textbf{85}: 166--193 (1952).

\bibitem{Cho75} Choi, M.: 
  Completely Positive Linear Maps on Complex Matrices. 
  \textit{Linear Algebra and its Applications} \textbf{10}: 285--290 (1975).

\bibitem{CDT05}
  Colin, S., Durt, T., Tumulka, R.:
  On Superselection Rules in Bohm--Bell Theories.
  \textit{J. Phys. A: Math. Gen.} \textbf{39}: 15403--15419 (2006).
  arXiv:quant-ph/0509177.
  
\bibitem{grw3C}
	Cowan, C. W., Tumulka, R.:
	Epistemology of Wave Function Collapse in Quantum Physics.
	In preparation.

\bibitem{Dav76}
  Davies, E. B.: 
  \textit{Quantum Theory of Open Systems}. 
  Academic Press (1976).

\bibitem{Dio89} Di\'osi, L.:
	Models for universal reduction of macroscopic quantum fluctuations.
	\textit{Physical Review A} \textbf{40}: 1165--1174 (1989).

\bibitem{Fay02} Dowker, F., Henson, J.: Spontaneous Collapse Models on a
  Lattice. \textit{Journal of Statistical Physics} \textbf{115}: 1327--1339 (2004).
  arXiv:quant-ph/0209051.

\bibitem{Fay03} Dowker, F., Herbauts, I.: 
  Simulating Causal Wave-Function Collapse Models. 
  \textit{Classical and Quantum Gravity} \textbf{21}: 1--17 (2004). 
  arXiv:quant-ph/0401075.

\bibitem{Fay04} Dowker, F., Herbauts, I.: 
  The Status of the Wave Function in Dynamical Collapse Models. 
  \textit{Foundations of Physics Letters} \textbf{18}: 499--518 (2005). 
  arXiv:quant-ph/0411050.

\bibitem{scatt} D\"urr, D., Goldstein, S., Teufel, S., Zangh\`\i, N.:
  Scattering theory from microscopic first principles.
  \textit{Physica A} \textbf{279}: 416--431 (2000).  arXiv:quant-ph/0001032.

\bibitem{dm} D{\"u}rr, D., Goldstein, S., Tumulka, R.,
  Zangh{\`{\i}}, N.: On the Role of Density Matrices in Bohmian Mechanics.
  \textit{Foundations of Physics} \textbf{35}: 449--467 (2005). 
  arXiv:quant-ph/0311127.

\bibitem{DGZ92} D\"urr, D., Goldstein, S., Zangh\`\i, N.: Quantum
  Equilibrium and the Origin of Absolute Uncertainty. \textit{Journal of
  Statistical Physics} \textbf{67}: 843--907 (1992). arXiv:quant-ph/0308039.

\bibitem{DGZ04} D\"urr, D., Goldstein, S., Zangh\`\i, N.: Quantum
  Equilibrium and the Role of Operators as Observables in Quantum 
  Theory. \textit{Journal of Statistical Physics} \textbf{116}: 959--1055 (2004).
  arXiv:quant-ph/0308038.

\bibitem{DGZ08} D\"urr, D., Goldstein, S., Zangh\`\i, N.: 
  On the Weak Measurement of Velocity in Bohmian Mechanics.
  \textit{Journal of Statistical Physics} \textbf{134}: 1023--1032 (2009).
  arXiv:0808.3324.

\bibitem{FT12} Feldmann, W., Tumulka, R.:
  Parameter Diagrams of the GRW and CSL Theories of Wave Function Collapse.
  \textit{Journal of Physics A: Mathematical and Theoretical} \textbf{45}: 065304 (2012).
  arXiv:1109.6579.

\bibitem{GPR90} Ghirardi, G. C., Pearle, P., Rimini, A.: Markov 
  processes in Hilbert space and continuous spontaneous localization 
  of systems of identical particles.  
  \textit{Physical Review A (3)} \textbf{42}: 78--89 (1990).
  
\bibitem{GRW86} Ghirardi, G. C., Rimini, A., Weber, T.: Unified
  Dynamics for Microscopic and Macroscopic Systems. \textit{Physical Review
    D} \textbf{34}: 470--491 (1986).

\bibitem{Gol98} Goldstein, S.: Quantum Theory Without Observers.
  \textit{Physics Today}, Part One: March 1998, 42--46. 
  Part Two: April 1998, 38--42.

\bibitem{GTTZ05b}
  Goldstein, S., Taylor, J., Tumulka, R., Zangh\`\i, N.:
  Are all particles real?
  \textit{Studies in History and Philosophy of Modern
      Physics} \textbf{36}: 103--112 (2005).  arXiv:quant-ph/0404134.

\bibitem{JPR04}
  Jones, G., Pearle, P., Ring, J.:
  Consequence for Wavefunction Collapse Model of the Sudbury 
  Neutrino Observatory Experiment.
  \textit{Foundations of Physics} \textbf{34}: 1467--1474 (2004). arXiv:quant-ph/0411019.

\bibitem{kent} Kent, A.: ``Quantum Jumps'' and Indistinguishability.
  \textit{Modern Physics Letters A} \textbf{4(19)}: 1839--1845 (1989).

\bibitem{kraus} Kraus, K.: \textit{States, Effects, and Operations}.
   Berlin: Springer (1983).

\bibitem{Mau05} Maudlin, T.: Non-Local Correlations in Quantum Theory:
  Some Ways the Trick Might Be Done. 
  In W.~L.~Craig and Q.~Smith (ed.s),
  \textit{Einstein, Relativity, and Absolute Simultaneity},
  London: Routledge (2008).

\bibitem{Pe89} Pearle, P.: Combining stochastic dynamical state-vector 
   reduction with spontaneous localization. \textit{Physical Review A} \textbf{39}: 
   2277--2289 (1989).

\bibitem{PS94} Pearle, P., Squires, E.: Bound State Excitation, Nucleon 
   Decay Experiments and Models of Wave Function Collapse.
   \textit{Physical Review Letters} \textbf{73}: 1--5 (1994).
   
\bibitem{Pen00} Penrose, R.: Wavefunction Collapse As a Real
   Gravitational Effect. In A. Fokas, T. W. B. Kibble, A. Grigoriou, 
   B. Zegarlinski (editors), \textit{Mathematical Physics 2000},
   pp. 266--282.
   London: Imperial College Press (2000).

\bibitem{Pen04} Penrose, R.: \textit{The Road to Reality.}
	London: Random House (2004).

\bibitem{PR84} 
Penrose, R., Rindler, W.:  
{\em Spinors and space-time. Vol. I: Two-spinor calculus and
  relativistic fields.} 
Cambridge: University Press (1984). 

\bibitem{Rae90} Rae, A.I.M.:
	Can GRW theory be tested by experiments on SQUIDS?
	\textit{Journal of Physics A: Mathematical and General} 
	\textbf{23}: L57--L60 (1990)

\bibitem{Tum04} Tumulka, R.: A Relativistic Version of the
  Ghirardi--Rimini--Weber Model.  
  \textit{Journal of Statistical Physics} \textbf{125}: 821--840 (2006). 
  arXiv:quant-ph/0406094.

\bibitem{Tum05} Tumulka, R.: 
  On Spontaneous Wave Function Collapse and Quantum Field Theory. 
  \textit{Proceedings of the Royal Society A} \textbf{462}:  
  1897--1908 (2006). arXiv:quant-ph/0508230.

\bibitem{Tum06c} 
  Tumulka, R.: 
  Collapse and Relativity.
  In A. Bassi, D. D\"urr, T. Weber, and N. Zangh{\`{\i}} (eds.), 
  \textit{Quantum Mechanics: Are there Quantum Jumps? and
  On the Present Status of Quantum Mechanics}, 
  AIP Conference Proceedings \textbf{844}, 340--352. American Institute of Physics
  (2006). 
  arXiv:quant-ph/0602208.

\bibitem{Tum06d}
  Tumulka, R.:
  The `unromantic pictures' of quantum theory.
  \textit{Journal of Physics A: Mathematical and Theoretical} 
  \textbf{40}: 3245--3273 (2007). arXiv:quant-ph/0607124.

\bibitem{Tum07b} Tumulka, R.: A Kolmogorov Extension Theorem for POVMs.
	\textit{Letters in Mathematical Physics} \textbf{84}: 41--46 (2008).
	arXiv:0710.3605.
	
\bibitem{Tum07}
  Tumulka, R.:
  The Point Processes of the GRW Theory of Wave Function Collapse.
	\textit{Reviews in Mathematical Physics} \textbf{21}: 155--227 (2009).
  arXiv:0711.0035.

\bibitem{Vac07} Vacchini, B.:
	On the precise connection between the GRW master equation and 
	master equations for the description of decoherence.
	\textit{Journal of Physics A: Mathematical and Theoretical} 
	\textbf{40}: 2463--2473 (2007).

\end{thebibliography}
\end{document}